\documentclass[12pt,twoside,openany,notitlepage]{extbook}  
\usepackage[text={144mm,218mm},centering,papersize={169mm,259mm},dvips]{geometry}
\lccode`\-=`\-         
\usepackage[cp1251]{inputenc}
\usepackage[T2A]{fontenc}

\usepackage[T1]{fontenc}

\usepackage[english,russian]{babel}
\usepackage{citehack}
\usepackage{epsf}
\usepackage{eqalign}
\usepackage{rotating}

\usepackage{amssymb}
\usepackage{amsbsy}
\usepackage{indntfst}
\usepackage[shorttoc]{miscdoublecorr}
\usepackage{lastpage}
\usepackage[small]{caption2}   

\usepackage{esint}

\usepackage{epsfig}

\usepackage{graphics}
\usepackage{graphicx}

\usepackage{color}
\usepackage{hyperref}
\usepackage{colordvi}

\usepackage{wrapfig}

\usepackage{literat} 

\def\url#1{\mbox{{#1}{\tt #1}}}

\usepackage{fancyhdr}
\renewcommand{\chaptermark}[1]{\markboth{\footnotesize\sc\chaptername \ \thechapter.\ #1}{\quad\quad \footnotesize\sc\chaptername \ \thechapter.\ #1}}

\renewcommand{\headrulewidth}{0.5pt}
\renewcommand{\footrulewidth}{0pt}
\fancypagestyle{plain}{%
   \fancyhead{} %
   \renewcommand{\headrulewidth}{0pt} %
}
\renewcommand{\captionlabeldelim}{.} 

\RequirePackage{ifthen}
\newboolean{mik@@@}\setboolean{mik@@@}{true}
\newboolean{upg@@@}\setboolean{upg@@@}{false}
\def\enc@@@{0}
\def\fnt@@@{0}
\newboolean{mak@@@}
\relax

\ifthenelse{\boolean{mik@@@}}{\def\ot@lru{OT1}}{\def\ot@lru{LRU}}
\begingroup
\catcode`\+\active\gdef+{\mathchar8235\nobreak\discretionary{}%
{\usefont{\ot@lru}{cmr}{m}{n}\char43}{}}
\catcode`\-\active\gdef-{\mathchar8704\nobreak\discretionary{}%
 {\usefont{OMS}{cmsy}{m}{n}\char0}{}}
\catcode`\=\active\gdef={\mathchar12349\nobreak\discretionary{}%
{\usefont{\ot@lru}{cmr}{m}{n}\char61}{}}
\endgroup
\def\cdot{\mathchar8705\nobreak\discretionary{}%
 {\usefont{OMS}{cmsy}{m}{n}\char1}{}}
\def\times{\mathchar8706\nobreak\discretionary{}%
 {\usefont{OMS}{cmsy}{m}{n}\char2}{}}
\mathcode`\==32768 \mathcode`\+=32768 \mathcode`\-=32768

\begin{document}
\renewcommand{\figurename}{\cyr\CYRR\cyri\cyrs.}
\renewcommand{\theequation}{\arabic{chapter}.\arabic{equation}}
\renewcommand{\thefigure}{\arabic{chapter}.\arabic{figure}}
\renewcommand{\thesection}{\arabic{chapter}.\arabic{section}}
\renewcommand{\contentsname}{\cyr\CYRO\cyrg\cyrl\cyra\cyrv\cyrl\cyre\cyrn\cyri\cyre}
\newcommand{\grad}{{\rm {grad}}}
\newcommand{\rot}{{\rm {rot}}}
\newcommand{\sgn}{{\rm {sgn}}}
\newcommand{\Alpha}{{\rm {A}}}
\newcommand{\Beta}{{\rm {B}}}

\language=0

\newcommand{\beq}{\begin{eqnarray}}
\newcommand{\eeq}{\end{eqnarray}}

\newcommand{\bea}{\begin{eqnarray}}
\newcommand{\eea}{\end{eqnarray}}

\renewcommand{\f}{\frac}
\newcommand{\y}{_\infty}
\newcommand{\p}{\partial}
\def\tb{\textBlack}
\def\tr{\textBlue}
\def\trr{\textCyan}
\renewcommand{\captionlabeldelim}{.}
\newcommand{\ff}{\frac}

\def\emline#1#2#3#4#5#6{%
       \put(#1,#2){\special{em:moveto}}%
       \put(#4,#5){\special{em:lineto}}}
\def\newpic#1{}

\global\long\def\pp{\vec{p}}
\global\long\def\lm{\lim_{a\to0}}
\global\long\def\R{\mathbb{R}}
\global\long\def\d{\mathrm{d}}
\global\long\def\i{\mathrm{i}}

\newpage
\thispagestyle{empty}

\null
$\,$
\vspace{-22.5mm}

\begin{center}
MINISTRY OF EDUCATION AND SCIENCE
OF
RUSSIAN FEDERATION\\[1mm]
SAMARA STATE UNIVER\-SITY
\vspace{10mm}

\vspace{19mm}

{\LARGE\bf{
MATHEMATICAL PHYSICS 
}}\\[5mm]
 {\Large\bf{PROBLEMS AND SOLUTIONS}}\\[5mm]

\vspace{32mm}
\vspace{-91mm}
\vspace{70mm}

{\sc The Students Training Contest Olympiad\\
in Mathematical and Theoretical Physics\\
({\small on May 21st -- 24th, 2010})\\[26mm]}
\end{center}
\begin{center}
{\bf Special Issue № 3 of the Series\\[1mm]
«Modern Problems of Mathematical Physics»}
\end{center}

\vspace{15mm}

\vspace{13mm}


\vspace{15mm}
\begin{center}
Samara\\
Samara University Press\\
2010
\end{center}

\newpage
\thispagestyle{empty}
$\,$
\vspace{-1.3cm}

$\,$\hspace{-18mm}УДК 51-7+517.958

$\,$\hspace{-18mm}ББК 22.311\\
         М 34\\

{\bf Authors}:\\
G.S. Beloglazov, A.L. Bobrick, S.V. Chervon,
B.V.~Danilyuk, M.V.~Dolgopolov,\\
M.G.~Ivanov, O.G. Panina, E.Yu. Petrova, I.N. Rodionova, E.N. Rykova,\\ M.Y.~Shalaginov, I.S. Tsirova,  I.V. Volovich, A.P. Zubarev\\[-1mm]

$\,$\hspace{-18mm}М 34 \hspace{1.9mm}\,\phantom{y} {\bf Mathematical Physics
:} Problems and Solutions of The Students Training Contest Olympiad
in Mathematical and Theoretical Physics
(May 21st -- 24th, 2010) / [G.S.\,Be\-loglazov et al.].
 --
Ser. «Modern Problems of Mathematical Physics».
 -- Spec. Iss. № 3.  -- Samara : Samara University Press, 2010. -- 68\,p.:~il.

ISBN 978-5-86465-494-1

\vspace{3mm}

{\footnotesize


The present issue of the series <<Modern Problems in Mathematical Physics>>
represents the Proceedings of the Students Training Contest Olympiad in Mathemati\-cal and Theoretical Physics and includes
the statements and solutions of the problems offered to the
participants. The contest Olympiad was held on May 21st-24th, 2010 by Scientific Research Laboratory of Mathe\-ma\-ti\-cal Physics of Samara State University, Steklov Mathematical Institute of Russia's Academy of Sciences, and Moscow Institute of Physics and Technology (State University) in cooperation.}

{\footnotesize The subjects covered by the problems include classical mechanics, integrable
nonlinear systems, probability, integral equations, PDE, quantum and particle
physics, cosmology, and other areas of mat\-hematical and theoretical physics.}

{\footnotesize The present Proceedings is intended to be
used by the students of physical and mechanical-ma\-the\-ma\-tical departments
of the universities, who are interested in acquiring a deeper knowledge of the~met\-hods of mathemati\-cal and theoretical physics, and could be also useful for the persons involved in teaching mathemati\-cal and theoretical physics.}
\vspace{-4mm}
\begin{flushright}
УДК 51-7+517.958\\
ББК  22.311~\,\phantom{yyyyyy}
\end{flushright}

\vspace{-1mm}

 {\bf Editors:}
B.V. Danilyuk,
M.V. Dolgopolov,
M.G. Ivanov,
I.S. Tsirova, I.V. Volovich\\[-1mm]

{\bf Invited reviewers}:
\, M.\,N. Dubinin, Skobeltsyn Institute of Nuclear Physics of Moscow State University, \, and \, Yu.\,N. Radayev, Institute for Problems in Mechanics of the Russian Academy of Sciences\\
\begin{center}
\vspace{-1mm}
{\footnotesize\it The Olympiad and the given edition are supported by the grants\\ ADTP~№~\,3341, 10854 and FTP~№~5163\\ of the Ministry of Education and Science of the
Russian Federation,\\ and by Training and retrainings of
specialist center\\
of Samara State University.}\\[3mm]
{\it Information support on the
website}
~ www.labmathphys.samsu.ru/eng\\
\end{center}
\setcounter{chapter}{0} \setcounter{section}{0}
\setcounter{equation}{0}


\vspace{0.1cm}
$\,$\hspace{-18mm}{\bf\footnotesize ISBN 978-5-86465-494-1
}
{\small
~\, \hspace{1mm}
~~~~\,\,~\,\,\,\,\,\hspace{-0.7mm} © Authors, 2010

\qquad \qquad \qquad \qquad \qquad © Samara State University, 2010

\qquad \qquad \qquad \qquad \qquad © Scientific Research Laboratory of
Mathematical Physics, 2010

\qquad \qquad \qquad \qquad \qquad ©
Registration.  Samara University Press,
2010}

\def\contentsname{Contents}
\tableofcontents

\fancyhead[LO]{\footnotesize\it
Математическая физика:
~ оглавление
}
\fancyhead[RE]{\footnotesize\it
Сер. <<Современные проблемы математической физики>>.
Спец. вып. № 3}

\section*{ }
\fancyhead[LO]{\footnotesize\it
\center Mathematical Physics:
~ Introduction
}
\fancyhead[RE]{\footnotesize\it
\center Modern Problems of Mathematical Physics. \,
Special Issue № 3}

\def\figurename{Рис}
\renewcommand{\theequation}{\arabic{equation}}
\renewcommand{\thefigure}{\arabic{figure}}

\def\chaptername{ }
\renewcommand{\thefigure}{\arabic{figure}}
\renewcommand{\theequation}{\arabic{equation}}

\language=0

\chapter*{Introduction}
\addcontentsline{toc}{chapter}{Introduction}
\vspace{-10mm}
\begin{center}
\rule{11cm}{0.3mm}
\end{center}
\vspace{-0.1cm}





1 ~ Regulations on The Olympiad\\

Regulations on holding The Olympiad contest for students on Mathematical and Theoretical Physics were developed in April 2010 [see Special Issue No. 2]. They~were signed by the three parties: Samara State University (hereinafter referred to as SamGU),
Steklov Mathematical Institute (SMI RAS), and Moscow Institute
of Physics and Technology (MIPT).
The text of the regulations is given below.\\

2 ~ Carrying out The Olympiad\\

On May 21-24th, 2010, All-Russian Student Training Olympiad in Mathematical and Theoretical Physics "Mathematical Physics" \,with International Participation has been held. It was the second in the series of Olympiads.  It is planned that in future such Olympiads will take place annually.

The organizers of the series of Olympiads on Mathematical \& Theoretical Physics "Mathematical Physics" \,are:

Aleksander Anatolyevitch Andreyev (staff member of the Scientific Research Laboratory of
Mathematical Physics of SamGU),

Georgiy Sergeyevitch Beloglazov (The University of Dodoma - UDOM, Tanzania;
Perm State Pharmaceutical Academy),

Boris Vasilyevitch Danilyuk (staff member of the Scientific Research Laboratory of~Mathematical Physics of SamGU),

Mikhail Vyacheslavovitch Dolgopolov (Head of the Scientific Research Laboratory of
Mathematical Physics of SamGU),

Vitaliy Petrovitch Garkin (Vice-rector for Academic Affairs,
Chairman of the Local Organizing Committee, SamGU),

Mikhail
Gennadievich Ivanov (Associate Professor, MIPT),

Yuri Nikolayevitch Radayev (staff member of the Scientific Research Laboratory of~Mathematical Physics of SamGU),

Irina Nikolayevna Rodionova (staff member of the Scientific Research Laboratory of
Mathematical Physics of SamGU),

Yuri Aleksandrovitch Samarskiy (Deputy Vice Chancellor on Education, MIPT),

Irina Semyonovna Tsirova (docent, SamGU),

Igor Vasilyevitch Volovich (scientific leader of the Scientific Research Laboratory of
Mathematical Physics of SamGU, head of the department of Mathematical Physics of MIAN),

Aleksander Petrovitch Zubarev (staff member of the Scientific Research Laboratory of
Mathematical Physics of SamGU).











The Olympiad has been held as a team competition. Number of participants of~each team -- from 3 to 10 students of 2nd to 6th courses (years) of higher educational establishments of Russia, CIS, and other countries. It was allowed that more than one team participates on behalf of any organization.
Order of the Olympiad:

The participants have been offered to solve 14 problems.
Time to start solving problems of the contest was
11:00\,pm
Moscow time on May 20th, 2010.
The statements of the contest tasks are published in *.pdf format at the webpage of the Olympiad
www.labmathphys.samsu.ru/eng/content/view/29/36/ of the website of the Scientific Research Laboratory of
Mathematical Physics of SamGU
\begin{center}www.labmathphys.samsu.ru/eng\end{center}
and have been sent to the registered participants of The Olympiad.

The deadline to send the scanned (or photographed) solutions to the E-mail address of the Mathematical Physics Laboratory: slmp@ssu.samara.ru was
11\,pm
Mos\-cow time on May 24th, 2010.
All participants of The Olympiad who had sent their solutions by E-mail, have received confirmation
that
their solutions
had been accepted.

It was allowed that the participants solve any problems from the number of the proposed ones which they find affordable for the own level of knowledge digestion in different units of mathematics and physics thus participating in the topical scoring nomination (for purpose of this scoring nomination, the problems are aggregated into groups 1 to 3 problems in each).

In the application letter, the name of organization hosting the team should be stated together with the surname, name,
for each participant of the team, Department (speciality), course/year; contact E-mail address.

The Nominations of The Olympiad:

1) The overall team scoring based on the three best team participants performance (3 prize-winning team places). In the present Olympiad, it is possible to submit only one solution on behalf of a team; it is advised to mention the author(s) of every solution or solution method [stating also the year(s)/course(s) of
studying] at the end of~each solution (or method of solution). The winner is the team which participants have solved correctly maximum number of different problems. Any participant of a team has the right to send a solution separately. Within the team scoring, the correct solutions will be considered and accounted. The maximum possible number of points in a team scoring is 14 (because the total number of problems offered is 14).

2) It is possible for a student to participate in the overall personal contest (within the framework of the Olympiad by correspondence) ON CONDITION OF THE PRESENCE OF AN INDIVIDUAL APPLICATION (REQUEST) from a participant of The Olympiad (3 prize-winning places).

3) Overall team topic scoring (1 -- 3 prize-winning places on each subject).

4) Separate team scoring among each of the years (second through sixth courses).

5) Best team among the technical specialities of the institutes of higher education.

6) Other nominations.
Separate nomination is supported by the Center on Advanced Training and Professional Development at Samara State University.

The winners of The Olympiad held by correspondence
participated in the day competition Olympiad
held in Samara
in September - 2010 (at the same time with the Second International Conference and School on Mathematical Physics and its Applications). For the above said winners, their travel and/or accommodation expenses
were reimbursed.\\

3 ~ Contents of the problems for the Olympiad contest\\

The topic range of our 'Olympiad' is related to mathematical methods in describing physical phenomena based on the following units of mathematics and theoretical physics:

theory of differential, integral equations, and boundary-value problems;

theory of generalized functions, integral transform, theory of functions of complex variable;

functional analysis, operational calculus, spectral analysis;

probability theory, theory of random processes;

differential geometry and topology;

theoretical mechanics, electrodynamics, relativity theory, quantum mechanics, and gravitation theory.








New scientific methodological approach to composing the statements of the problems for  The Olympiad was first introduced in the sense that about a half of the problems offered to the participants for the solution supposed that certain stage of research (taken from original modern academic research in mathematical physics and its applications) is involved. On the basis of the above mentioned approach, the recommendations on composing statements of the problems for The Olympiad are developed.

In the present issue, we quote the statements of problems offered to the participants of All-Russia Students Training Olympiad in Mathematical and Theoretical Physics "Mathematical Physics" \,with International Participation (held on May 21-24th, 2010).\\

4 ~ Results and resume of The Olympiad\\

In The Olympiad, the teams from the following institutes of higher education and other organizations have participated:

Belarusian State university,

Moscow Institute of Physics and Technology (State University),

National University of Singapore,

Department of Theor. Phys. named after I.E.Tamm of FIAN (the Institute of Phy\-sics of Academy of Sciences of Russia) - postgraduate,

Samara State University of Architecture and Construction (two teams),

Samara State Aerospace University (SGAU),

Samara State University,

The Federal University of Siberia,

Ulyanovsk State Pedagogical University, UlGPU (two teams),

The University of Dodoma (UDOM, Tanzania),

Yaroslavl State University (the team of the Physics Department).

The jury has positively assessed the works by the following participants of the teams:

Belarusian State University: Alexey Bobrick.

Moscow Institute of Physics and Technology (State University): Kostjukevich Yury, and the fourth course team: Nikolai Fedotov, Anton Fetisov, Mikhail Shalaginov, Aleksander Shtyk.

Department of Theor. Phys. named after I.E.Tamm of FIAN (the Institute of Physics of Academy of Sciences of Russia): Andrey Borisov.

Samara State University of Architecture and Construction, SGASU (two teams of the students of the 5th year): leader -- Sergey Zinakov.

Samara State Aerospace University: Mikhail Malyshev, Yekaterina Pudikova.

Samara State University: team of theoretical physicists -- Tatiana Volkova, Matvei Mashchenko, Maksim Nefedov, Yelena Petrova.

The Federal University of Siberia: Artyom Ryasik, Polina Syomina, Anton Sheykin.

Ulyanovsk State Pedagogical University (two teams): 3rd year -- Yuri Antonov, Aleksandra Volkova, Oksana Rodionova;

5th year: Maria Vasina, Artyom Ovchinnikov. Aleksander Chaadayev, Aleksander Ernezaks.

The diploma of Laureates or diploma of the winners in nominations have been sent to all above mentioned participants. All participants of The Olympiad have been invited to attend the School-2010 on Applied Mathematical Physics (PMF) from July 1st till July 14th, and the scientific Conference together with another School \& Olympiad (August 29th - September 9th, 2010).

The Winners of The Olympiad in the nominations:

The Overall Team Score:

1st Place, 5 problems solved correctly (means, 8 and more points per a problem, maximum 10 points per a problem), the total score is 102 points, - the team of the 4th year of MIPT. The winners are granted the prize - traveling costs be paid for them to participate in the scientific Conference together with School \& Olympiad (August 29th - September 7th, 2010).

2nd Place - The Federal University of Siberia

3rd Place - Samara State University

The Total Personal Score:

1st Place, 4 problems solved correctly, score is 126 points, -- Alexey Bobrick, Theoretical Physics magistracy  at the Department of Physics of Belarusian State University. The winner is granted the prize - either traveling or accommodation costs be paid for him to participate in the scientific Conference together with School \& Olympiad (August 29th - September 7th, 2010).

2nd Place, 3 problems solved correctly, score is 74 points, -- Yuri Kostyukevitch, the student of the 5th year of the Department of Molecular and Biological Physics, group No. 541, MIPT. Recommended for the Magistracy or (post)graduate school of MIPT.

3rd Place, 2 problems solved correctly, score is 70 points, -- Anton Sheykin, the student of the 4th year at Engineering Physical Department of IIFiRE (Physics and Radioelectronics) of Siberian Federal University. Recommended for the Magistracy or (post)graduate school of MIPT.

The best (complete and original) solutions of separate problems:
by Alexey Bobrik, Polina Syomina, Mikhail Shalaginov, Anton Sheykin.

1st Team Place among the 3rd year students -- SamGU;

2nd Team Place among the 3rd year students -- SGAU.

1st Place in
Personal contest among the 3rd year students -- Maksim Nefedov;

2nd Place in Personal contest among the students of the 3rd year students -- Mikhail Malyshev.

1st Place in Personal contest among the 4th year students -- Anton Sheykin;

2 -- 3 Places in Personal contest among the students of the 4th year -- Nikolai Fedotov, Anton Fetisov, and Mikhail Shalaginov.

Among the teams of Pedagogical, Engineering \& Technical institutes of higher education:

1st Place -- UlGPU, 3 year;

2nd Place -- UlGPU, 5 year;

3rd Place -- SGASU.

All winners and prize winners of The Olympiad are granted with the free of charge accommodation at the PMF School-2010.


\vspace{3mm}

{\it The information on the Olympiad,
formu\-la\-tion of the Problems-2010 statements, answers and
solutions of
the tasks-2010
are presented in
this
document.}
\vspace{-4mm}
\begin{flushright}
~ www.labmathphys.samsu.ru/eng ~ ~~ ~~ ~~ slmp@ssu.samara.ru
\end{flushright}


\begin{flushright}
{\small Organizers of a series of the Mathematical Physics Olympiads:
Alexander Andreev,
George~Beloglazov,
Boris Danilyuk,
Mikhail Dolgopolov,
Vitaliy Garkin,
Mikhail Ivanov,
Yury~Radaev,
Irina Rodionova,
Yury Samarsky,
Irina Tsirova,
Igor Volovich,
Alexander~Zubarev}
\end{flushright}


\chapter*{Problems and Solutions }
\addcontentsline{toc}{chapter}{Problems and Solutions}
\vspace{-5mm}
\begin{center}
\rule{11cm}{0.3mm}
\end{center}
\vspace{0.5cm}

\fancyhead[LO]{\footnotesize\sc\leftmark}
\fancyhead[RE]{\footnotesize\sc\leftmark}

\fancyhead[LO]{\footnotesize\it \center Mathematical Physics:
~ Problems and Solutions}
\fancyhead[RE]{\footnotesize
\it \center
Modern Problems of Mathematical Physics. \,
Special Issue № 3}


\begin{center}
{\bf Statements of the Problems\\ and Solutions at
Students Training Olympiad \\ on Mathematical \& Theoretical Physics\\ {\large MATHEMATICAL PHYSICS} \\ by Correspondence \\ with International Participation \\ May 21st -- 24th, 2010 \\[4mm]}
\end{center}

\section*{1. Virial for anharmonic oscillations}
\addcontentsline{toc}{section}{1. VIRIAL FOR ANHARMONIC OSCILLATIONS}
For
a particle
moving along the $x$ axis with Hamiltonian
$$ H=\frac{p^2}{2m}+\lambda x^{2n},$$
where $\lambda $ is a positive constant, $m$ is the mass of the particle, $p$ is
the momentum of the particle,
$n=1,\ 2,\ 3,\ \ldots
\ ,$ obtain the relationship between the average
values of kinetic~$\langle K\rangle$ and potential energy $\langle U\rangle$ using two methods:\\
\textbf{(a)} directly from the virial theorem {\it (see explanation below)};\\ 
\textbf{(b)} from the condition
$$\left\langle \frac{d}{dt}(xp)\right\rangle=0,$$
which is true due to the fact that the motion of the particle is finite.

\underline{Instruction}: when a particle moves in a potential field, its Hamiltonian $H$ and acting force $\vec F$ are defined by
$$H=K+U,\qquad \, \vec F =-\mbox{grad} \ U.$$

{\it{\bf\itshape Explanation to Problem 1}. In classical mechanics time average values of kinetic and potential
energies of the systems performing finite
motion  are in
rather simple relationship.

The average value for a physical quantity $G$ for a sufficiently large time interval
$\tau $ is defined in a standard way:
$$\langle G\rangle\,=\frac{1}{\tau}\int\limits_0^{\tau}G\ dt.$$

If $\langle K\rangle$ is the average (for a rather long time interval) kinetic energy of the~system of point
particles
(radius-vectors of the particles given as $\vec r_i$)
subjected to
forces
$\vec F_i$, then the following relation takes place:
$$\langle K\rangle\,=-\,\frac{1}{2}\left\langle \sum\limits_i\vec F_i \cdot \vec
r_i\right\rangle. \eqno{(\blacklozenge)}$$
The right hand side of
equation ($\blacklozenge$) is called
\textit{Clausius virial}, and the equation itself expresses the
so called the \textit{virial Theorem}. The proof of the theorem is given,
for example, in}~\cite{tms}.\\

{\bf SOLUTION}

\setcounter{equation}0

\textbf{(a)} According to the given statement the particle is moving in the field of a potential force
and possesses potential energy $U(x)=\lambda x^{2n}.$ The
force
equals
$$F_x=-\frac{dU}{dx}=-2\lambda nx^{2n-1}.$$
Substitute
it into the equation ($\blacklozenge$) which expresses the virial theorem:
$$\langle K\rangle =-\frac{1}{2}\langle\vec F \cdot \vec
r\rangle =-\frac{1}{2}\langle F_xx\rangle
=\frac{1}{2}\langle2\lambda nx^{2n-1}x\rangle =n\langle\lambda
x^{2n}\rangle =n\langle U\rangle .$$

\textbf{(b)} According to the given conditions,
$$ 0=\Big\langle\frac{d}{dt}(xp)\Big\rangle
=\Big\langle\frac{dx}{dt}p+x\frac{dp}{dt}\Big\rangle
=\Big\langle\frac{p}{m}p+xF_x\Big\rangle
=\Big\langle\frac{p^2}{m}+x(-\lambda 2nx^{2n-1})\Big\rangle =$$
$$=\Big\langle\frac{p^2}{m}\Big\rangle -2n\langle\lambda x^{2n}\rangle =2\langle K\rangle -2n\langle U\rangle,$$
that is why
$$\langle K\rangle =n\langle U\rangle.\\$$

\section*{2. Method of successive approximations}

\setcounter{equation}0
\addcontentsline{toc}{section}{2. METHOD OF SUCCESSIVE APPROXIMATIONS}

Solve the integral Volterra equation of 2nd kind
\beq
\varphi(x)=\frac{\alpha{\bf'}(x)}{1-\alpha(x)}\int\limits_0^x\varphi(t)dt+f(x),
\label{en2-z1}
\eeq
where $x\in[0,h]$, $f(x)$ is a given (known) continuous on $[0,h]$ function, $\alpha(x)\in C^{1}{[0,h]}$ (continuously differentiable function), and $\alpha(x)\neq1$, $\alpha{\bf'}(x)$ is the derivative. Perform your solution check.\\
\parbox{0.5\textwidth}{{\it\bf Vito Volterra} ~ (3 May 1860 –- 11 October 1940) ~ was an Italian mathematician and physicist, known for his contributions to mathematical biology and integral equations.}
\hspace{12mm}
\parbox{0.24\textwidth}{\includegraphics[scale=0.5]{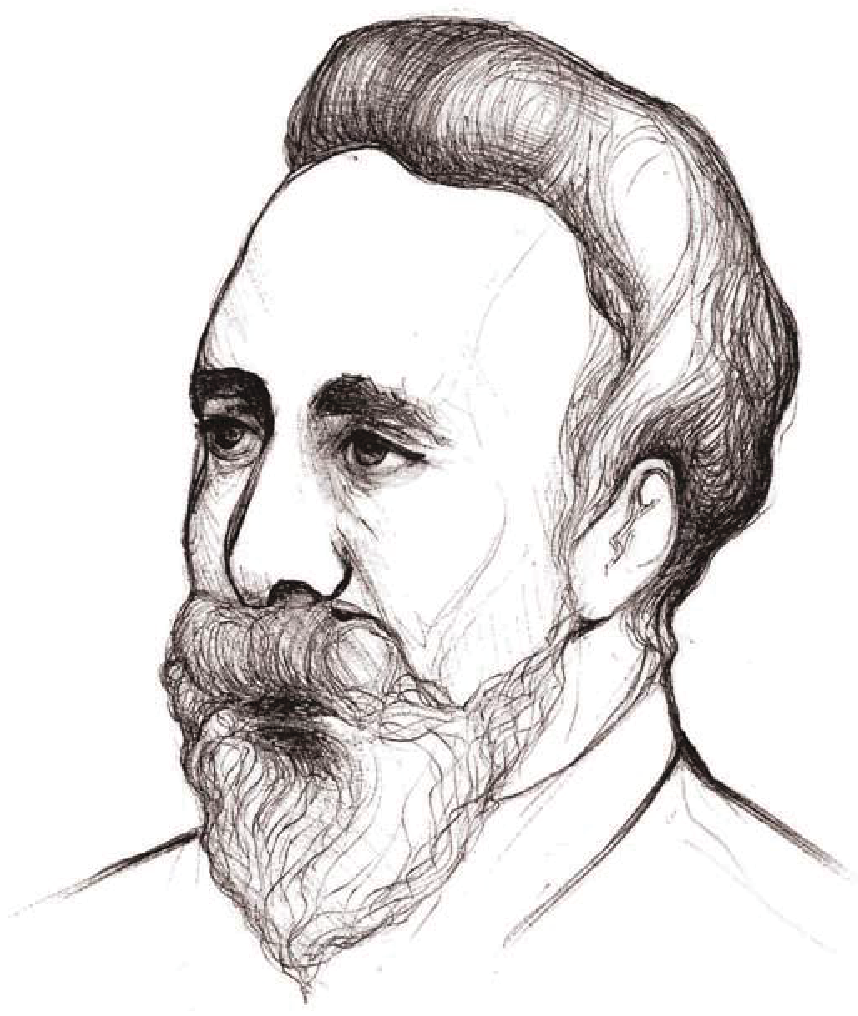}
}

\vspace{1mm}

{\bf SOLUTION}

Using method of successive approximations, we find the solution of the equation~(\ref{en2-z1})
through kernel resolvent $\displaystyle K(x,t)=\frac{\alpha'(x)}{1-\alpha(x)}$:
\beq
\varphi(x)=\int\limits_0^x R(x,t)f(t)dt+f(x),
\label{22e}
\eeq
\beq
R(x,t)=\sum_{n=1}^\infty K_n(x,t),
\label{23e}
\eeq
\beq
K_1(x,t)=K(x,t),
\label{24}
\eeq
\beq
K_n(x,t)=\int\limits_t^xK_1(x,s)K_{n-1}(s,t)ds.
\label{25e}
\eeq
Using
formula~(\ref{25e}) we find the
repeated
kernel $K_2(x,t)$, $K_3(x,t)$
\beq
K_2(x,t)=\int\limits_t^x\frac{\alpha'(x)}{1-\alpha(x)} \frac{\alpha'(s)}{1-\alpha(s)}ds=
\label{en2-26}
\eeq
$$
=\frac{\alpha'(x)}{1-\alpha(x)}[\ln(1-\alpha(t))-\ln(1-\alpha(x))],
$$
\beq
K_3(x,t)=\frac{\alpha'(x)}{1-\alpha(x)}\int\limits_t^x \frac{\alpha'(s)}{1-\alpha(s)}[\ln(1-\alpha(t))-\ln(1-\alpha(s))]ds=
\label{en2-27}
\eeq
$$
=\frac{\alpha'(x)}{1-\alpha(x)}\left[\frac{\ln^2(1-\alpha(t))}{2}-\ln(1-\alpha(t))
\ln(1-\alpha(x))+\frac{\ln^2(1-\alpha(x))}{2}\right]=
$$
$$
=\frac{\alpha'(x)}{2!(1-\alpha(x))}[\ln(1-\alpha(t))-\ln(1-\alpha(x))]^2.
$$

Similarly, using the formula~(\ref{25e}),
\beq
K_4(x,t) = \frac{\alpha'(x)}{3!(1-\alpha(x))}[\ln(1-\alpha(t))-\ln(1-\alpha(x))]^3,
\label{28}
\eeq
we come to the conclusion that
\beq
K_n(x,t) = \frac{\alpha'(x)}{(n-1)!(1-\alpha(x))}\ln^{n-1}\left(\frac{1-\alpha(t)}{1-\alpha(x)}\right).
\label{29e}
\eeq

Expression~(\ref{29e}) should be substituted into the formula~(\ref{23e}):
\beq
R(x,t)=\sum_{n=1}^\infty\frac{\alpha'(x)}{1-\alpha(x)}
\frac{\ln^{n-1}\left(\frac{1-\alpha(t)}{1-\alpha(x)}\right)}{(n-1)!}.
\label{210}
\eeq

Performing the transformation $n-1=m$ in (\ref{210}) and
recalling the expansion
\beq
e^z=\sum_{m=0}^\infty\frac{z^m}{m!},
\label{211}
\eeq
we obtain
\beq
R(x,t)=\frac{\alpha'(x)(1-\alpha(t))}{(1-\alpha(x))^2}.
\label{212}
\eeq
Substituting (\ref{212}) into formula~(\ref{22e}), we obtain
\beq
\varphi(x)=f(x)+\frac{\alpha'(x)}{(1-\alpha(x))^2}\int\limits_0^x f(t)
[1-\alpha(t)]dt.
\label{213}
\eeq

\underline{Checking}.
Let us show that the function~(\ref{213}) is the solution of eq.~(\ref{en2-z1}).
Designate
\beq
J(x)=\varphi(x)-\frac{\alpha'(x)}{1-\alpha(x)}\int\limits_0^x\varphi(t)dt,
\label{214}
\eeq
and substitute the function~(\ref{213}) into the right side of equation~(\ref{214}). As a result, we shall obtain
\beq
J(x)=f(x)+\frac{\alpha'(x)}{[1-\alpha(x)]^2}\int\limits_0^xf(t)[1-\alpha(t)]dt-
\label{215}
\eeq
$$
-\frac{\alpha'(x)}{1-\alpha(x)}\int\limits_0^xf(t)dt - \frac{\alpha'(x)}{1-\alpha(x)}
\int\limits_0^xdt\int\limits_0^t\frac{\alpha'(t)(1-\alpha(s))f(s)ds}{[1-\alpha(t)]^2}.
$$

In the last term of formula~(\ref{215}) let us change the integration order and calculate the inner integral:
\beq
\int\limits_s^x\frac{\alpha'(t)dt}{[1-\alpha(t)]^2}=\frac{1}{1-\alpha(x)}-\frac{1}{1-\alpha(s)}.
\label{216}
\eeq

The result should be substituted into the formula~(\ref{215}):
\beq
J(x)=f(x)+\frac{\alpha'(x)}{[1-\alpha(x)]^2}\int\limits_0^xf(t)[1-\alpha(t)]dt-
\label{217}
\eeq
$$
-\,\frac{\alpha'(x)}{1-\alpha(x)}\int\limits_0^xf(t)dt-\frac{\alpha'(x)}{(1-\alpha(x))^2}
\int\limits_0^x(1-\alpha(s))f(s)ds+\frac{\alpha'(x)}{1-\alpha(x)}\int\limits_0^xf(s)ds\equiv f(x).
$$

Our checking has shown that the function~(\ref{213}) is the correct
solution of the equation~(\ref{en2-z1}).

\section*{3. Evaluation for ultrametric diffusion}
\addcontentsline{toc}{section}{3. EVALUATION FOR ULTRAMETRIC DIFFUSION}
\setcounter{equation}0

When solving equations of the ultrametric diffusion type
(that have a relation to the description of conformational dynamics of complicated systems such as
biomacromolecules)
the results
can often be presented in the form of
series
of exponents. Two of such series are represented below:
\[R(t)=\mathop{\sum }\limits_{n=0}^{\infty } a^{-n} e^{-b^{-n} t} , \qquad
S(t)=\mathop{\sum }\limits_{n=1}^{\infty } \frac{1}{n^{k} } a^{-n} e^{-b^{-n} t} .\]
Here $t$ is time, $R(t)$ and $S(t)$ are probabilities that a system is
in some definite groups of states, $k$ is some
integer number, $a>1$, $b>1$ are some parameters.

Study the asymptotic behavior of functions $R(t)$ and $S(t)$ at $t\to \infty $ and
evaluate, if possible, their
asymptotics
using elementary functions depending on $t$.\\

{\bf SOLUTION}

\setcounter{equation}0

Let us explore $S(t)$
and $R(t)=S(t)|_{k=0}+e^{-t}$.
Note that the function
$\displaystyle\frac{1}{x^{k} } a^{-x} $ decreases, while the function $\displaystyle e^{-b^{-x} t} $ increases with the growth of $x$. Then
in the interval $x-1\le n\le x$ the
inequality takes place\textbf{}
\[\frac{1}{x^{k} } a^{-x} e^{-b^{-(x-1)} t} \le \frac{1}{n^k} a^{-n} e^{-b^{-n} t} \le \frac{1}{(x-1)^{k} } a^{-(x-1)} e^{-b^{-x} t} \]
takes place. Integrating it with respect to $x$ from $n$ to $n+1$ gives (for $n>1$):

\[a^{-1} \int _{n}^{n+1}\frac{1}{x^{k} } a^{-(x-1)} e^{-b^{-(x-1)} t} dx \le \frac{1}{n^k} a^{-n} e^{-b^{-n} t} \le a\int _{n}^{n+1}\frac{1}{(x-1)^{k} } a^{-x} e^{-b^{-x} t} dx .\]

Now, by summing over $n$ from $2$
to $\infty $, we obtain:
$$ \tilde{S}_{\min}(t) \leqslant \tilde{S}(t) \leqslant \tilde{S}_{\max}(t) $$
where
$\displaystyle\tilde{S}(t)\equiv\sum_2^\infty\frac{1}{n^{k} } a^{-n} e^{-b^{-n} t}$,
\[
\tilde{S}_{\min}(t)\equiv
a^{-1}\int_{2}^{\infty}\frac{1}{x^{k}}a^{-(x-1)}e^{-b^{-(x-1)}t}dx, \, \qquad
\tilde{S}_{\max}(t) \equiv a\int_{2}^{\infty}\frac{1}{(x-1)^{k}}a^{-x}e^{-b^{-x}t}dx.\]

By
switching to new variables, we have:
\[
\tilde{S}_{\min}(t)=a^{-1}\left(\ln b\right)^{k-1}\left(\ln t\right)^{-k}t^{-\frac{\ln a}{\ln b}}\int_{0}^{b^{-1}t}\left(1-\frac{\ln\left(b^{-1}y\right)}{\ln t}\right)^{-k}y^{\frac{\ln a}{\ln b}-1}e^{-y}dy,\]
 \[
\tilde{S}_{\max}(t)=a\left(\ln b\right)^{k-1}\left(\ln t\right)^{-k}t^{-\frac{\ln a}{\ln b}}\int_{0}^{b^{-2}t}\left(1-\frac{\ln\left(by\right)}{\ln t}\right)^{-k}y^{\frac{\ln a}{\ln b}-1}e^{-y}dy.\]

Let us designate the function
$${\displaystyle\gamma^{(k)}(z,t,\alpha,\beta)\equiv\int_{0}^{\alpha t}\left(1-\frac{\ln\beta y}{\ln t}\right)^{-k}y^{z-1}e^{-y}dy},$$
considering ${\alpha\beta<1}$ for convergence.

Note that the limit for this function is the Gamma-function (see the proof in~\cite{spec4}):
$$\displaystyle\mathop{\lim}\limits _{t\to+\infty}\gamma^{(k)}\left(z,t,\alpha,\beta\right)=
\int_{0}^{\infty}y^{z-1}e^{-y}dy=\Gamma(z).$$


Then~at $t\gg1$ it is possible to write
\[
\tilde{S}_{\min}(t)=a^{-1}\left(\ln b\right)^{k-1}\left(\ln t\right)^{-k}t^{-\frac{\ln a}{\ln b}}\Gamma\left(\frac{\ln a}{\ln b}\right)\left(1+o(t)\right),\]
 \[
\tilde{S}_{\max}(t)=a\left(\ln b\right)^{k-1}\left(\ln t\right)^{-k}t^{-\frac{\ln a}{\ln b}}\Gamma\left(\frac{\ln a}{\ln b}\right)\left(1+o(t)\right).\]
Notation $o(t)$ means that in the limit at $t\to\infty$
the value of $o(t)$ 
tends
to zero.

Since
\[
S(t) = \tilde{S}(t)+a^{-1}e^{-b^{-1}t},\]
the final asymptotic evaluation is of the form $\displaystyle\left(\ln t\right)^{\displaystyle-k}t^{\displaystyle-\frac{\ln a}{\ln b}}$:
\[
a^{-1}\left(\ln b\right)^{k-1}\Gamma\left(\frac{\ln a}{\ln b}\right)\left(\ln t\right)^{-k}t^{-\frac{\ln a}{\ln b}}\left(1+o(t)\right)\le S(t)\le\]
 \[
\le a\left(\ln b\right)^{k-1}\Gamma\left(\frac{\ln a}{\ln b}\right)\left(\ln t\right)^{-k}t^{-\frac{\ln a}{\ln b}}\left(1+o(t)\right).\]

Because
\[
R(t)=\tilde{S}(t)|_{k=0}+e^{-t}+a^{-1}e^{-b^{-1}t},\]
we have also
\[
a^{-1}(\ln b)^{-1}\Gamma\left(\frac{\ln a}{\ln b}\right)t^{-\frac{\ln a}{\ln b}}\left(1+o(t)\right) \le R(t)\le a(\ln b)^{-1}\Gamma\left(\frac{\ln a}{\ln b}\right)t^{-\frac{\ln a}{\ln b}}\left(1+o(t)\right).\]


\section*{4. Double effort}
\addcontentsline{toc}{section}{4. DOUBLE EFFORT}
\setcounter{equation}0

Solve the Volterra integral equation
\beq
\varphi(x)=x+\int_0^x (s-x)\varphi(s) ds. \label{4e00}
\eeq

\begin{center}
\includegraphics[width=5cm]{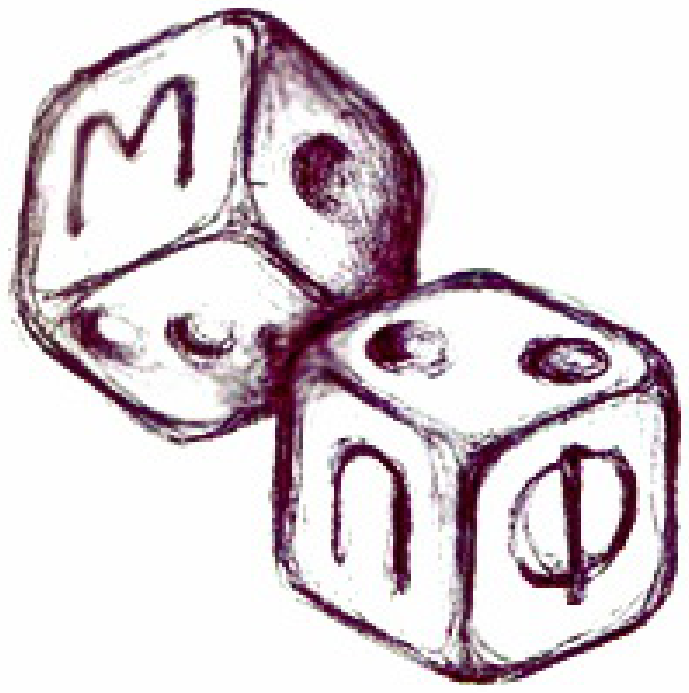}
\end{center}

{\bf SOLUTION}

The considered equation is in fact Volterra integral equation of $2^{nd}$ kind with continuous kernel. According to the theory of linear integral equations, it has a unique solution. To~find its solution, we reduce it to Cauchy problem for ordinary differential equation.~\footnote{Special Issue No.\,4 contains also another methods of solution (see \cite{spec4}).}

Let us assume that  $\varphi(x)$ is the solution of the equation~(\ref{4e00}). Double differentiation of the identity equation~(\ref{4e00}) by $x$ gives:
\begin{equation}
\varphi'(x)=1+\int_0^x \left(-
\varphi(s)\right)ds,\label{e11}\end{equation}
\begin{equation}
\varphi''(x)= - \,\varphi(x). \label{e12}
\end{equation}
\beq
\varphi(x)=C_1 \sin x + C_2 \cos x.
\eeq

Assuming~$x=0$ results in:
\begin{equation}
\varphi(0)=0. \label{e14}\end{equation}

At $x=0$ we have from eq.~(\ref{e11}):
\begin{equation}
\varphi'(0)=1. \label{e15}\end{equation}

Hence $C_1=1$, $C_2=0$, and
\begin{equation}
\varphi(x)=\sin x.\end{equation}


It is possible to perform a check:
\beq
J(x)=x+\int_0^x (s-x)\sin s \,ds=
\eeq
\beq
\left. = x-(s-x)\cos s \right|_0^x + \int_0^x \cos s \,ds,\eeq
$$\left.=x-x+\sin(s)\right|_0^x = \sin x.
$$

It is possible to propose other ways to solve this equation, such as with the help of Laplace transformation, or by building a resolvent of the kernel by successive approximations method. However, the technique developed above is the
simplest.


\section*{5. Random walk}
\addcontentsline{toc}{section}{5. RANDOM WALK}

A particle performs random walk on one-dimensional lattice situated on the $OX$~axis, the nodes of the lattice
have the coordinates
$m=0,\pm 1,\pm 2,...$. At the initial time moment $t_{0} =0$ the particle is
at the origin of the coordinates. At random time moments $t_{1} $, $t_{2} $, $t_{3}$,\dots the particle performs the jumps into adjacent lattice nodes with the probabilities of a jump leftwards and rightwards equal to $\displaystyle\frac{\alpha }{2}$, the probability of remaining
still being $\beta =1-\alpha$. The time intervals between the jumps $ $$t_{i+1} -t_{i} $, $i=0,1,2,...$ are independent random
quantities which have the
same exponential distribution
$\displaystyle\Phi(t)=\frac{1}{\tau } \exp (-t/\tau )$ with expectation $\tau$.
Find:

\textbf{(a)} dispersion of the location of the particle as time function $t$;

\textbf{(b)} probability that the particle is in $m$-th node at time moment $t$.\\

\setcounter{equation}0

{\bf SOLUTION}
\setcounter{equation}0

1. Let us find the
probability $p(t,n)$ that the particle during time interval $(0,t]$ would perform exactly $n$ jumps,
taking into account that this probability is the distribution function of the Poisson process.
Let $t_{1} ,t_{2} ,...$ be time instants of the jumps, so that
\beq
t_{0} =0<t_{1} <t_{2} <...<t_{n-1} <t_{n} <t<t_{n+1}.
\eeq

Probability $p(t,n)$ that the particle during the time interval $(0,t]$ would perform exactly $n$ jumps, can be represented as
\beq p(t,n)={\rm M} \left[I\left(t_{n} <t<t_{n+1} \right)\right],\eeq
where   ${\rm M} \left[...\right]$ is expectation and
\beq I\left(t_{n} <t<t_{n+1} \right)=\left\{\begin{array}{l} {1,{\rm \; \; \mbox{if}\; \; }t_{n} <t<t_{n+1} {\rm ,\; }} \\ {{\rm 0,\; \;  \mbox{if}\; \; }t_{n} \ge t{\rm \; \; \; \mbox{or}\; \; \; }t\ge t_{n+1} {\rm .\; }} \end{array}\right. \eeq

Let us perform Laplace transformation of the function $p(t,n)$:
$$\hat{p}(s,n)=\int _{0}^{\infty }dte^{-st}  p(t,n)={\rm M} \left[\int _{0}^{\infty }dte^{-st}  I\left(t_{n} <t<t_{n+1} \right)\right] = $$
\beq ={\rm M} \left[\frac{e^{-st_{n} } -e^{-st_{n+1} } }{s} \right].\eeq
Due to the fact that $\displaystyle t_{n} =\sum _{i=1}^{n}\tau _{i}  $ is the sum of independent random variables, then
\beq{\rm M} \left[e^{-st_{n} } \right]={\rm M} \left[\exp \left(-s\sum _{i=1}^{n}\tau _{i}  \right)\right]=\prod _{i=1}^{n}{\rm M} \left[\exp \left(-s\tau _{i} \right)\right] =\eeq
\[=\prod _{i=1}^{n}\int _{0}^{\infty }d\tau _{i} \exp \left(-s\tau _{i} \right) \Phi (\tau _{i} ) =\hat{\Phi }^{n} (s),     \]
where $\displaystyle\hat{\Phi }(s)=\int _{0}^{\infty }d\tau \exp \left(-s\tau \right) \Phi (\tau )$ is Laplace image of $\Phi (\tau )$. From this,
\beq\hat{p}(s,n)={\rm M} \left[\frac{e^{-st_{n} } -e^{-st_{n+1} } }{s} \right]=\hat{\Phi }^{n} (s)\frac{1-\hat{\Phi }(s)}{s} .\eeq
Due to the fact that $\displaystyle\Phi (t)=\frac{1}{\tau } \exp (-t/\tau )$, it follows that
\beq\displaystyle\hat{\Phi }(s)=
\frac{1/\tau }{s+1/\tau } ,\eeq
and
\beq\hat{p}(s,n)=\frac{\left(1/\tau \right)^{n} }{(s+1/\tau )^{n+1} }. \eeq
Transforming from Laplace image to the original, we obtain the distribution function for the Poisson process
\beq p(t,n)=\frac{t^{n} e^{-t/\tau } }{\tau ^{n} n!}. \eeq

2. Let us find the dispersion $D(t)$ of the location of the particle as a function of time~$t$. Location of the particle after $n$ jumps is defined by
the random variable $X_{n} (t)\equiv \xi _{1} +\xi _{2} +...+\xi _{n} $, where  $\xi _{i} $, $i=1,...,n$ are independent random variables,
possessing the~values $\pm 1$ with the probability $\displaystyle{\frac{{\alpha}}{2}}$ and 0 with the probability $\beta=1-\alpha$.
Due to the fact that ${\rm M}\left[\left(\xi _{i} \right)^{2} \right]=\alpha $, ${\rm M}(\xi_{i})=0$ and ${\rm M}\left[\xi _{i} \xi _{j} \right]=0$ for
 $i\ne j$,
the~dispersion
equals to
\beq D(t)=\sum _{{\rm n}={\rm 0}}^{\infty }p(t,n){\rm M}\left[\left(X_{n} (t)\right)^{2} \right]=\sum _{{\rm n}={\rm 0}}^{\infty }p(t,n)\alpha n. \eeq
Substituting into this formula the expression for
$p(t,n)$, we obtain
\beq D(t)=\sum _{{\rm n}={\rm 1}}^{\infty }\frac{t^{n} e^{-t/\tau } }{\tau ^{n} n!} \alpha n =\alpha e^{-t/\tau } \frac{t}{\tau } \sum _{{\rm n}={\rm 1}}^{\infty }\frac{t^{n-1} }{\tau ^{n-1} \left(n-1\right)!}  =\alpha \frac{t}{\tau }. \eeq

3. Let us find the probability that the particle is located at node $m$ at time moment~$t$.
Let $h_{n} (m)$ designate the probability that the particle is located at the point with coordinate $m$ after $n$ jumps (transitions). Then the
probability $f(m,t)$ that the particle is at~location $m$ at time moment $t$ will be given by formula
\beq f(m,t)=\sum _{n=0}^{\infty }p(t,n)h_{n} (m) .     \eeq
Function $h_{n} (m)$ equals to
\beq h_{n} (m)={\rm M}\left[\delta _{m,\xi _{1} +\xi _{2} +...+\xi _{n} } \right],\eeq
where $\delta _{m,n}$ is the Kronecker delta.
Due to the fact that $\displaystyle\delta _{m,n} =\frac{1}{2\pi } \int _{-\pi }^{\pi }e^{i(m-n)\varphi } d\varphi  $, it is possible to write
\beq h_{n} (m)=\frac{1}{2\pi } {\rm M}\left[\int _{-\pi }^{\pi }e^{i(m-\xi _{1} -\xi _{2} -...-\xi _{n} )\varphi } d\varphi  \right]=\frac{1}{2\pi } \int _{-\pi }^{\pi }e^{im\varphi }  \prod _{j=1}^{n}{\rm M}\left[e^{-i\xi _{j} \varphi } \right] d\varphi .\eeq
Note that
\beq {\rm M}\left[e^{-i\xi _{j} \varphi } \right]=\frac{\alpha }{2} \left(e^{-i\varphi } +e^{i\varphi } \right)+1-\alpha =\alpha \cos \varphi +1-\alpha. \eeq
Therefore
\beq h_{n} (m)=\frac{1}{2\pi } \int _{-\pi }^{\pi }e^{im\varphi }  \left(\alpha \cos \varphi +1-\alpha \right)^{n} d\varphi. \eeq
Substituting this equation for $h_{n} (m)$, and earlier found expression for $p(t,n)$ into the formula for $f(m,t)$,
we obtain
\beq f(m,t)=\sum _{n=0}^{\infty }\frac{t^{n} e^{-t/\tau } }{\tau ^{n} n!} \frac{1}{2\pi } \int _{-\pi }^{\pi }e^{im\varphi }  \left(\alpha \cos \varphi +1-\alpha \right)^{n}
d\varphi  =\eeq
\[ = \frac{1}{2\pi } e^{-t/\tau } \int _{-\pi }^{\pi }e^{im\varphi }  d\varphi \sum _{n=0}^{\infty }\left(\alpha \cos \varphi +1-\alpha \right)^{n}
\frac{t^{n} }{\tau ^{n} n!}  =\]
\[=\frac{1}{2\pi } e^{-t/\tau } \int _{-\pi }^{\pi }e^{im\varphi } e^{t\left(\alpha \cos \varphi +1-\alpha \right)/\tau } d\varphi  =\]
\[ = \frac{1}{2\pi } e^{-t/\tau } \left(\int _{0}^{\pi }e^{im\varphi } e^{t\left(\alpha \cos \varphi +1-\alpha \right)/\tau } d\varphi  +\int _{0}^{\pi }e^{-im\varphi } e^{t\left(\alpha \cos \varphi +1-\alpha \right)/\tau } d\varphi  \right)=\]
\[=\frac{1}{\pi } e^{-t/\tau } \int _{0}^{\pi }\cos (m\varphi )e^{t\left(\alpha \cos \varphi +1-\alpha \right)/\tau } d\varphi  =
\frac{1}{\pi } e^{-\alpha t/\tau } \int _{0}^{\pi }\cos (m\varphi )e^{t\alpha \cos \varphi /\tau } d\varphi  .\]
The last equation can be reproduced in a different form, using integral representation of Bessel function $J_{m} (z)$:
\beq J_{m} (z)=\frac{i^{-m} }{\pi } \int _{0}^{\pi }\cos \left(m\varphi \right)e^{iz\cos \varphi } d\varphi  .\eeq
As a result, we obtain
\beq f(m,t)=i^{m} e^{-\alpha t/\tau } J_{m} \left(-i\frac{\alpha t}{\tau } \right).\eeq
Taking into consideration that
\beq J_{m} (iz)\equiv i^{m} I_{m} (z),\eeq
where
\beq I_{m} (z)=\sum _{k=0}^{\infty }\frac{\left(\frac{z}{2} \right)^{2k+m} }{k!\left(k+m\right)!}  \eeq
are the modified Bessel functions, we write the final result for the probability of location of the particle at node $m$ at time moment $t$:
\beq f(m,t)=\left(-1\right)^{m} e^{-\alpha t/\tau } I_{m} \left(-\frac{\alpha t}{\tau } \right)=e^{-\alpha t/\tau } I_{m} \left(\frac{\alpha t}{\tau } \right).\eeq

Asymptotics of $f(m,t)$ at $t\to \infty $
with the asymptotic behavior of modified Bessel functions
\beq I_{m} (z)=\frac{e^{z} }{\sqrt{2\pi z} } \left(1+O\left(z^{-1} \right)\right) \qquad \mbox{at} \quad  z\to \infty \eeq
considered,
is
of the form:
\beq f(m,t)=e^{-\alpha t/\tau } \frac{e^{\alpha t/\tau } }{\sqrt{2\pi \alpha t/\tau } } \left(1+O\left(t^{-1} \right)\right)=\frac{1}{\sqrt{2\pi \alpha t/\tau } } \left(1+O\left(t^{-1} \right)\right).\eeq

\section*{6. Thermal equations of the Universe evolution}
\addcontentsline{toc}{section}{6. THERMAL EVOLUTION EQUATIONS of the UNIVERSE}

\setcounter{equation}0

It is assumed that at high temperature
(at early stage of the evolution of the Universe) it
is possible to describe
matter
using field theory. Equation of state
with good approximation corresponds
to ideal quantum gas of massless particles
(in the general case, it can be a mixture of ideal Bose- and Fermi-gases).
In this theory,
under the condition that the temperature $T$ is far from mass threshold yet (radiation\footnote{By radiation here, we mean any relativistic object, including relativistic matter as well as photons.} dominance, $\rho=3p$),
thermodynamic functions are given by the formulae:
\begin{eqnarray}
 \rho = 3 p = \frac{\pi^2}{30}N(T)T^4, \label{5e}
\end{eqnarray}
\begin{eqnarray}
\label{6e} s = \frac{2\pi^2}{45}N(T)T^3,
\end{eqnarray}
where $ N(T) $ is the
function related to the
number of bosonic and fermionic degrees of freedom
($\displaystyle N(T)=N_b(T)+\f{7}{8}N_f(T)$), $\rho$ and $p$
are equilibrium energy density and pressure of the matter, $s$ is specific entropy.
All
expressions are written in
the unified system of units $c = \hbar =1$.
%

Formulate the
dynamic equations of the evolution of the Universe in terms of temperature.

\underline{Note}. The required equations are not Einstein's equations in the standard form (see~Einstein equations, for example,
in~\cite{wei72,wald84}).
It is proposed to write the equations of~the evolution of the Universe in Friedmann model using thermodynamic functions and
temperature as function of time.
It is possible to do this
with the
use of energy conservation law and condition of adiabatic expansion of the Universe in the framework of the standard cosmologic model.

\underline{Instruction}. Introduce
auxiliary function
\begin{equation}
\label{7e} \epsilon (T) = \frac{k}{a^2 T^2},
\end{equation}
where $k = 0, \pm 1$ respectively for flat, open and closed models of the Universe with
time-dependent scale factor of $a\equiv a(t)$.\\
\parbox{0.5\textwidth}{{\it\bf
Guth}, {\it\bf Alan Harvey}
-- American physicist and cosmologist who has first proposed the idea of cosmological inflation.
In 2004, Guth together with Andrew Linde
were awarded cosmological prize named after Peter Gruber for his work on the theory of~inflation Universe.}
\hspace{1mm}
\parbox{0.24\textwidth}{\includegraphics[scale=0.5]{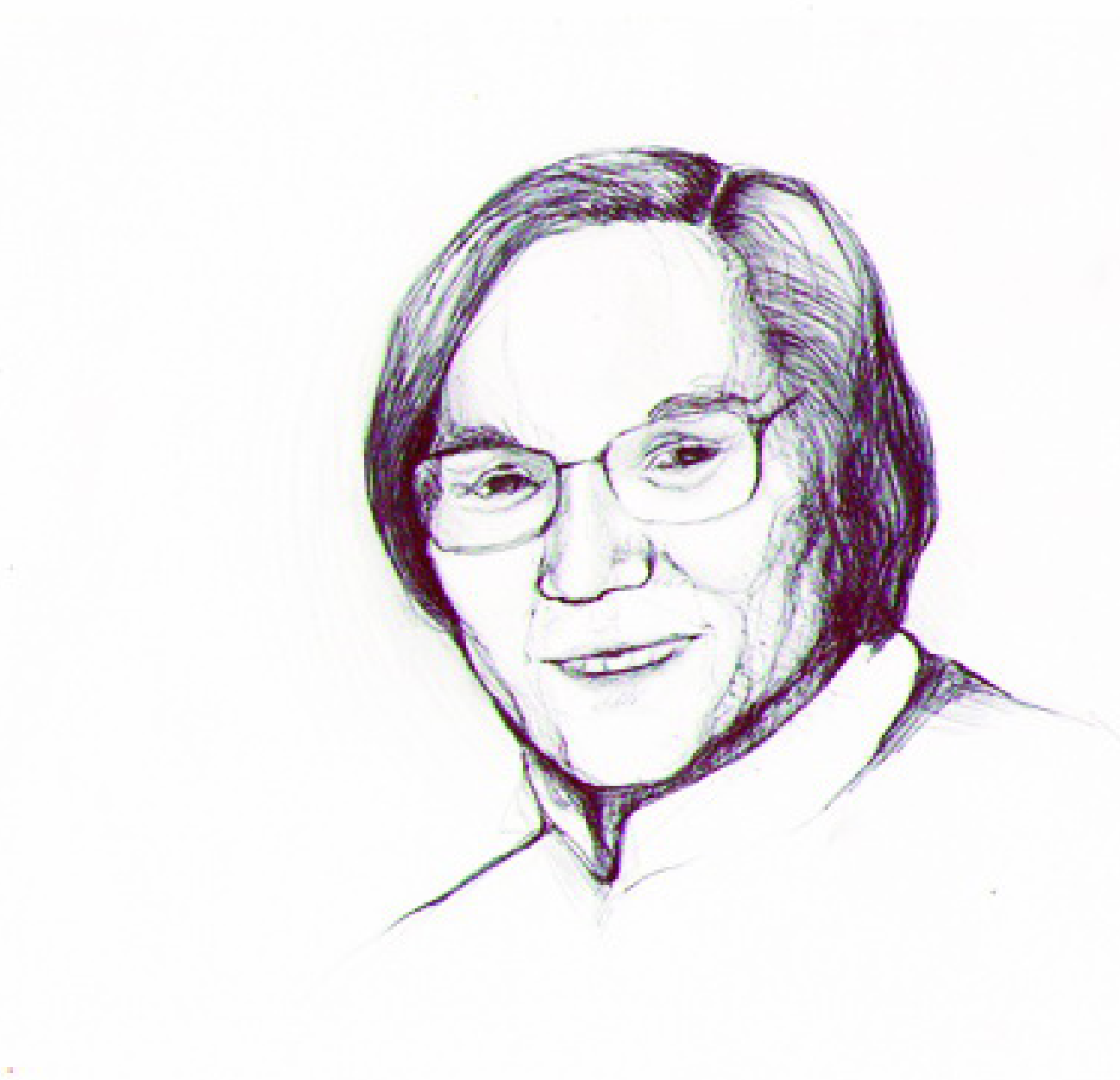}
}

\vspace{1mm}

{\bf SOLUTION}

This problem
of finding the time dependence of the temperature of the Universe was first formulated by A. Guth
in
\cite{guth81}.
To solve this problem, we would need to rewrite Einstein--Friedmann equation
\begin{eqnarray}
\label{2} \left(\frac{\dot{a}}{a} \right)^2 + \frac{k}{a^2} =
\frac{8\pi}{3}G\rho
\end{eqnarray}
in terms of temperature
~\cite{wald84} (here $G$ is the gravitation constant).
It is considered that stress energy
tensor
of the Universe takes the form of
energy
momentum
tensor for ideal liquid
\cite{wei72}.
We shall use the energy conservation law
\begin{equation}
\label{e3} \frac{d}{dt}(\rho a^3) = -\, p \frac{d}{dt}(a^3)
\end{equation}
(the change in energy in a comoving volume element, $d\,(\rho a^3)$, is equal to minus the pressure times the change in volume, $-p\,d\,(a^3)$) and the fact that in the standard cosmological model
it is supposed that the Universe undergoes adiabatic expansion
\begin{equation}
\label{e4} \frac{d}{dt}(s a^3) = 0
\end{equation}
(the entropy per comoving volume element remains constant).

Let us write Einstein--Friedmann equation
(\ref{2}) in terms of
temperature
supposing that temperature value is far from the mass threshold
(see e.\,g. \cite{wei72}).
We deal with matter which is found in thermodynamical equilibrium at almost all the time stages during cosmological
expansion, so the chemical potential is considered to be zero.

Taking into consideration (\ref{5e}) and auxiliary function~(\ref{7e})
let us represent eq. (\ref{2}) in the form
\begin{equation}
\label{e2a} \left(\frac{\dot{a}}{a} \right)^2+\epsilon(T)T^2 =
\frac{4\pi^3}{45}G N(T)T^4.
\end{equation}

Taking into account equation of state $ \rho = 3p $, from the energy conservation law
(\ref{e3}) we obtain the relationship
\begin{equation}
\label{3a} \frac{\dot{a}}{a} = - \,\frac14 \frac{\dot{\rho}}{\rho},
\end{equation}
which in agreement
with
(\ref{5e})
would take the form
\begin{equation}
\label{e3b} \frac{\dot{a}}{a} = -\frac{\dot{T}}{T} -\frac14
\frac{\dot{N}(T)}{N(T)}.
\end{equation}

Using the condition of adiabatic expansion of the Universe (\ref{e4}),
we can find
\begin{equation}
\label{4aen} \frac{\dot{a}}{a} = - \,\frac13 \frac{\dot{s}}{s}.
\end{equation}

Substituting into (\ref{4aen}) the expression
for specific entropy~(\ref{6e}),
we obtain
\begin{equation}
\label{e4b} \frac{\dot{a}}{a} = -\,\frac{\dot{T}}{T} -\,\frac13
\frac{\dot{N}(T)}{N(T)}.
\end{equation}

Comparing (\ref{e3b}) and (\ref{e4b}), we come to the conclusion that $ \dot{N}(T) = 0
$.
So the relation between
the scaling factor and temperature should have the form
\begin{equation}
\label{b} \frac{\dot{a}}{a} = -\frac{\dot{T}}{T}.
\end{equation}
Substituting this equation into
eq.
(\ref{e2a}),
we obtain one of the dynamic equations of the evolution of the Universe:
\begin{equation}
\label{6p10en} \left(\frac{\dot{T}}{T} \right)^2+\epsilon(T)T^2 =
\frac{4\pi^3}{45}G N(T)T^4.
\end{equation}
To write the second required equation, let us
multiply both parts of the eq.~(\ref{6e})
by $ a^3 $,
express
$\displaystyle\frac{1}{a^2 T^2}$ and substituting it
into the auxiliary~eq.~(\ref{7e}), we obtain the equation
\begin{equation}
\label{6p11en} \epsilon(T) = k\left[\frac{2\pi^2}{45} \frac{N(T)}{S}
\right]^{2/3},
\end{equation}
where $ S \equiv sa^3 $ is the total entropy in the volume defined by the radius
of curvature $ a $. We need to note that $N$, $S$ hence $\epsilon$ are constant in the considered temperature (or time) range (however $a$ is not constant) due to the particle counting by allowed particles degrees of freedom thresholds.

As the result, the dynamic equations of the evolution of the Universe in terms of temperature
and entropy are the equations (\ref{6p10en}) and (\ref{6p11en}).


\section*{7. Trapped
electron}
\addcontentsline{toc}{section}{7. ELECTRON THAT IS NOT GOING TO LEAVE}

Consider an isolated conducting sphere of radius $R$ carrying the total charge
$Q$. At~the distance
$a>R$ from its center, there is a point charge $q$ ($qQ>0$). Find potential of the
system $\varphi(\vec r)$ and the force $\overrightarrow{F}(a)$ acting on the
point charge. Analyze
the limit $\displaystyle\lim_{a\to R+0}F(a)$, explain the obtained result.\\

\setcounter{equation}0

{\bf SOLUTION}

The present problem can be solved using method of image charges.


It is known that for any two point electric charges of opposite sign, it is always possible to
find such a spherical surface that
the resulting potential on it would be~zero.
Radius of the sphere and the distance from its center to the charges is determined uniquely if the values of the charges and the distance between them are known. So,~the~system under discussion (the point charge and the conducting sphere) is equivalent to the set of point charges.
Thus let us place a charge $q_1$ on a line connecting the center of the sphere $O$ with the $q$ charge at the distance of $d$ in the direction of the charge $q$. One more charge, $q_0=Q-q_1$, we shall place at the $O$ point, see
fig.\,1.

Let us place the origin of the reference frame also in $O$ point and direct $OX$ axis to the point charge $q$ (leftwards). Let us write the condition that the total potential (due to the charges $q$ and~$q_1$) is zero at points ($R,0,0$) and ($-R,0,0$).

Hence we obtain that $d=R^2/a$, $q_1=-qR/a$, thus $q_0=Q+qR/a$.

The total potential inside the full sphere equals $q_0/R$ and outside is given by the expression:
\beq
\varphi(\vec{r}_0)=\frac{q}{r}+\frac{q_1}{r_1}+\frac{q_0}{r_0}.
\label{7new5}
\eeq
The same in the Cartesian coordinates:
\begin{equation}
\varphi(\vec r_0)=\frac{q}{\sqrt{(x-a)^2+y^2+z^2}}-\frac{R}{a}\frac{q}{\sqrt{(x-\frac{R^2}{a})^2+y^2+z^2}}+
\frac{Q+q\frac{R}{a}}{\sqrt{x^2+y^2+z^2}},
\label{7new6}
\end{equation}

\begin{center}
\includegraphics[width=11cm]{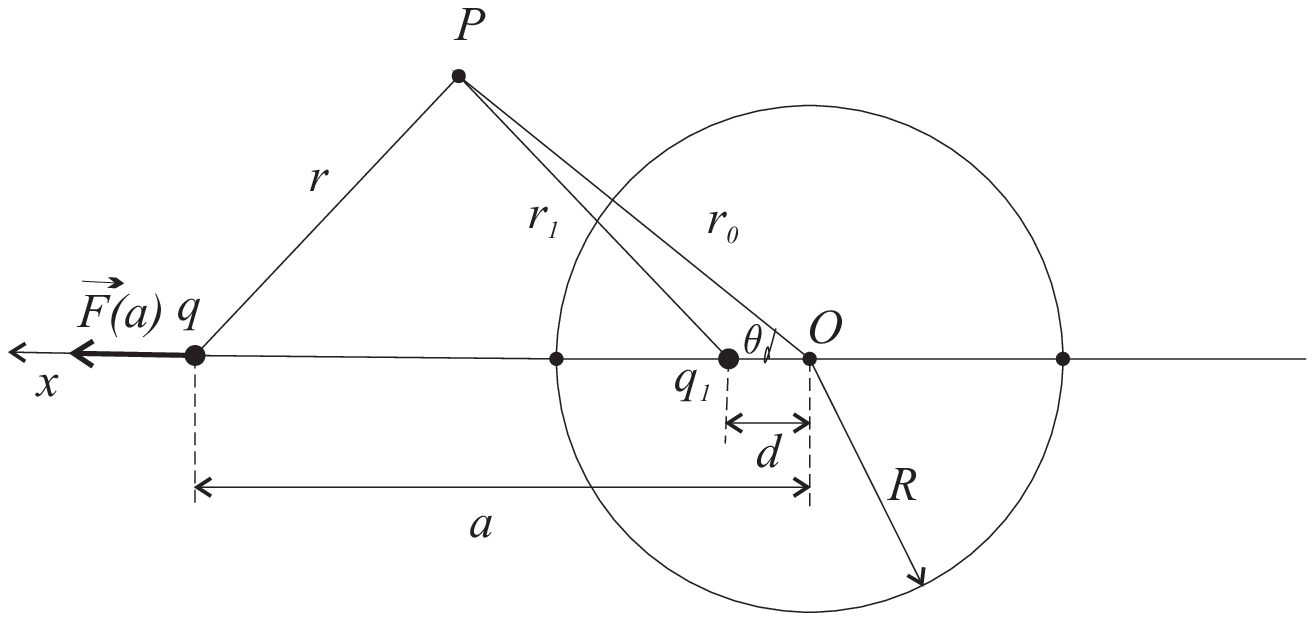}\\[1mm]
{\small
Fig. 1 {\it \\[2mm]}}
\end{center}

The force  on $q$
acts
along axis
$OX$.
Its projection is given by
\begin{equation}
F=\frac{qQ}{a^2}+\frac{q^2 R}{a^3}\left(1-\frac{1}{\left(1-\displaystyle\frac{R^2}{a^2}\right)^2}\right).\label{eq:7forse}
\end{equation}

It is easy to check that if we represent $a$ as $R+\Delta$, in the limit at $\Delta\to0$ the~expression for the force would become $F\sim-q^2/(2\Delta)^2$ which corresponds to the interaction force of a point charge with non-charged conducting plane.

At $a\to\infty$, $\displaystyle F\to\frac{q Q}{a^2}$ (Coulomb force, as expected).

At $a\to R$, no matter how large $Q$ is, even if $Qq>0$, we have $F\to-\infty$, i.\,e.~the~force becomes very strong and attractive!

Let us find the distance at which the effect of attraction starts.
Let us write the interaction force~(\ref{eq:7forse}) in a dimensionless form:
\beq
f(s)=\frac{\alpha}{s^2}-\frac{2s^2-1}{s^3(s^2-1)^2},
\label{7newdimlessforse}
\eeq
where
\beq
\alpha=\frac{Q}{q}>0, \,\,\, ~ ~ ~ ~ s=\frac{a}{R}>1, \,\,\, ~ ~ ~ ~ f(s)=\frac{R^2}{q^2}F_x(a).
\label{7new9}
\eeq
The force occurs to be zero at a distance $a=s_0R$,
where $s_0>1$ satisfying the equation
\beq
\frac{2s^2-1}{s(s^2-1)^2}=\alpha.\label{7new10}
\eeq

Function $f(s)$ graphs are shown in fig.\,2 at $\alpha=\{2; 1;
0.5\}$. In these cases $s_0=\{1.43; 1.62; 1.88\}$,
and force $f(s)$ reaches its maximum $\{0.43, 0.15, 0.05\}$ when
$s=s_{max}=\{1.79, 2.07, 2.46\}$.

This explains why electrons can't escape from metals, even though they are repelled by the other electrons. If an electron manages to escape and gets to a small distance
from the surface of metal, the other electrons conspire to bring it back by rearranging themselves in such a way
to create a huge image charge which attracts the electron back to the metal with a strong electric force!

\begin{center}
\includegraphics[width=12cm]{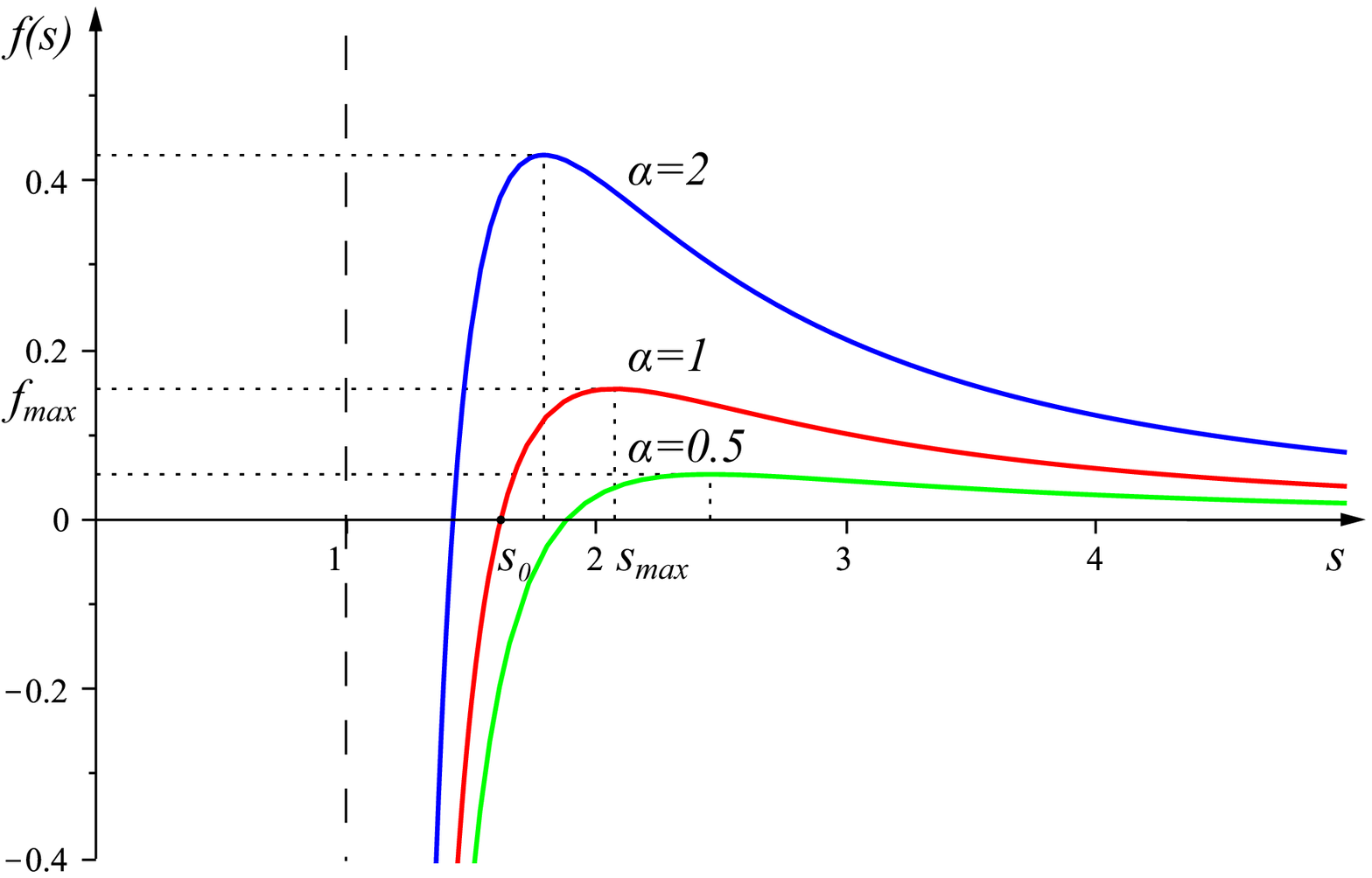}\\[1mm]
{\small {
Fig.\,2 \\[4mm]}}
\end{center}




\section*{8. $\mbox{\sc\bf Х}$-sector~\footnote{Some extension of Peskin \& Schroeder~\cite{pands} problem 20.5.}}\setcounter{equation}0
\addcontentsline{toc}{section}{8. $\mbox{\sc Х}$-SECTOR}

Consider the following model of Higgs sector
with two doublet scalar fields $\phi_1$ and~$\phi_2$ transformable as $SU(2)$ doublets, with the weak hypercharge generator
of $Y_W=1$,
and with each component of the doublet being a
complex scalar field.
Suppose both fields
acquire parallel vacuum averages (vevs) of the type
\beq
\langle\phi_i\rangle=\frac{1}{\sqrt{2}}\left(\begin{array}{c}  0 \\ v_i\end{array}\right) \qquad ( i = 1 , 2 ),\label{eq:6-vevs}
\eeq
with the values $v_1$, $v_2$
(these vacuum averages lead to gauge bosons mass matrix as in the Standard Model with the
replacement \footnote{Recall that in the Standard Model (SM),
the gauge bosons masses come
from the term $|D\phi|^2$ in the Lagrangian, where we set $\phi$ equal to its vacuum expectation value
$v$.}
$v^2=v_1^2+v_2^2$).
The most general form of the potential energy function (potential) for
a model with two Higgs doublets is rather complicated\,\footnote{It is only for reasons of simplicity that the
SM contains just a single Higgs doublet. Supersymmetric extensions of the
SM typically contain two or more Higgs
doublets, and singlets.}.
However, the model hermitian potential possessing the main properties can be written in the following form:
$$V(\phi_1,\phi_2) = \nonumber - \, \mu_1^2
(\phi_1^\dagger\phi_1) - \, \mu_2^2 (\phi_2^\dagger \phi_2) - \, \mu_{12}^2
(\phi_1^\dagger\phi_2) - \, \left(\mu_{12}^2\right)^{\bf*} (\phi_2^\dagger \phi_1) \,+ $$
$$ + \lambda_1
(\phi_1^\dagger \phi_1)^2
      +\lambda_2(\phi_2^\dagger \phi_2)^2
+ \lambda_3 (\phi_1^\dagger \phi_1)(\phi_2^\dagger \phi_2) +
\lambda_4 (\phi_1^\dagger \phi_2)(\phi_2^\dagger \phi_1) + \f{\lambda_5}{2}(\phi_1^\dagger \phi_2)^2+
\f{{\lambda_5}^{\bf*}}{2}(\phi_2^\dagger \phi_1)^2,$$
where $\mu_{12}^2$ and $\lambda_5$ may be
the complex numbers.

\textbf{(a)} Obtain the conditions that for the direction in field space at given configuration of
vevs (\ref{eq:6-vevs})
the potential is bounded
below
at large field values. (The analogue of $\lambda>0$ for the theory with a single Higgs doublet.)

\textbf{(b)} Find the conditions
to
impose on
the parameters $\mu$ and $\lambda$,
so that the configuration of vacuum averages (\ref{eq:6-vevs})
gives strictly local (locally stable) minimum of this potential.

\textbf{(c)} In the unitary gauge (rotation to
canonical form),
one linear combination
of the upper components
$\phi_1$ and $\phi_2$ nulls,
while another
becomes a physical field. Show that charged physical Higgs field is of the form:
\beq
H^+=\phi_1^+\sin\beta-\phi_2^+\cos\beta,
\eeq
where $\beta$ is defined by the relation
\beq
\tg\beta=\frac{v_2}{v_1}.
\eeq

\textbf{(d)} Investigate if the CP invariance breaks in the given potential. Substantiate the obtained results.\\

{\bf SOLUTION}

\textbf{(a)} First of all, note that $\mu_{i=1, 2}^2$ and $\lambda_{i=1...4}$ are all real due to the fact that 
Lagrangian is Hermitian. But $\mu_{12}^2$ and $\lambda_5$ may be the complex numbers.

Second, find the condition
under which both $\phi_1$ and $\phi_2$ have parallel non-zero vevs (recall that in the case of a single Higgs doublet the gauge symmetry allows the vacuum expectation value (vev) to be
taken in the form $(0, v)$, with $v$ real).
Now use a~rotation to
make
$\displaystyle\langle\phi_1\rangle=\frac{1}{\sqrt{2}}\left(\begin{array}{c}  0 \\ v_1\end{array}\right)$, where $v_1$ is~real. Note that after we have done this we have used all of our gauge rotation freedom, so the vevs of $\phi_2$ are still completely general, i.\,e. $\displaystyle\langle\phi_2\rangle=\frac{1}{\sqrt{2}}\left(\begin{array}{c}  v_2' \\ v_2''\end{array}\right)$. So we rewrite the potential in terms of the two (complex) vevs of $\displaystyle\phi_2$, and of the $v_1$. Letting $\displaystyle v_2^2={v_2'}^2+{v_2''}^2$, we~find
\beq
V = F(v_1,v_2) \, - \,\frac{\mu_{12}^2}{2} v_1 {v_2''} - \frac{\left(\mu_{12}^2\right)^{\bf*}}{2} v_1 {{v_2''}^*} + \frac{\lambda_4}{4} v_1^2 ({v_2''}^*{v_2''}) + \frac{\lambda_5}{8} v_1^2 {v_2''}^2 +
\frac{{\lambda_5}^{\bf*}}{8} v_1^2 {{v_2''}^*}^2,
\eeq
where we denote $F(v_1,v_2)$ the function dependent on values $v_1$ and $v_2$ only.

First we want to answer two questions: (i) What is the condition that $v_2''$ is real? (ii)~What is the condition that $v_2'$ is zero?
If we enforce these conditions, it ensures that the vevs take
the form of (\ref{eq:6-vevs}),
where both $v_1$ and $v_2$ are real. This is what it means for the vevs to be 'aligned'. So, how to find these conditions? It
becomes clear if we rewrite $\displaystyle v_2''\equiv ae^{i\theta}$. The potential written in terms of $a$ and $\theta$ is given by:
$$ V(a,\theta) = (\mbox{Stuff not dependent  on} ~ \theta ~ \mbox{or} ~ a)
- \mbox{Re}\mu_{12}^2v_1a\cos\theta + \mbox{Im}\mu_{12}^2v_1a\sin\theta \,+ $$\beq + \frac{1}{4}\lambda_4v_1^2a^2 + \frac{1}{4}\mbox{Re}\lambda_5v_1^2a^2\cos2\theta \,- \,\frac{1}{4}\mbox{Im}\lambda_5v_1^2a^2\sin2\theta. \label{eq:en8-st}
\eeq
The reality of $\displaystyle v_2''$ is ensured by forcing $\theta=0,\,\pi$. We can see that $\theta=0$ or $\pi$ will be a~stable minimum of the potential if the second derivative of expression (\ref{eq:en8-st}) is positive at these values of phase $\theta$, i.\,e.
\beq
\mbox{Re}\lambda_5   \,- \, \mbox{Re}\mu_{12}^2 \frac{1}{v_1a} < 0, \label{eq:8en-23}
\eeq
and in the limit
of large field values
\beq
\mbox{Re}\lambda_5 < 0. \label{eq:8en-23largefield}
\eeq
In this case the
Eqn.~(\ref{eq:en8-st}) will be minimized for $\cos2\theta=1$. We must show that $v_2'=0$. Recall that the sum ${v_2'}^2+{v_2''}^2$ is fixed to be $v_2^2$. So if we can arrange the potential so that it is energetically advantageous for all the vevs to go into ${v_2''}^2$, we are done. This is equivalent to saying that we want $V(a,\theta)$ to be minimized for $a\to\infty$. This is accomplished if
\beq
\lambda_4 + \lambda_5 < 0. \label{eq:8en-24}
\eeq
So, together, eqns. (\ref{eq:8en-23largefield}), (\ref{eq:8en-24}) guarantee that the
vacuum expectation values can be align\-ed.




\textbf{(b)} Now what is still required, is to show that this
is a stable minimum.
Using aligned forms for the vevs, rewrite the potential in terms of $v_1$ and $v_2$. What additional conditions on the parameters are necessary to guarantee a stable minimum? Stability is equivalent to saying that there is a positive mass squared for fluctuations about the minimum. In~other words, we examine the mass matrix Hessian
\beq
H = \displaystyle \left(\begin{array}{cc}  \displaystyle \frac{\partial^2V(v_1,v_2)}{\partial v_1^2} & \displaystyle \frac{\partial^2V(v_1,v_2)}{\partial v_1\partial v_2} \\ \displaystyle \frac{\partial^2V(v_1,v_2)}{\partial v_2\partial v_1} & \displaystyle \frac{\partial^2V(v_1,v_2)}{\partial v_2^2} \end{array}\right)_{\displaystyle in \, \, minimum}\hspace{-21mm}.
\eeq
In order to have a stable minimum the matrix of second derivatives needs to be positively definite. This does not mean that all second derivatives need to be positive.
Both eigenvalues, $e_i$, need to be positive to ensure the stability of the minimum. The~abo\-ve matrix is to be evaluated at the minimum, setting to zero the derivatives evaluated at the vevs (i.\,e. where $\displaystyle\frac{\partial V}{\partial v_1}=\frac{\partial V}{\partial v_2}=0$). Since invariant ${\rm{Tr}}(H)=e_1+e_2$ and ${\rm{Det}}(H)=e_1e_2$, we can simply require $$\displaystyle\frac{\partial^2V(v_1,v_2)}{\partial v_1^2}+\frac{\partial^2V(v_1,v_2)}{\partial v_2^2}>0, \quad {\rm and}  ~ ~ ~ ~ {\rm{Det}}(H)\geqslant0.$$ Straight-forward algebra gives:
\beq
H = \left(\begin{array}{cc} \displaystyle 2\lambda_1v_1^2 + {\tt Re} \mu_{12}^2\f{v_2}{v_1} & - \,{\tt Re} \mu_{12}^2 + \lambda_{345}v_1v_2 \\  - \,{\tt Re} \mu_{12}^2 + \lambda_{345}v_1v_2 & 2\lambda_2v_2^2 + \displaystyle {\tt Re} \mu_{12}^2\f{v_1}{v_2}\end{array}\right).
\eeq
We denote $\lambda_{345} = \lambda_{3} +\lambda_{4}+{\tt Re} \lambda_{5}$.

This shows
that stability conditions are equivalent to:
\beq
2\lambda_1v_1^2+ 2\lambda_2v_2^2+ \left(\f{v_2}{v_1}+\f{v_1}{v_2}\right){\tt Re}\mu_{12}^2 >0,\label{eq:traseposit}\eeq
and
\beq 4\lambda_1\lambda_2 + 2\lambda_1 \f{v_1}{v_2^3}{\tt Re}\mu_{12}^2 + 2\lambda_2 \f{v_2}{v_1^3}{\tt Re}\mu_{12}^2 \, \geqslant \, (\lambda_{345})^2 - 2\lambda_{345}{\tt Re}\mu_{12}^2, \label{eq:8en-minv1v2}
\eeq
by taking first derivatives with respect to all the scalar fields, and setting them equal to~zero (i.\,e. in minimum).

In
particular case ${\tt Re}\mu_{12}^2=0$ we see simply
\beq
4\lambda_1\lambda_2 \geqslant (\lambda_{345})^2, \, \quad {\rm and}  ~ ~ ~ ~ \lambda_1> 0, \, \lambda_2 > 0.
\eeq

In addition, we will requite concavity, which is implied to
be positive, which assures us that we're at a minimum and not at a saddle point. The tricky part of the problem is to show that indeed it is possible to have the vevs parallel, i.\,e. of the form of~(\ref{eq:6-vevs}).\,\footnote{Thus, in fact, condition (\ref{eq:8en-23}) is equivalent to positive sign of squared mass for third Higgs (pseudoscalar) boson $\displaystyle m_3^2=-v_2^2\mbox{Re}\lambda_5 + \mbox{Re}\mu_{12}^2 \frac{v_2}{v_1}$, and condition (\ref{eq:8en-24}) at large field values is equivalent to positive sign of squared mass for charge Higgs boson $\displaystyle m_{H^\pm}^2=-\,\frac{v_2^2}{2}(\lambda_4+\mbox{Re}\lambda_5) + \mbox{Re}\mu_{12}^2 \frac{v_2}{v_1}$.} To~do this, the best way is to use the $SU(2)$ rotation to force the vev of $\phi_1$ to have the~right form. Then all we need to show is that the potential is minimized (i.\,e.~the~appropriate derivatives satisfy the conditions stated above) when $\phi_2$ takes
the right form.

Thus the conditions for a local stable minimum of the potential in this problem are~(\ref{eq:8en-23largefield}), (\ref{eq:8en-24}), (\ref{eq:traseposit}), (\ref{eq:8en-minv1v2}), see also \cite{SherNie}.

\textbf{(c)} In the SM, there are 4 degrees of freedom in the Higgs doublet,
three of which
are consumed
by the $W^+$, $W^-$ and $Z^0$. When there are 2 Higgs
doublets, they
contain in total 8~degrees of freedom, so the 5 remaining after goldstones are
consumed. Now, the vev given breaks the~$SU(2)\times U(1)_Y$ symmetry to $U(1)_{em}$, therefore in general there
would be 3~goldstone bosons, and 5 physical Higgs fields. Two of these remaining degrees of freedom are charged, and three are  neutral.
The task in this problem is to determine which two charged degrees of freedom are eaten, and which two charged degrees of freedom remain. The simplest way to do this is to consider the components of the two Higgs doublets as being part of a larger vector.
An orthogonal transformation will rotate the~different components amongst themselves.
In particular, we can find the basis where the~vev is entirely in one neutral component.
The charged piece associated with this neutral component is the would-be goldstone boson that
is eaten.
The~charged Higgs is the piece which is orthogonal to this goldstone boson that
is eaten.
So,~considering the neutral components (which have the vevs), we have:
\beq
\left(\begin{array}{c}  \phi'_1 \\ \phi'_2\end{array}\right) = \left(\begin{array}{cc}  \cos\beta & \sin\beta \\ - \sin\beta & \cos\beta \end{array}\right)
\left(\begin{array}{c}  \phi_1 \\ \phi_2\end{array}\right).
\eeq
So, putting all of the vacuum in one Higgs field\footnote{Once one goes to the proper basis for describing the Higgs mechanism, there is really only one doublet (in the above case it is $\phi'_1$) that acts as the Higgs and has three of four degrees of freedom that are eaten. So, we can say that it is a model with "two-complex scalars".}
(say $\phi'_1$), we get
\beq
\left(\begin{array}{cc}  \cos\beta & \sin\beta \\ - \sin\beta & \cos\beta \end{array}\right) \left(\begin{array}{c}  v_1 \\ v_2\end{array}\right) =
\left(\begin{array}{c}  v \\ 0\end{array}\right).\label{eq:8_14}
\eeq
It follows from (\ref{eq:8_14}) that $\displaystyle\mbox{tg}\beta=\f{v_2}{v_1}$.
So, the
whole vev lives in the $\phi{\bf'}_1$ field, which means that the charged component of $\phi{\bf'}_1$ is the Goldstone Boson eaten by the $W$. That means $\phi'_1 =\phi_1\cos\beta+\phi_2\sin\beta$ is the Goldstone, and
\beq
\phi'_2 = H^+=\phi_1^+\sin\beta-\phi_2^+\cos\beta
\eeq
is the physical charged Higgs field.

\textbf{(d)} In the models with two doublets of scalar fields {\it CP} invariance can be violated by the terms of the potential containing $(\phi_1^\dagger\phi_2)$ or $(\phi_2^\dagger\phi_1)$ with the complex parameters $\mu_{12}^2$ and $\lambda_5$. In the case of real parameters {\it CP} invariance is not broken.

\section*{9. By the cradle of {\sc\bf LHC \footnote{\it Large Hadron Collider.}}}
\addcontentsline{toc}{section}{9. BY THE CRADLE OF {\sc LHC}}
\setcounter{equation}0

Generally, it is possible to describe a scattering experiment in the following way\\ (see fig.\,3):\\
\parbox{0.75\textwidth}{1) sufficiently wide uniform beam of particles is prepared so that it is possible to assume
the momentum of each particle be equal to $\vec
p_0=\hbar \vec k$, where $\vec k$ is wave vector, $\hbar$ -- Planck constant;

2) this beam of particles is directed to stationary target consisting of
identical particles;

3) at certain distance from the target, products of reaction of the particles from the
beam with the particles forming the target, are registered at different angles.}\hfill\hfill\hspace{2mm}
\parbox{0.24\textwidth}{\includegraphics[scale=0.4]{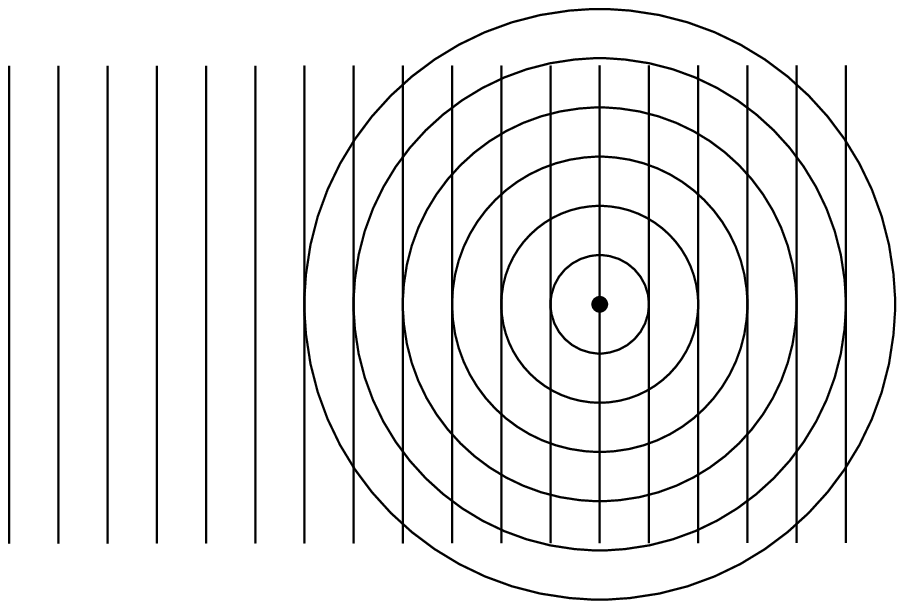}
\vspace{-10mm}    \center{\small
Fig. 3}}

\vspace{1mm}
Thus at sufficiently
large distances from the target
the wave function of the particles is the superposition of plane incident wave
$\psi_{||||}=\exp \{i \vec k\ \vec r\}$ and spherical scattered wave $\psi_\bigodot=\exp \{i  k r\}/r$. Here, $\vec r=x\vec {\mbox{i}} +y \vec {\mbox{j}} +z \vec {\mbox{k}}$, $r=|\vec r|$.

\textbf{(a)} Calculate flux density of the probability $\displaystyle\vec j=\frac{\hbar
}{2\,m\,i}(\psi ^*\vec \nabla \psi - \psi\vec \nabla \psi ^* )$ for
the wave functions $\displaystyle\psi_\rightsquigarrow=e^{i\vec k\cdot \vec x}$ and
$\displaystyle\psi_\bigodot$.
Here,
$m$ is the mass of the particle, and $\vec x =
\vec i x$.

\textbf{(b)} Illustrate the obtained result with the help of the graph: draw the pattern of the vector field $\vec j$ (lines of the $\vec j$ vector) in both cases.
To build such graph, use any available computer software suitable for building graphics (plots).

\textbf{(c)} Prove that $\vec \nabla \cdot \vec j_\rightsquigarrow=0,$ ~  $\vec \nabla \cdot
\vec j_\bigodot\sim \delta (\vec r)$.\\

{\bf SOLUTION}
\setcounter{equation}0

\textbf{(a)}
$$ \vec j_\rightsquigarrow=\frac{\hbar
}{2mi}(\psi_\rightsquigarrow^* \vec \nabla \psi_\rightsquigarrow - \psi_\rightsquigarrow\vec \nabla \psi
^*_\rightsquigarrow) = \frac{\hbar}{2mi} \ \vec{\mbox{i}}\ [e^{-i\vec k \vec x}\ i k_x \ e^{i\vec k \vec x} - e^{i\vec k
\vec x}(-ik_x) \ e^{-i\vec k \vec x}\ ]=\vec{\mbox{i}} \,
\frac{\hbar k_x}{m}, $$
$$\vec j_\bigodot=\frac{\hbar
}{2mi}(\psi_\bigodot^*\vec \nabla \psi_\bigodot- \psi_\bigodot\vec \nabla \psi ^*_\bigodot)=$$
$$ = \frac{\hbar }{2mi} \ \vec n_r\Bigl[\frac{e^{-i kr}}{r}\
\frac{d}{dr} \ \frac{e^{i kr}}{r}- \Bigl(\frac{d}{dr} \
\frac{e^{-i kr}}{r}\Bigr)\ \frac{e^{ikr}}{r}\Bigr]=\frac{\hbar
k}{m}\frac{\vec n_r}{r^2}.$$

Here, $\vec n_r =\vec r/r. $

\vspace{1mm}
\textbf{(b)}

The field in both cases is represented by a graph in $x$--$y$ plane at
$z=0$ (in \textsl{Mathema\-ti\-ca} realization, see fig.\,4).
\begin{center}
\includegraphics[scale=0.73]{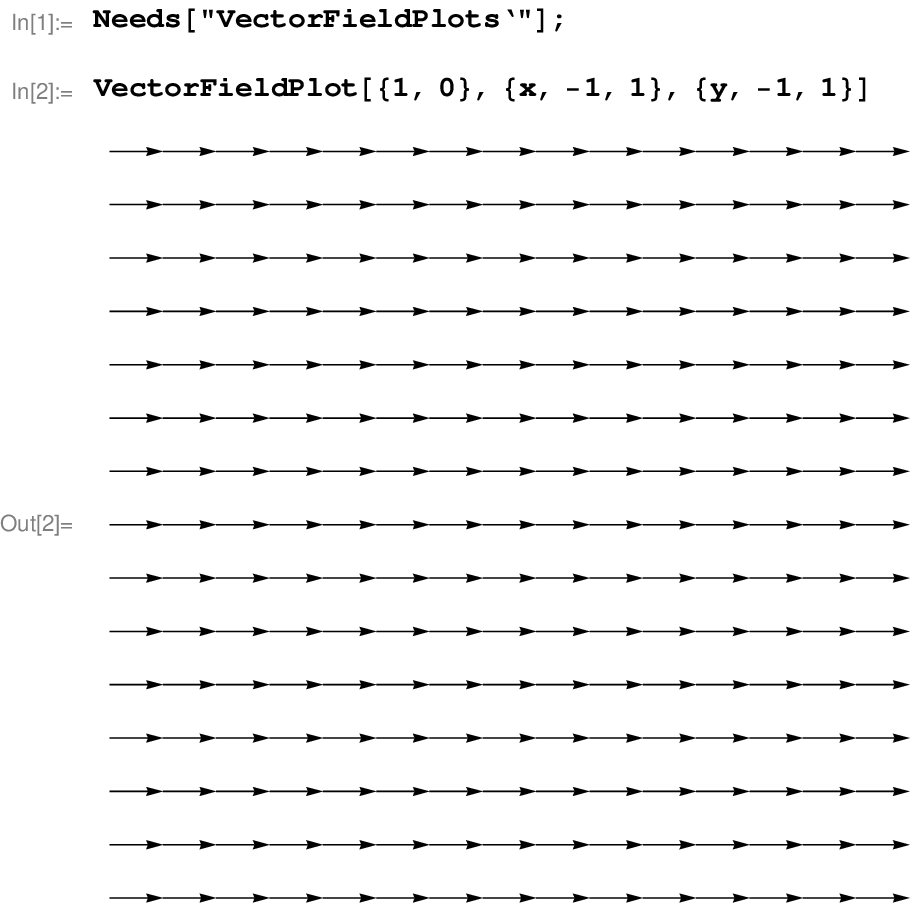}\hspace{6mm}
\includegraphics[scale=0.73]{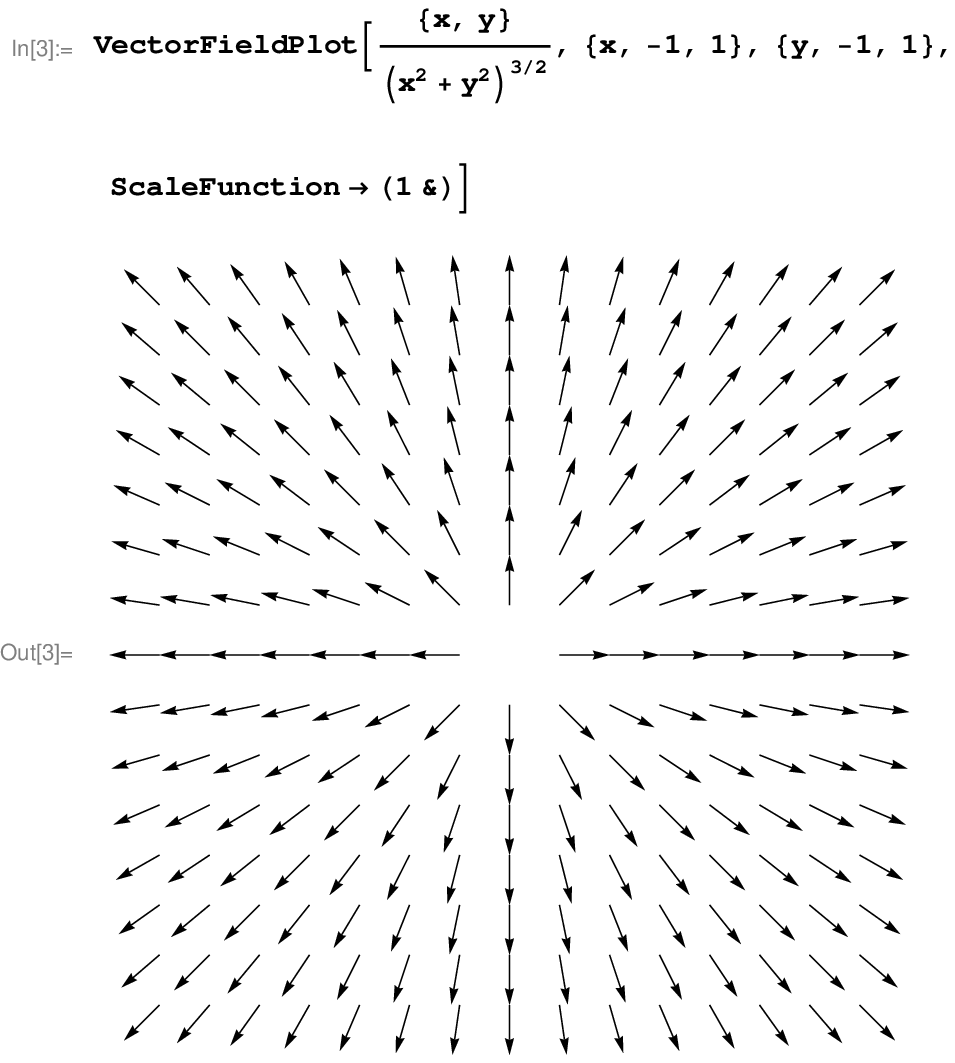}
\vspace{-2mm}    \center{\small {\it a} ~~ \phantom{yyyyyyyyyyyyyyyyyyyyyyyyyyyyyyyyyyyy} ~~ {\it b}\\Fig.
4}
\end{center}

It is possible to see that the spherical wave $\psi_\bigodot$ is diverging from
the origin of coordinates.

\vspace{1mm}
\textbf{(c)}

Vector field $\vec j_\rightsquigarrow$ is uniform, $\displaystyle\vec j_\rightsquigarrow=\vec {\mbox i}\
\frac{\hbar  k_x}{m}=\vec{\rm const}$, so that $\vec \nabla \cdot \vec j_\rightsquigarrow=0$.

It is possible to represent vector $\vec j_\bigodot$ in the form:
$$\vec j_\bigodot=\frac{\hbar
k}{m}\frac{\vec n_r}{r^2}=\frac{\hbar k}{m}\frac{\vec
r}{r^3}=-\frac{\hbar k}{m}\vec \nabla \frac{1}{r}.$$ Hence
$$\vec \nabla \cdot \vec j_\bigodot=-\frac{\hbar k}{m}\Delta \frac{1}{r}=\frac{\hbar
k}{m}4\pi \delta(\vec r).$$

\section*{10. ‘Whipping Top-Toy' from Samara
\footnote{ Place in Russia where this competition is held and assessed.}}
\addcontentsline{toc}{section}{10. ‘WHIPPING Top-Toy' from SAMARA}

Rigid ball of the mass $m$ with radius $R$ rests on smooth rigid horizontal surface. Center of mass $C$ of the ball is at the distance of $l$ from its geometric center~$O$. Mass~of the material is symmetrically distributed along the volume of the ball relatively to $OC$ axis, and also any plane containing that axis. Moments of inertia of the ball relatively to $OC$ axis and any axis passing the center of mass and perpendicular to $OC$ axis, are equal respectively to $J_0$ and $J$. During certain time period, the ball is accelerated around the static vertical axis passing its center~$O$. The moment of time when the action of the "accelerating" \,forces is finished, is chosen as the time origin. At this moment the ball has the angular velocity $\overrightarrow{\omega}_0$, directed vertically up, and $OC$ axis makes
some angle $\varepsilon$ with the rotation axis of the ball. The center of mass of the ball is lower than its mechanical center (see fig.\,5).

\begin{center}
\includegraphics[width=6cm,height=5.4cm]{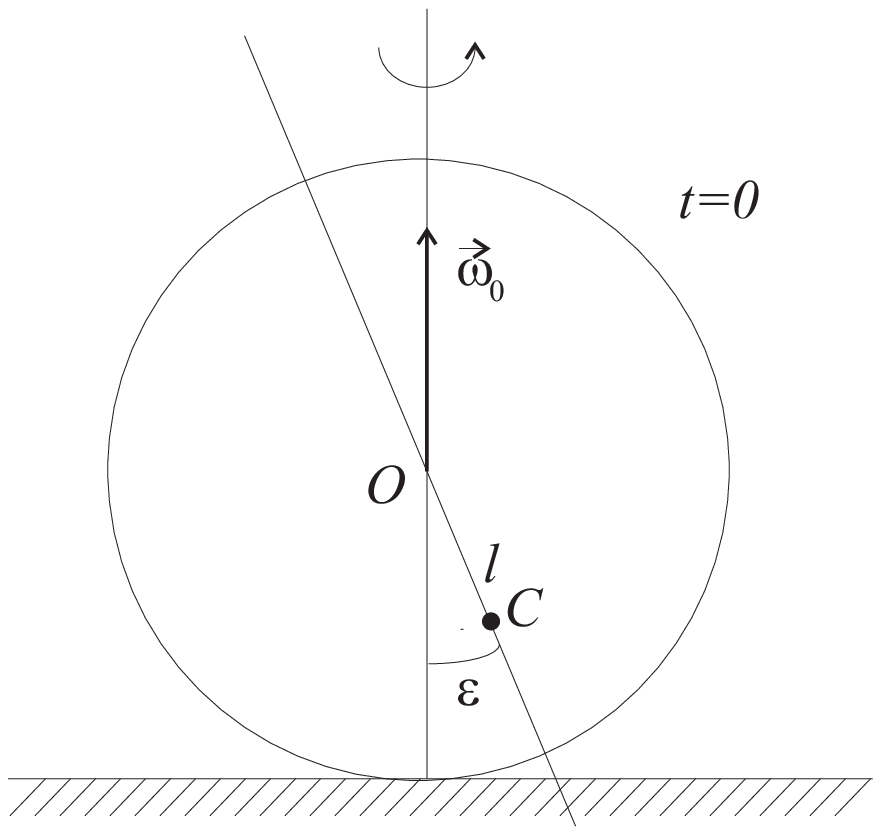}\\
{\small Fig. 5. {\it Rotating ball at time moment $t=0$\\[2mm]}}
\end{center}


Study the motion of the ball at $t>0$. Obtain the equation of motion of the ball and find the integrals of motion (conserved values). Show that the center of mass of the ball is lifting up (at a certain relationship between the parameters), and find out, whether it can approach its top position at which the $\overrightarrow{OC}$ vector is directed vertically up.\\

\setcounter{equation}0

{\bf SOLUTION}

In order to describe the motion of the ball we introduce an inertial frame of reference $S$ (with axes $X$, $Y$, $Z$),
that is at rest relatively to the horizontal surface, and an noninertial frame $S'$ (with axes $X'$, $Y'$, $Z'$),
that is rigidly bound to the ball. The origin of the frame $S$ coincides with the initial position of the center of the ball, the axis $Z$ is directed vertically and the plane $YOZ$ is chosen so that it contains the axis $OC$ at the initial moment $t=0$. The origin of the frame $S'$ coincides with the mass center of the ball and the axis $Z'$ --
with the axis $OC$, so that at the moment $t=0$ the axis $Z'$ makes an angle $\varepsilon$ with the axis $Z$. Additionally, the axis $X'$ at this moment has the direction similar to that of the horizontal axis $X$, and the axis $Y'$ belongs to the plane $YOZ$ (see~fig.\,6).

\begin{center}
\includegraphics[width=7cm,height=5.4cm]{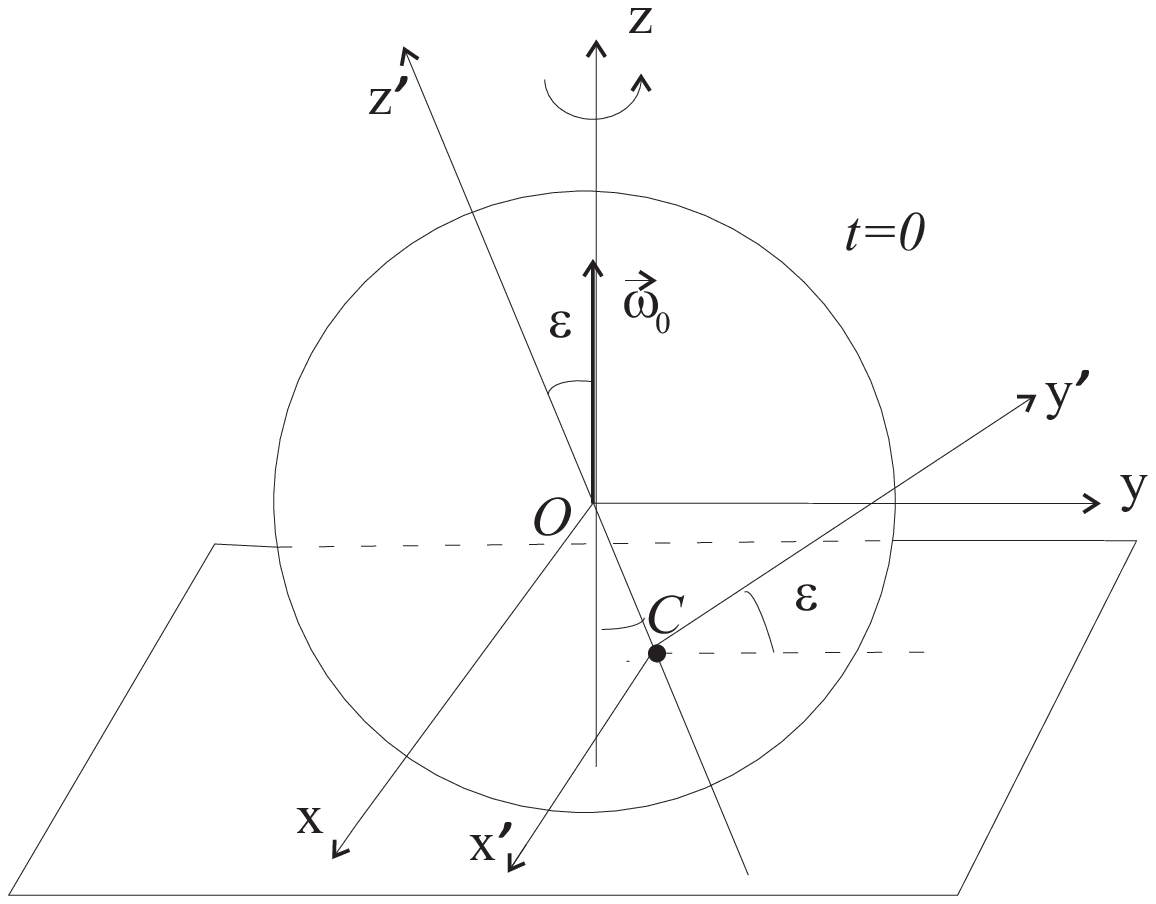}\\[1mm]
{\small Fig. 6. {\it Coordinate axes and the origins of $S$, $S'$ at the moment $t=0$\\[2mm]}}
\end{center}

The reference frames $S$ and $S'$ being chosen, the Euler angles $\varphi$, $\theta$, $\psi$ that define the orientation
of the ball (and the frame $S'$) relatively to the frame $S$, are given by the following values at the initial moment:
\beq
\varphi(0)=0, \,\,\, ~ \theta(0)=\varepsilon, ~ \,\,\, \psi(0)=0.
\label{e101}
\eeq

Components of the angular velocity $\vec{\omega}$ of the ball in the frames $S$ and $S'$ at any instant of time are given by the cinematic Euler's formulas:
\beq
\left\{\begin{array}{l} {\omega_x=\dot{\theta}\cos\varphi+\dot{\psi}\sin\theta\sin\varphi, } \\ {\omega_y=\dot{\theta}\sin\varphi-\dot{\psi}\sin\theta\cos\varphi, } \\
{\omega_z=\dot{\varphi}+\dot{\psi}\cos\theta,}\end{array}\right.
\label{e102}\eeq
\beq \left\{\begin{array}{l} {\omega_{x'}=\dot{\varphi}\sin\theta\sin\psi+\dot{\theta}\cos\psi, } \\ {\omega_{y'}=\dot{\varphi}\sin\theta\cos\psi-\dot{\theta}\sin\psi, } \\
{\omega_{z'}=\dot{\varphi}\cos\theta+\dot{\psi}.}\end{array}\right.
\label{e103}
\eeq

For the initial values of the components of the angular velocity in $S$ the equations~(\ref{e102}), with the account of~(\ref{e101}), give
\beq
\left\{\begin{array}{l} {\omega_x(0)=\dot{\theta}(0), } \\ {\omega_y(0)=-\dot{\psi}(0)\sin\varepsilon, } \\
{\omega_z(0)=\dot{\varphi}(0)+\dot{\psi}(0)\cos\varepsilon.}\end{array}\right.
\label{e104}
\eeq

On the other hand, it is known that the initial angular velocity $\vec{\omega}(0)=\vec{\omega}_0$ has the same direction as the axis $Z$ and, hence,
\beq
\omega_x(0)=\omega_y(0)=0, \,\,\, \omega_z(0)=\omega_0.
\label{e105}
\eeq

Comparing~(\ref{e104}) and~(\ref{e105}), one finds the initial values for the time derivatives of the Euler angles:
\beq
\dot{\varphi}(0)=\omega_0, \,\,\, \dot{\theta}(0)=0, \,\,\, \dot{\psi}(0)=0.
\label{e106}
\eeq

As seen from the pictures, the cartesian coordinates $x_m$, $y_m$, $z_m$ of the mass center of the ball in $S$ at the initial moment of time have the following values:
\beq
x_m(0)=0, \,\,\, y_m(0)=l\sin\varepsilon, \,\,\, z_m(0)=-l\cos\varepsilon.
\label{e107}
\eeq

For $t<0$ (until the rotating forces action has stopped) the center of mass of the ball moves in the horizontal plane along the circle of the radius $l\sin\varepsilon$ with the axis $Z$ passing through its center. Therefore, the velocity $\vec{v}_m$ of the mass center of the ball in the frame $S$ at the initial instant of time is given by
\beq
\vec{v}_m(0)=[\vec{\omega}_0\,\,\vec{r}_m(0)],
\label{e108}
\eeq
where $\vec{r}_m(0)$ is the initial value of the radius-vector $\vec{r}$ of the mass center of the ball in $S$ (that is the vector with the components $x_m(0)$, $y_m(0)$, $z_m(0)$). From ~(\ref{e105}), (\ref{e107}) and~(\ref{e108}) it follows that the initial velocity of the mass center of the ball in $S$ is directed opposite to the axis $X$ and is equal to
\beq
v_m(0)=\omega_0l\sin\varepsilon,
\label{e109}
\eeq
or
\beq
\dot{x}_m(0)=-\omega_0l\sin\varepsilon, \,\,\, \dot{y}_m(0)=\dot{z}_m(0)=0.
\label{e1010}
\eeq

The equations of motion of the ball for $t>0$ can be derived as the Lagrange equations of the 2-nd kind. The ball is a mechanical system with five degrees of freedom. One can use the Euler angles $\varphi$, $\theta$, $\psi$ and coordinates $x_m$, $y_m$ of the mass center in $S$ as the independent generalized coordinates of the ball, the coordinate $z_m$ at any instant of time being defined by
\beq
z_m=-l\cos\theta.
\label{e1011}
\eeq

The Lagrange function has the form
\beq
L=T-U,
\label{e1012}
\eeq
where $T$ is the kinetic energy of the ball and $U$ is its gravitational potential energy (these values are defined in the frame $S$).

The kinetic energy of the ball may be expressed in the form
\beq
T=\frac{mv_m^2}{2}+\tau,
\label{e1013}
\eeq
where $\tau$ is its kinetic energy of rotation. As the axes $x'$, $y'$, $z'$ are directed along the principal axes of inertia of the ball, then it follows
\beq
\tau=\frac{1}{2}(J_{x'}\omega_{x'}^2+J_{y'}\omega_{y'}^2+J_{z'}\omega_{z'}^2),
\label{e1014}
\eeq
where $J_{x'}$, $J_{y'}$, $J_{z'}$ are the moments of inertia of the ball relatively to the axes $x'$, $y'$, $z'$, with, according to the problem statement,
\beq
J_{x'}=J_{y'}=J, \,\,\, J_{z'}=J_0.
\label{e1015}
\eeq

Inserting the expressions~(\ref{e103}) and~(\ref{e1015}) into~(\ref{e1014}) one gets
\beq
\tau=\frac{J}{2}(\dot{\varphi}^2\sin^2\theta+\dot{\theta}^2)
+\frac{J_0}{2}(\dot{\varphi}\cos\theta+\dot{\psi})^2.
\label{e1016}
\eeq

The expression for the velocity's square of the mass center of the ball with the account of~(\ref{e1011}) gets the form
\beq
v_m^2=\dot{x}_m^2+\dot{y}_m^2+l^2\sin^2\theta\dot{\theta}^2.
\label{e1017}
\eeq

From~(\ref{e1013}), (\ref{e1016}) and~(\ref{e1017}) it follows that
$$ T=\frac{m}{2}(\dot{x}_m^2+\dot{y}_m^2)+\frac{m}{2}l^2\sin^2\theta\dot{\theta}^2+
\frac{J}{2}(\dot{\varphi}^2\sin^2\theta \,+
\dot{\theta}^2)+$$
\beq + \frac{J_0}{2}\left(\dot{\varphi}^2\cos^2\theta+2\dot{\varphi}\dot{\psi}\cos\theta+\dot{\psi}^2\right).
\label{e1018}
\eeq

The potential energy of the ball with the account of~(\ref{e1011}) is given by the expression
\beq
U =-mgl\cos\theta.
\label{e1019}
\eeq

Inserting the equations~(\ref{e1018}) and~(\ref{e1019}) into~(\ref{e1012}) one gets the following explicit expression for the Lagrange function of the ball:
$$ L=\frac{m}{2}(\dot{x}_m^2+\dot{y}_m^2)+\frac{1}{2}(ml^2\sin^2\theta+J)\,\dot{\theta}^2 +\frac{1}{2}(J\sin^2\theta+J_0\cos^2\theta)\dot{\varphi}^2+$$ \beq + \frac{J_0}{2}\dot{\psi}^2+J_0\cos\theta\dot{\varphi}\dot{\psi}+mgl\cos\theta.
\label{e1020}
\eeq

It is known that the Lagrange equations of the 2-nd kind for a system with ideal holonomic constraints and $S$ degrees of freedom under the absence of dissipative forces have the form
\beq
\frac{d}{dt}\left(\frac{\partial L}{\partial \dot{q}_j}\right) -\frac{\partial L}{\partial q_j}=0 \,\,\, (j=1,2,..., S),
\label{e1021}
\eeq
where $q_j (j=1,2,..., S)$ are independent generalized coordinates of the system.

If $q_j$ is a cyclic coordinate of the system (that is, the Lagrange function doesn't depend on it), then the corresponding generalized momentum $\displaystyle p_j=\frac{\partial L}{\partial\dot{q}_j}$ is the integral of motion.

According to~(\ref{e1020}), $x_m$, $y_m$, $\varphi$, $\psi$ are cyclic coordinates of the ball. Therefore, the generalized momentums
\beq
\frac{\partial L}{\partial \dot{x}_m}=m\dot{x}_m,
\label{e1022}
\eeq
\beq
\frac{\partial L}{\partial \dot{y}_m}=m\dot{y}_m,
\label{e1023}
\eeq
\beq
\frac{\partial L}{\partial \dot{\varphi}}=J\sin^2\theta\dot{\varphi}+ J_0\cos\theta(\cos\theta\dot{\varphi}+\dot{\psi}),
\label{e1024}
\eeq
\beq
\frac{\partial L}{\partial \dot{\psi}}=J_0(\cos\theta\dot{\varphi}+\dot{\psi})
\label{e1025}
\eeq
are the integrals of motion of the ball that are constant and are defined by the initial conditions~(\ref{e101}), (\ref{e106}), (\ref{e107}) and~(\ref{e1010}). Writing down the corresponding conservation laws, one gets the following system of the 1-st order differential equations:
\beq
\dot{x}_m=-\omega_0l\sin\varepsilon,
\label{e1026}
\eeq
\beq
\dot{y}_m=0,
\label{e1027}
\eeq
\beq
J\dot{\varphi}\sin^2\theta+J_0\omega_0\cos\varepsilon\cos\theta=\omega_0(J\sin^2\varepsilon
+J_0\cos^2\varepsilon),
\label{e1028}
\eeq
\beq
\dot{\varphi}\cos\theta+\dot{\psi}=\omega_0\cos\varepsilon
\label{e1029}
\eeq
(when deriving the equation~(\ref{e1028}), that presents the conservation law for the generalized momentum $\displaystyle p_\varphi=\frac{\partial L}{\partial\dot{\varphi}}$, the equation~(\ref{e1029}) was used).

From ~(\ref{e1026}) and~(\ref{e1027}) and the initial conditions~(\ref{e107}) it follows that
\beq
x_m=-\omega_0 t l\sin\varepsilon,
\label{e1030}
\eeq
\beq
y_m=l\sin\varepsilon.
\label{e1031}
\eeq

Hence, the mass center of the ball moves in the vertical plane that is normal to the axis $y$ and that passes through the initial position of the mass center. The horizontal component of the ball's velocity stays constant and equal to the initial velocity $\vec{v}_m(0)$.

Note that the equations~(\ref{e1026}) -- (\ref{e1028}) may be also derived from the conservation laws for the components of momentum and angular momentum of the mechanical system in an inertial frame. In order to do this one has to consider the directions of external forces, applied to the ball. These forces are the reaction force $\vec{N}$ that is directed vertically up and the gravitational forces that are applied to every element of the ball and are directed vertically down, their resultant force $m\vec{g}$ being applied to the mass center of the ball.

As the sum of the external forces that are applied to the ball, is directed vertically, the $X$ and $Y$ components of the ball's momentum
\beq
\vec{P}=m\vec{v}_m
\label{e1032new}
\eeq
are conserved for $t\geq0$. The corresponding conservation laws, with the account of the initial conditions~(\ref{e1010}), give the equations~(\ref{e1026}) and~(\ref{e1027}).

The angular momentum $\vec{M}$ of the ball in the frame $S$ may be presented in the form
\beq
\vec{M}=[\vec{r}\vec{P}]+\vec{\mu},
\label{e1033new}
\eeq
where $\vec{\mu}$ is the
intrinsic angular momentum
momentum of the ball (that is the angular momentum of the ball in the mass center reference frame that moves translatory to $S$). The components of $\vec{\mu}$ in $S'$ are defined by
\beq
\mu_{x'}=J\omega_{x'}, \,\,\, \mu_{y'}=J\omega_{y'}, \,\,\, \mu_{z'}=J_0\omega_{z'}.
\label{e1034new}
\eeq
As the moments of all external forces in $S$ that are acting on the ball are directed horizontally for $t\geq 0$, the projection $M_z$ of the angular momentum on the axis $Z$ is the integral of motion. Taking into account~(\ref{e1030}), (\ref{e1031}), (\ref{e1011}), (\ref{e1026}) and~(\ref{e1027}), it easy to check that $Z$-component of the vector $[\vec{r}\vec{P}]$ has the constant value equal to $m\omega_0l^2\sin^2\varepsilon$. Hence, with the account of~(\ref{e1033new}), it follows that
\beq
M_z=m\omega_0l^2\sin^2\varepsilon+\mu_z,
\label{e1035new}
\eeq
or that $Z$-component of the ball's
rotate momentum $\mu_z$ is an integral of motion.

Using the well-known linear transformation law for the components of an arbitrary vector for a frame rotation and expressing all the coefficients of such transformation as the functions of Euler angles,
one may show that
\beq
\mu_z=\sin\theta\sin\psi\mu_{x'}+\sin\theta\cos\psi\mu_{y'}+\cos\theta\mu_{z'}.
\label{e1036new}
\eeq

Inserting the equations~(\ref{e1034new}), where the angular velocity components are given by the formulas~(\ref{e103}),  into~(\ref{e1036new}), one gets
\beq
\mu_z=(J\sin^2\theta+J_0\cos^2\theta)\dot{\varphi}+J_0\cos\theta\dot{\psi}.
\label{e1037new}
\eeq

Comparing the equations~(\ref{e1037new}) and~(\ref{e1024}), one sees that $\mu_z$ coincides with the generalized momentum $p_\varphi$. Hence the conservation law for $\mu_z$, written with the account of the initial conditions~(\ref{e101}) and~(\ref{e106}) and of the equation~(\ref{e1029}), is the equation~(\ref{e1028}).


%
%
%


Inserting the equation~(\ref{e1020}) into~(\ref{e1021}) and letting $q_j=\theta$, the Lagrange equation for the ball that corresponds to the generalized coordinate $\theta$ is obtained:
\beq
(ml^2\sin^2\theta+J)\ddot{\theta}+ml^2\sin\theta\cos\theta\dot{\theta}^2+ (J_0-J)\sin\theta\cos\theta\dot{\varphi}^2+
\label{e1032}
\eeq
$$
+J_0\sin\theta\dot{\varphi}\dot{\psi}+mgl\sin\theta=0.
$$

Note that the equations~(\ref{e1028}) and~(\ref{e1029}) form a system of linear algebraic equations relatively to $\dot{\varphi}$ and $\dot{\psi}$ with the coefficients that depend only on the angle $\theta$. Having solved this system, one gets the dependencies of these generalized velocities on $\theta$:
\beq
\dot{\varphi}=\frac{\omega_0}{J\sin^2\theta}(J\sin^2\varepsilon+ J_0\cos^2\varepsilon-J_0\cos\varepsilon\cos\theta),
\label{e1033}
\eeq
\beq
\dot{\psi}=\frac{\omega_0}{J\sin^2\theta}[\cos\varepsilon(J\sin^2\theta+ J_0\cos^2\theta)-\cos\theta(J\sin^2\varepsilon+ J_0\cos^2\varepsilon)].
\label{e1034}
\eeq

After having inserted the expressions~(\ref{e1033}) and~(\ref{e1034}) into~(\ref{e1032}) one may get the nonlinear 2-nd order differential equation that defines the dependency $\theta(t)$. In order to get this dependency one may also use the conservation law for the total mechanical energy of the ball
\beq
E=T+U.
\label{e1035}
\eeq
This quantity coincides with the generalized energy of the ball and is conserved due to the absence of both dissipative forces and explicit dependency of the Lagrange function on time.

Setting the explicit expression for the energy of the ball that  follows from~(\ref{e1018}), (\ref{e1019}) and~(\ref{e1035}), equal to its initial value, that is found from~(\ref{e101}), (\ref{e106}) and~(\ref{e1010}), and taking into account the conservation laws~(\ref{e1026}) and~(\ref{e1027}), one gets
$$ (ml^2\sin^2\theta+J)\dot{\theta}^2+(J\sin^2\theta+J_0\cos^2\theta)\dot{\varphi}^2+ J_0\dot{\psi}^2+2J_0\cos\theta\dot{\varphi}\dot{\psi}-2mgl\cos\theta=$$
\beq=(J\sin^2\varepsilon+J_0\cos^2\varepsilon)\omega_0^2-2mgl\cos\varepsilon.
\label{e1036}\eeq

Substituting the expressions~(\ref{e1033}) and~(\ref{e1034}) into the last equation, one
obtains the following nonlinear 1-st order differential equation that defines the dependency $\theta(t)$:
\beq
\sin^2\theta (1+\beta\sin^2\theta) \,
\dot{\theta}^2 =\omega_0^2 \, (\cos{\varepsilon}-\cos{\theta})\Theta(\theta),
\label{e1037}
\eeq
where $\Theta(\theta)$ is quadratic in $\cos\theta$ and has the form:
\beq
\Theta(\theta)=\mathfrak{a}_0+2\mathfrak{a}_1\cos\theta+\mathfrak{a}_2\cos^2\theta,
\label{e1038}
\eeq
where $\mathfrak{a}_0-\mathfrak{a}_2$ are given by:
\beq
\mathfrak{a}_0=-2\beta\gamma-\frac{1}{4}\left((\alpha -1)^2\cos{3\varepsilon}+(\alpha +1)(3\alpha-1)\cos{\varepsilon}\right),
\label{e1039}
\eeq
\beq
\mathfrak{a}_1=\frac{1}{4}\left((\alpha ^2-1)\cos{2\varepsilon}+\alpha ^2+1\right),
\label{e1040}
\eeq
\beq
\mathfrak{a}_2=2\beta\gamma,
\label{e1041}
\eeq

Dimensionless parameters $\alpha$, $\beta$, $\gamma$ in equations~(\ref{e1039}) -- (\ref{e1041}) and equation~(\ref{e1037}), are related to the known parameters in the following way:
\beq
\alpha=\frac{J_0}{J},
\label{e1045}
\eeq
\beq
\beta=\frac{ml^2}{J},
\label{e1046}
\eeq
\beq
\gamma=\frac{g}{l\omega_0^2}=\frac{\omega_m^2}{\omega_0^2}
\label{e1047}
\eeq
(here $\omega_m=\sqrt{g/l}$ is the cyclic frequency of a flat simple pendulum of length $l$).

The region of motion of the ball top is defined by the sign and the roots of the function~$\Theta(\theta)$. In order to investigate them one has to specify the range of the parameters that are present in the function. As $J,J_0\geq0$ and as for a particular case of the ball being a rotator the moments of inertia have the values $J_0=0,J\neq0$, we conclude that $\alpha$ is not negative. On the other hand, it is known that none
of the principle moments of inertia is
bigger than the sum of the other two, which gives $J_0\leq2J$. Therefore, we conclude that $0\leq\alpha\leq2$. Further, as it is easy to notice, $0\leq\beta<\infty$ and $0<\gamma<\infty$.

Taking into account the equations~(\ref{e1039}) -- (\ref{e1041}), one finds that:
\beq
\label{e104701}
0\leq\mathfrak{a}_1\leq2,
\eeq
\beq
\label{e104702}
0\leq\mathfrak{a}_2<\infty.
\eeq

Concerning $\mathfrak{a}_0$, one may notice that this parameter is not bound from below and approaches its maximum value with respect to $\beta\gamma$ at $\beta\gamma=0$. Further analysis shows that for $0\leq\alpha\leq2,0\leq\varepsilon\leq\pi$ the parameter $\mathfrak{a}_0$ reaches its maximum value, that is equal to $\mathfrak{a}_0=4$, at $\alpha=2, \varepsilon=\pi$. For small $\varepsilon$ the parameter $\mathfrak{a}_0$ is maximized with respect to $\alpha$ at $\alpha=0$ and increases monotonically with $\varepsilon$. Hence, we finally conclude that
\beq
\label{e104703}
-\infty<\mathfrak{a}_0\leq4.
\eeq

From the form of the equation~(\ref{e1037}) one can conclude that for
$\Theta(\varepsilon)>0$ the motion of the top will always go on in the area where
$\cos{\varepsilon}\geq\cos{\theta}$ or, that is, where $\theta\geq\varepsilon$.
Correspondingly, for $\Theta(\varepsilon)<0$ the mass center will always be lower than
in its initial position during the motion. Finally, for $\Theta(\varepsilon)=0$ the
motion
will be going on at constant $\theta$, that is, the mass center of the top will be moving at constant height (however, this regime is unstable with respect to small variations in
$\alpha, \beta, \gamma$).
The last result follows from the fact that for $\Theta(\varepsilon)=0$, as we shall see further, the right-hand side of~(\ref{e1037}) is negative everywhere
except for at $\theta=\varepsilon$.

From the equations~(\ref{e1038}) and~(\ref{e1039})\,--\,(\ref{e1041}) it follows:
\beq
\Theta(\varepsilon)=-2 \sin^2\varepsilon((\alpha -1)\cos\varepsilon+\beta\gamma)
\label{e10471}
\eeq

Hence, the condition that the mass center lifts up during the initial stages of motion~is:
\beq
\Theta(\varepsilon)>0\Longleftrightarrow(\alpha -1)\cos\varepsilon+\beta\gamma<0
\label{e10472}
\eeq

Note, in particular, that for the tops with $\alpha>1$ the mass center will always be moving downwards under the condition $\displaystyle\varepsilon<\frac{\pi}{2}$ irrespectively to the kinematic parameters of the problem.

\underline{Remark}. Note that in \cite{spec4} the case for small $\varepsilon$ is considered in details and analytical solutions are obtained, and the analysis of trajectories is carried out also.

Let us now study the behavior of the roots of the parabola, defined by the dependency $\Theta(\cos{\theta})$. As $\mathfrak{a}_2\geq0$, the parabola is open upwards (the case $\mathfrak{a}_2=0$ will be understood as the limiting one). The minimum of the parabola corresponds to $\cos{\theta}$ equal to $\displaystyle z_m~\equiv~-~\frac{\mathfrak{a}_1}{\mathfrak{a}_2}$. As from the equations~(\ref{e1039}) -- (\ref{e1041}), (\ref{e1045}) -- (\ref{e1047}) it follows that $\mathfrak{a}_1,\mathfrak{a}_2$ contain independent parameters, one concludes that
\beq
\label{e10473}
-\infty<z_m\leq0.
\eeq
Further, from~(\ref{e1038}) one gets that:
\beq
\label{e10474}
\Theta_m\equiv\Theta(z_m)=-\frac{\mathfrak{a}_1^2}{\mathfrak{a}_2}+\mathfrak{a}_0.
\eeq

Substitution of~(\ref{e1039}) -- (\ref{e1041}) leads to the expression:
\beq
\label{e10475}
\Theta_m=-2\beta\gamma-\frac{(\alpha^2\cos^2\varepsilon+\sin^2\varepsilon)^2}{8\beta\gamma}-\cos{\varepsilon}(\alpha^2-\sin^2\varepsilon(1-\alpha)^2).
\eeq

As the parameters $\alpha,\beta\gamma$ may be considered as independent, one can say that
\beq
\label{e10476}
\Theta_m\leq-(\alpha^2\cos^2\varepsilon+\sin^2\varepsilon)-\cos{\varepsilon}(\alpha^2-\sin^2\varepsilon(1-\alpha)^2).
\eeq
The last expression becomes the equality when $\displaystyle\beta\gamma=\frac{1}{4}(\alpha^2\cos^2\varepsilon+\sin^2\varepsilon)$. After elementary transformations one arrives at
\beq
\label{e10477}
\Theta_m\leq-2\cos^2\frac{\varepsilon}{2}(1+(\alpha-1)\cos{\varepsilon})^2.
\eeq

Therefore, it is seen that the minimum of the parabola defined by the dependency $\Theta(\cos{\theta})$ is a strictly nonpositive quantity that achieves its maximum value equal to~zero, under $\varepsilon=\pi$ (and, simultaneously, $\displaystyle\beta\gamma=\frac{1}{4}\alpha^2$). Hence, the binomial $\Theta(\cos{\theta})$ does always have two real roots with respect to $\cos{\theta}$. Let us introduce the notation $z_+,z_-$ for the bigger and the smaller root correspondingly:
\beq
\label{e104772}
z_\pm=-\frac{\mathfrak{a}_1}{\mathfrak{a}_2}\pm\sqrt{(\frac{\mathfrak{a}_1}{\mathfrak{a}_2})^2-\frac{\mathfrak{a}_0}{\mathfrak{a}_2}}
\eeq

Further, notice that:
\beq
\label{e10478}
\left.\frac{\mathrm{d}\Theta(\cos{\theta})}{\mathrm{d\cos{\theta}}}\right|_{\theta=\varepsilon}=4\beta\gamma\cos{\varepsilon}+\alpha^2\cos^2\varepsilon+\sin^2\varepsilon.
\eeq

The last equation is strictly nonnegative under $\displaystyle\varepsilon\leq\frac{\pi}{2}$. Hence, for a motion that starts from the position\footnote{Further in the text, the condition $\displaystyle\varepsilon\leq\frac{\pi}{2}$ will always be considered true.} where $\displaystyle\varepsilon\leq\frac{\pi}{2}$, one can state that the region of motion will be defined by
\beq
\label{e10479}
\max[-1,z_+]\leq\cos{\theta}\leq\cos{\varepsilon}
\eeq
when the condition~(\ref{e10472}) is satisfied and
\beq
\label{e104711}
\cos{\varepsilon}\leq\cos{\theta}\leq\min[z_+,1]
\eeq
when it is not satisfied.

Considering the equations~(\ref{e10479}) -- (\ref{e104711}) and the equation~(\ref{e104772}), it is easy to get the condition that at its highest position the mass center belongs to the same horizontal plane as the center of the ball. In order for this to happen the condition~(\ref{e10472}) and the equation $z_+=0$ must hold, which is equivalent to $\mathfrak{a}_0=0$ or
\beq
\label{e104712}
8\beta\gamma+\left((\alpha -1)^2\cos{3\varepsilon}+(\alpha +1)(3\alpha-1)\cos{\varepsilon}\right)=0.
\eeq

Note that the last condition can be satisfied only if it admits real roots for $\alpha$ (the~other parameters being fixed) and only if one of the roots is in the range $0\leq\alpha\leq1$. The~roots for $\alpha$ may be transformed to the form:
\beq
\label{e104712}
\alpha_{1,2}=\frac{1}{\cos^2\varepsilon}(-\sin^2\varepsilon\pm\sqrt{\sin^2\varepsilon-2\beta\gamma\cos{\varepsilon}})
\eeq

The condition of the existence of two real roots implies the inequality
\beq
\label{e104713}
\sin^2\varepsilon-2\beta\gamma\cos{\varepsilon}\geq0
\eeq

The condition $\alpha\geq0$ implies than only the bigger root may be a physically meaningful value and that the following inequality holds:
\beq
\label{e104714}
\sin^2\varepsilon\cos^2\varepsilon-2\beta\gamma\cos{\varepsilon}\geq0,
\eeq
Note  that it already includes the inequality~(\ref{e104713}).  The condition $\alpha\leq1$ leads  to   the inequality
\beq
\label{e104715}
\cos^2\varepsilon+2\beta\gamma\cos{\varepsilon}\geq0,
\eeq
which is satisfied automatically.

We can finally conclude that for a top that was initially at the position where $\displaystyle\varepsilon<\frac{\pi}{2}$, the highest position for its mass center will be $\displaystyle\theta=\frac{\pi}{2}$ if the conditions~(\ref{e104712}),~(\ref{e10472}) hold, the necessary condition for the first one being~(\ref{e104715}).

Let us now study the values of the root $z_+$.
When~(\ref{e10472})
holds (the top lifts up), it~is always true that $z_+\leq\cos{\varepsilon}$. The lower bound for $z_+$ may be derived after inserting the~equations~(\ref{e1039})\,--\,(\ref{e1041}) into~(\ref{e104772}), which gives, after transformations:
\beq
z_+=\frac{-\lambda^2+\sqrt{(\mu+\tau)^2+\lambda^4-\tau^2}}{\mu},
\label{e104716}
\eeq
where
\beq
\lambda^2=\sin^2\varepsilon+\alpha^2\cos^2\varepsilon,
\label{e104717}
\eeq
\beq
\tau=\cos{\varepsilon}(\alpha^2\cos^2\varepsilon+(-1+2\alpha)\sin^2\varepsilon),
\label{e104718}
\eeq
\beq
\mu=4\beta\gamma.
\label{e104719}
\eeq

As in~(\ref{e104716}) the parameter $\mu$ is independent with respect to the other ones, one can minimize this expression with the account of $\mu\geq0$. It is easy to show that the expression for $z_+$ is minimal in the limit $\mu\rightarrow0$, converging to the value
\beq
z_{+\min}=\frac{\tau}{\lambda^2}
\label{e104720}
\eeq

Now, considering $\alpha$ as a variable, it is easy to see that the equation $z_{+\min}$ is minimal for $\alpha=0$ and in this case is
equivalent to $\left.z_{+\min}\right|_{\alpha=0}=-\cos{\varepsilon}$. Therefore, we conclude that, irrespectively to the parameters of the problem, if the mass center lifts up at the beginning of motion, it is always in the range $[\varepsilon,\pi-\varepsilon]$.

Similarly, for the downwards motion the minimal value for $z_+$ is $\cos{\varepsilon}$. The maximum value may be derived from the equation~(\ref{e104716}). Maximizing this expression with respect to $\mu$, one obtains that $z_+$ reaches its maximum value that is equal to unity in the limit $\mu\rightarrow\infty$. Therefore, one can conclude that for any initial position of the top the parameters of the problem may be chosen in such a way that the top may approach arbitrarily close to the position $\theta=0$. This result, in particular, corresponds to a simple case $\omega_0=0$.

In conclusion, notice that the equation~(\ref{e1037}) may be easily integrated in elementary functions by the use of the variable change $\tau=\cos{\theta}$.



\section*{11. Laplacian $\bf\Delta_{}$ spectrum on a {\itshape{doughnut}}}
\addcontentsline{toc}{section}{11. LAPLACIAN $\Delta_{}$ SPECTRUM ON A {\itshape{DOUGHNUT}}}
\setcounter{equation}0

\begin{flushleft}
\parbox{0.25\textwidth}{\includegraphics[scale=0.47]{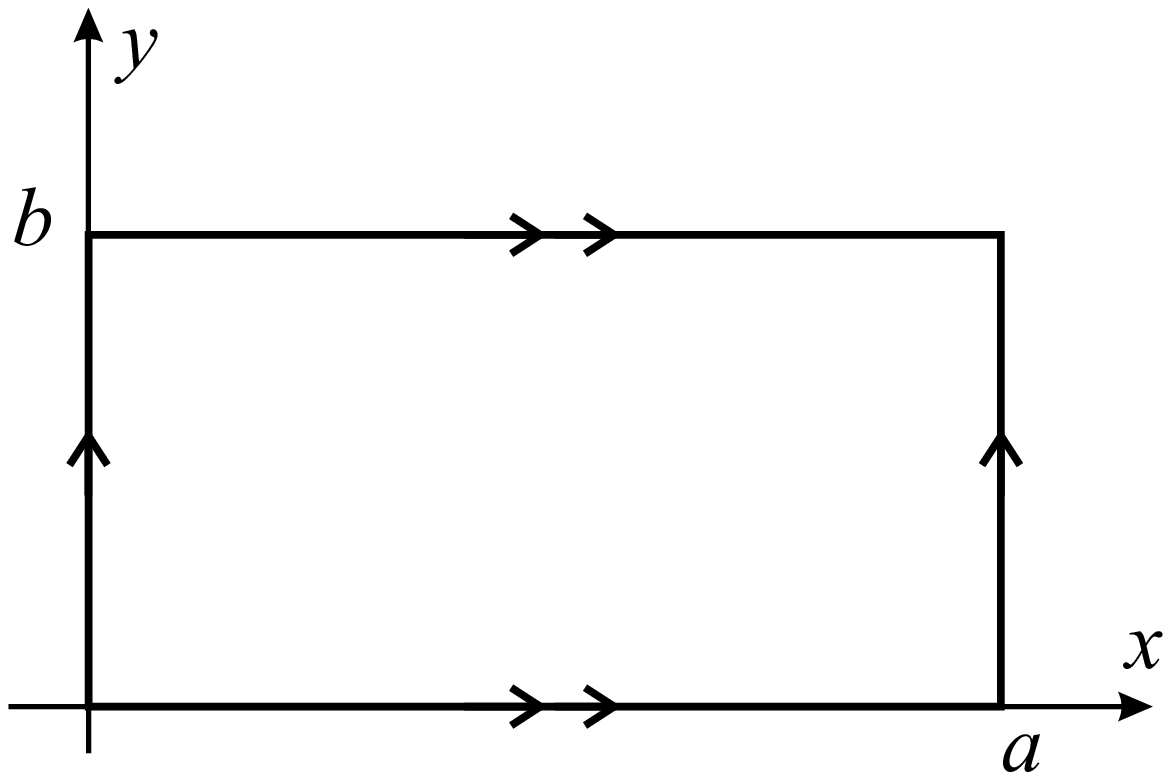}\vspace{-2mm}    \center{\small Fig. 7}}\hfill\hspace{2mm}\parbox{0.56\textwidth}{Consider a torus made of a rectangular block \\ $$0\leq
x\leq a, \qquad 0\leq y\leq b$$ with glued opposite sides
(see fig.\,7) where the identical arrows mark the sides to be glued together.
For sufficiently big ratio $a/b$, it is possible to implement such torus nearly without deformations, as the surface of a
doughnut
in 3D~space.}
\end{flushleft}

\vspace{-5mm}
\begin{flushleft}
\parbox{0.73\textwidth}{If inside the doughnut
magnetic field with the flux
$\Phi_1$ is created, and also magnetic flux
$\Phi_2$ is passing through the
hole of the doughnut (see
fig.\,8), then the wave function of the stationary state of the charged particle with the charge $e$ and mass $m$,
on the surface of the torus, is the eigenfunction 
$\Psi(x,y)$ of the operator}
\hfill
\parbox{0.26\textwidth}{\includegraphics[scale=0.32]{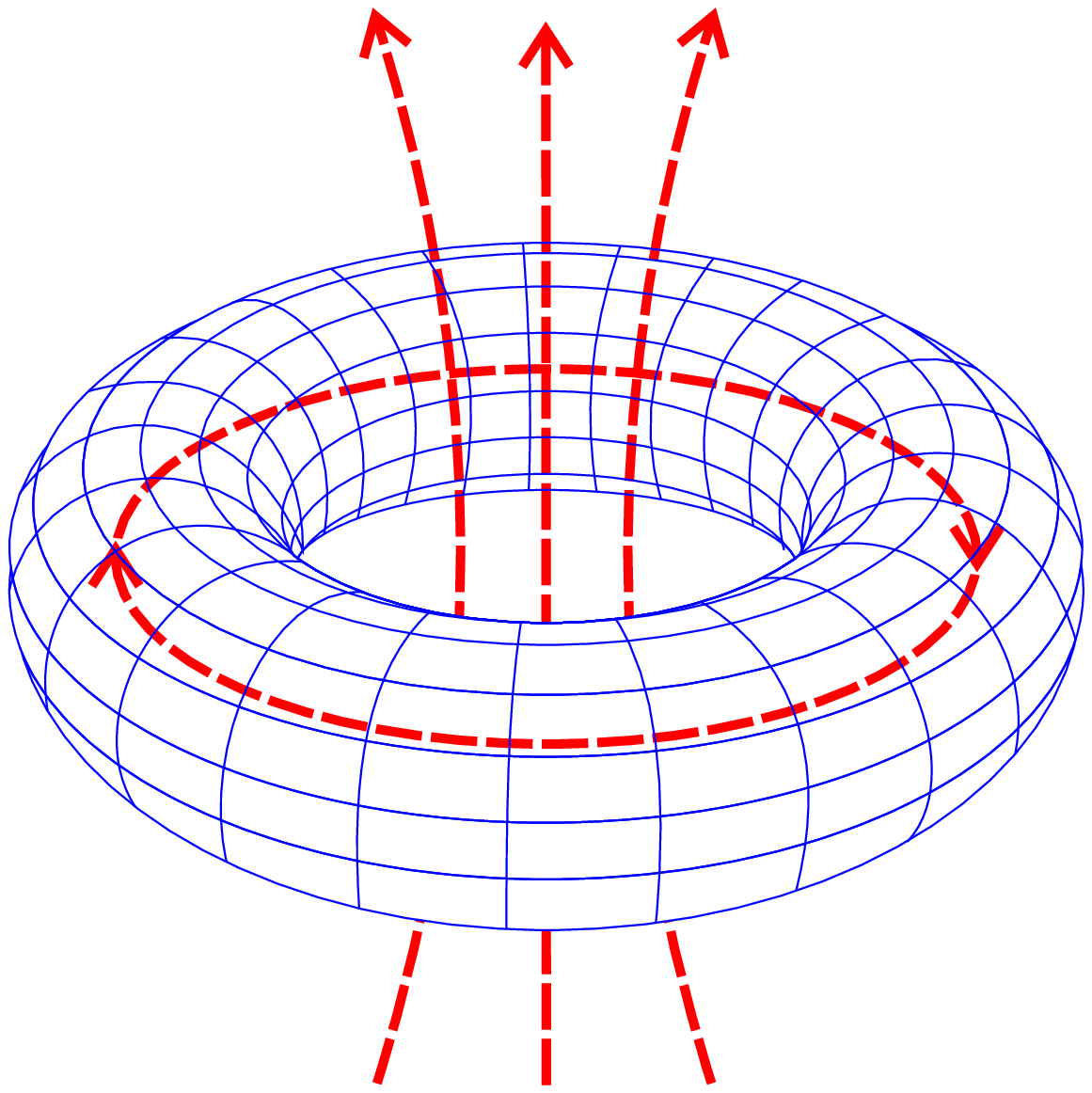}\vspace{-2mm}    \center{\small Fig. 8}}
\end{flushleft}
\vspace{0mm}
\setcounter{equation}0
\begin{equation}\label{H_0}
 \hat H_0=-\frac{\hbar^2}{2m}
\left[
 \left(\frac{\partial~}{\partial x}-i\frac{e}{c\hbar}A_x(x,y)\right)^2
+\left(\frac{\partial~}{\partial y}-i\frac{e}{c\hbar}A_y(x,y)\right)^2
\right],
\end{equation}
where $\mathbf{A}(x,y)$ is a vector potential, $\hbar$ is Planck constant,
and $c$ is speed of light. Function $\Psi(x,y)$ satisfies to
periodic boundary conditions:
\begin{eqnarray}\label{bc old 1}
  \Psi(x,0)=\Psi(x,b),\qquad
  \Psi(0,y)=\Psi(a,y),
\end{eqnarray}
\begin{eqnarray}\label{bc old 2}
  \Psi'_y(x,0)=\Psi'_y(x,b),\quad ~
  \Psi'_x(0,y)=\Psi'_x(a,y).
\end{eqnarray}

Suppose the magnetic field turns to zero on the~surface,
we can make cuts of the~surface and, using
gauge transformation, nullify $\mathbf{A}$ on the surface of the torus.
Then the wave function would be discontinuous at the cuts (such cuts are possible to make exactly on the border of the rectangle). It is also clear that $|\Psi(x,y)|^2$ should not change.

After the gauge transformation, the task to find stationary states is modified and requires to find eigenfunctions
$\psi(x,y)$ of the operator
\begin{equation}\label{H_1}
 \hat H_1=-\,\frac{\hbar^2}{2m}
\left[
  \left(\frac{\partial~}{\partial x}\right)^2
+\left(\frac{\partial~}{\partial y}\right)^2 \right],
\end{equation}
with phase shift (lagging) boundary conditions:
\begin{eqnarray}\label{bc new 1}
  \psi(x,0)=e^{i\varphi_1}\psi(x,b),\qquad
  \psi(0,y)=e^{i\varphi_2}\psi(a,y),
\end{eqnarray}
\begin{eqnarray}\label{bc new 2}
  \psi'_y(x,0)=e^{i\varphi_1}\psi'_y(x,b),\qquad
  \psi'_x(0,y)=e^{i\varphi_2}\psi'_x(a,y),
\end{eqnarray}

\textbf{(a)} Find eigenfunctions and eigenvalues of the operator $\hat H_1$.

\textbf{(b)} What is the relationship between the
fluxes $\Phi_1$, $\Phi_2$ and
phase displacements $\varphi_1$, $\varphi_2$ ?\\


{\bf SOLUTION} 

Let's make mentioned above cuts of doughnut surface and nullify vector potential~$\bf{A'}$ with the help of
gauge
transformation
\begin{equation}\label{A}
{\bf A'}(x,y)={\bf A} - {\bf\triangledown} f(x,y) = 0.
\end{equation}
Thus the Hamiltonian $\hat H_0$ is simplified.
In spite of the fact that the wave function would undergo certain changes, it is clear that the probability $|\Psi(x,y)|^2$ of particle location should not be changed by gauge (gradient) transformation, i.\,e. the new wave function $\psi(x,y)$ differs from the old one $\Psi(x,y)$ only in phase factor~\cite{lanIII}:
\begin{equation}\label{psi}
\psi(x,y)=\Psi(x,y)\exp\left(-\,\frac{ie}{\hbar c}f(x,y)\right),
\end{equation}
where $f(x,y)$ is the function of gauge
transformation, $\psi(x,y)$ is the eigenfunction of the new modified Hamiltonian $\hat{H_1}$, that is in fact Laplace operator.
\par\textbf{a)} Let us find eigenfunctions $\psi(x,y)$ of the transformed Hamiltonian $\hat{H_1}$. On $\mathbb{R}^2$, eigenfunctions of Laplace operator can be chosen in the form of plane waves $\exp(i\bf{kr}):$
\begin{equation}
\psi(x,y)=\exp(ik_x x)\exp(ik_y y).
\end{equation}
It is clear that any linear combination of eigenfunctions, for which the value
$k_x^2+k_y^2$ is the same, will also be an eigenfunction.
Let us find $k_x$ and $k_y$ from the boundary conditions (\ref{bc new 1}), (\ref{bc new 2}):
\begin{eqnarray}\label{k_x, k_y}
k_x=-\frac{\varphi_2}{a}+\frac{2\pi n_2}{a}, \qquad
k_y=-\frac{\varphi_1}{b}+\frac{2\pi n_1}{b},
\end{eqnarray}
where $n_1, n_2 \in\mathbb{Z}$.
\par Eigenvalues of $\hat{H_1}$ are found from the equation $\hat{H_1}\psi(x,y)=E\psi(x,y)$ and have the form:
\begin{equation}\label{E}
E=\frac{\hbar^2}{2m}\left[\left(-\frac{\varphi_1}{b}+\frac{2\pi n_1}{b}\right)^2+\left(-\frac{\varphi_2}{a}+\frac{2\pi n_2}{a}\right)^2\right].
\end{equation}
\par Let us prove that we have found the complete basis of the $\hat{H_1}$ eigenfunctions.
If~the~torus is without phase lagging, then we obtain the regular Fourier series. With~pha\-se lagging,
not quite regular Fourier series
is obtained, however it is reducible to the regular one:
$$
   \psi(x,y)=\exp\left[-
   i\left(\frac{\varphi_{2}x}{a}+
   \frac{\varphi_{1}y}{b}\right)\right]\Psi(x,y).
$$
Here, $\Psi$ is
a regular periodic function for which the normal Fourier series is written as an expansion
using the basis in $L_2$ for which the completeness has been already proved.
Multiplication by $\displaystyle\exp\left[-i\left(\frac{\varphi_{2}x}{a}+
\frac{\varphi_{1}y}{b}\right)\right]$
in terms of space $L_2$
is unitary transformation.
\\
\par\textbf{b)} Let us find the relationship between the fluxes of magnetic field $\Phi_1,\Phi_2$ and phase lagging $\varphi_1, \varphi_2$.
The flux of magnetic field is:
$\displaystyle\Phi=\oint_S {\bf H}d{\bf S}=\oint_S\rot{\bf A}d{\bf S}$
or, after a~transformation using Stokes theorem:
$\displaystyle\Phi=\oint_l {\bf A}d{\bf l}$.
Then the fluxes of magnetic field inside the torus and through the hole of the
doughnut are respectively equal to:
\begin{equation}\label{Phi_1}
\Phi_1=\int_0^b {A}_y(x,y)dy,
\end{equation}
\begin{equation}\label{Phi_2}
\Phi_2=\int_0^a {A}_x(x,y)dx.
\end{equation}
Recalling expression (\ref{A}) and substituting it into (\ref{Phi_1}) and (\ref{Phi_2}):
\begin{equation}
\Phi_1=f(x,b)-f(x,0),
\end{equation}
\begin{equation}
\Phi_2=f(a,y)-f(0,y).
\end{equation}
It is possible to find the differences
in values of $f(x,y)$ in these points using expression~(\ref{psi}) and boundary conditions
(\ref{bc old 1}), (\ref{bc old 2}), (\ref{bc new 1}), (\ref{bc new 2}). As a result we obtain:
\begin{equation}
\Phi_i=\frac{\hbar c}{e}\varphi_i=\Phi_0\frac{\varphi_i}{\pi},
\end{equation}
where $i=1,2$; $\Phi_0$ is the quantum of magnetic flux.
It is necessary to note that in this case the magnetic flux is not quantized.

\section*{12. ~ 3D Delta function}
\addcontentsline{toc}{section}{12. 3D DELTA FUNCTION}

\setcounter{equation}0

Coulomb wave function of the
ground state has the form
$$ \varphi_c({\vec p}) = 8\pi\alpha\mu|\varphi_c(r=0)|\varphi_p^2,
     \qquad
    |\varphi_c(r=0)|^2 = \f{\alpha^3\mu^3}{\pi}, \quad \varphi_p =
      ({\vec p}^2+\mu^2\alpha^2)^{-1} $$
and satisfies Schroedinger equation in the momentum representation.

The values of hyperfine splitting of the
ground level
of hydrogen-like atom with accuracy up to $\alpha^5$ found on the basis of quasipotential
built from the diagrams of the order of $\alpha^2$ and higher, are found most easily
if to assume
$$ \varphi_c(\vec p) \approx (2\pi)^3\delta(\vec p)|\varphi_c(r=0)|. \eqno{(\bigstar)} $$
(Infrared singularities in the elements of the amplitude of scattering that appear in the framework of this approach,
are normally eliminated by cutting off the value of virtual 3-dimensional
momentum.)

Derive the relation ($\bigstar$)
by proving the following statements:

\textbf{(a)} $\displaystyle\delta(x) = \lim\limits_{a\to 0}\f{1}{\pi}\f{a}{a^2+x^2}$, $x\in R^1$; ~
\textbf{(b)} $\displaystyle\f{\pi\delta(x)}{2x^2} = \lim\limits_{a\to 0}\f{a}{(a^2+x^2)^2}$, $x\in R^1$;\\
\textbf{(c)} $\displaystyle\delta(\vec p)=\f{\delta(p)}{2\pi p^2}$ , $p=|\vec p|, p\geqslant0$; ~
\textbf{(d)} $\displaystyle\delta(\vec p) = \lim\limits_{a\to 0}\f{a}{\pi^2(a^2+p^2)^2}$,
$p=|\vec p|, p\geqslant0$.

\underline{Direction}:
equations (a) -- (d). containing generalized
functions,
of the class $D^{\bf\prime}$
should be proved on the space
of the basic functions of the class
${\cal D}$.\\

{\bf SOLUTION}

The above statement ($\bigstar$) is based on the fact that the value of the square of module of the wave
function (present in a matrix element) in coordinate space at
$r=0$ has the order\,\footnote{Description of hyperfine splitting is based on computation of quantum
distributions being squaresd by the wave function. From the explicit form of the wave function in the momentum representation it turns that the terms of the
lowest order
in the expansion of a matrix element are $\alpha^5$.}
of $\alpha^3$.

Let us prove the \textbf{(a)} statement.

Let $x\in R^1, \varphi(x)\in {\cal D}(R^1)$ is the
test function. Then
$$
\left(\lim\limits_{a\to 0} \f{a}{\pi}\f{1}{x^2+a^2} ,
 \varphi(x)\right)
=
\int\limits_{-\y}^{\y} dx\lim\limits_{a\to 0} \f{a}{\pi}\f{1}{x^2+a^2}
 \varphi(x) = $$
$$ = \lim\limits_{a\to 0} \f{a}{\pi}\int\limits_{-\y}^{\y} \f{dx
 \varphi(x)}{x^2+a^2}
  = \lim\limits_{a\to 0} \f{a}{\pi} 2\pi i {\rm Res}_{x=ia}
 \f{\varphi(x)}{x^2+a^2}
  = $$
$$ = \lim\limits_{a\to 0} \f{a}{\pi} 2\pi i\f{\varphi(ia)}{2ia}
  = \varphi(0) =
\left(\delta(x),\varphi(x)\right). $$
The first and the last parts are underlined to focus on equality of the functionals hence the generalized functions themselves.

Derivation of the \textbf{(с)} relationship:
$$ \delta(\vec p) = \delta(p_x)\delta(p_y)\delta(p_z) = (2\pi)^{-3}\int
    e^{i\vec k\vec p}d\vec k = $$
$$ = (2\pi)^{-3}\int\limits_{0}^{\pi}\int\limits_{0}^{2\pi}\int
\limits_{0}^{\y} e^{ikp\cos\theta}k^2dk\sin\theta d\varphi
d\theta = (2\pi)^{-2}\int\limits_{0}^{\y}\int\limits_{-1}^1e^{ikpx}dxk^2dk
 = $$
$$ \left. = (2\pi)^{-2}\int\limits_{0}^{\y}\f{e^{ikpx}}{ikp}\right|_{-1}^1k^2dk = \f{1}
 {(2\pi)^2ip}\int\limits_{0}^{\y}\left(e^{ikp}-e^{-ikp}\right)kdk = \f{1}
 {(2\pi)^2ip}\int\limits_{-\y}^{\y}e^{ikp}kdk = $$
   $$ = \f{1}
 {(2\pi)^2ip}\f{\p}{i\p p}\int\limits_{-\y}^{\y}e^{ikp}dk = \f{-1}{2\pi p}
\delta'(p) = \f{\delta(p)}{2\pi p^2}. $$
where the known relation $p\delta'(p)=- \,\delta(p)$
 is utilized (the proof is partial integration).

To check \textbf{(d)}, we consider the
test function $\Phi(\vec p)\in {\cal D}(R^3)$ and perform the following transformations:
$$
\left(\lim\limits_{a\to 0} \f{a}{\pi^2}\f{1}{(p^2+a^2)^2} ,
 \Phi(\vec p)\right)
=
\int d\vec p\lim\limits_{a\to 0} \f{a}{\pi^2}\f{1}{(p^2+a^2)^2}
 \Phi(\vec p) = $$
$$ = \int d\Omega \lim\limits_{a\to 0} \f{a}{\pi^2}\int\limits_0^{\y}
\f{p^2\Phi(p,\Theta,\varphi)}{(p^2+a^2)^2}dp = $$
$$ = \int d\Omega \lim\limits_{a\to 0} \f{a}{\pi^2}\left(\f{1}{2}\int
\limits_0^{\y}
\f{p^2\Phi(p,\Theta,\varphi)}{(p^2+a^2)^2}\d p - \f{1}{2}\int
\limits_0^{-\y}
\f{p'^2\Phi(-p',\Theta,\varphi)}{(p'^2+a^2)^2}dp'\right) = $$
$$ \left\{ \mbox{let} ~ ~ \Phi(p,\Theta,\varphi) = \Phi(-p,\Theta,\varphi)
\right\} $$
$$ = \int d\Omega \lim\limits_{a\to 0} \f{a}{2\pi^2}\int\limits_{-\y}^{\y}
\f{p^2\Phi(p,\Theta,\varphi)}{(p^2+a^2)^2}dp = $$
$$ = \int d\Omega \lim\limits_{a\to 0} \f{a}{2\pi^2}2\pi i {\rm Res}_{p=ia}
\f{p^2\Phi(p,\Theta,\varphi)}{(p^2+a^2)^2} ~\, ~ (\mbox{residue in the pole of the 2nd order}) = $$
$$ = \int d\Omega \lim\limits_{a\to 0} \f{ai}{\pi}\lim\limits_{p\to ia}
\f{d}{dp}\f{p^2\Phi(p,\Theta,\varphi)}{(p+ia)^2} = $$
$$ = \int d\Omega \lim\limits_{a\to 0} \f{ai}{\pi}\lim\limits_{p\to ia}
\left\{\f{2p\Phi+p^2\Phi'}{(p+ia)^2}-\f{2p^2\Phi}{(p+ia)^3}\right\} = $$
$$ = \int d\Omega \lim\limits_{a\to 0} \f{ai}{\pi}
\left\{\f{2ia\Phi(ia,\Theta,\varphi)}{(2ia)^2}+\f{(ia)^2\Phi'
(ia,\Theta,\varphi)}{(2ia)^2}-\f{2(ia)^2\Phi
(ia,\Theta,\varphi)}{(2ia)^3}\right\} = $$
$$ = \int d\Omega \lim\limits_{a\to 0} \f{ai}{\pi}
\left\{\f{\Phi(ia)}{2ia}+\f{\Phi'(ia)}{4}-\f{\Phi
(ia)}{4ia}\right\} = $$
$$ = \int d\Omega \lim\limits_{a\to 0} \f{ai}{\pi}
\left\{\f{\Phi(ia)}{4ia}+\f{\Phi'(ia)}{4}\right\} = \int d\Omega\f{\Phi(0)}
{4\pi} = \Phi(0) =
\left(\delta(\vec p),\Phi(\vec p)\right). $$

The truth of the statement \textbf{(b)} follows particularly from the proven \textbf{(с)} and \textbf{(d)}.
It~is also possible to obtain the relationship \textbf{(b)}
using
the representation of the $\delta$-func\-tion from the statement \textbf{(a)}.

It is possible to prove the relations \textbf{(a)} -- \textbf{(d)} in another way.

Using Sokhotsky's formula
\[
\frac{1}{x+i0}=\mathrm{P}\frac{1}{x}-\mathrm{i}\pi\delta(x),\]
which is correct in $D^{\bf\prime}$,
it is easy to derive eq. \textbf{(a)}:
\begin{eqnarray*}
\lim_{a\to0}\frac{1}{\pi}\frac{a}{a^{2}+x^{2}} & = & \lim_{a\to0}\frac{1}{2\mathrm{i}\pi}\left(\frac{1}{x-\mathrm{i}a}-\frac{1}{x+\mathrm{i}a}\right) =\\
 & = & \frac{1}{2\pi\mathrm{i}}\left(\mathrm{P}\frac{1}{x}+\mathrm{i}\pi\delta(x)-\mathrm{P}\frac{1}{x}+
 \mathrm{i}\pi\delta(x)\right) \, = \,
\delta(x).\end{eqnarray*}

It is possible to transform eq. \textbf{(b)} into the following form:
\begin{equation}
\delta(x)=\lim_{a\to0}\frac{2}{\pi}\frac{ax^{2}}{\left(a^{2}+x^{2}\right)^{2}}.\label{eq:edelta/x}\end{equation}

Transforming the right part of the last equation, we obtain by recalling eq. \textbf{(a)}:
\begin{eqnarray}
\lim_{a\to0}\frac{2}{\pi}\frac{ax^{2}}{\left(a^{2}+x^{2}\right)^{2}} & = & \lim_{a\to0}\frac{2}{\pi}\frac{a}{a^{2}+x^{2}}\left(1-\frac{a^{2}}{a^{2}+x^{2}}\right)= \nonumber \\
 & = & 2\delta(x)-\lim_{a\to0}\frac{2a^{3}}{\pi\left(a^{2}+x^{2}\right)^{2}}.\label{eq:eprelim equation}\end{eqnarray}

For any function $\varphi(x)\in {\cal D}$, the following relations are true:
\begin{eqnarray*}
\left(\lm\frac{2a^{3}}{\pi\left(a^{2}+x^{2}\right)^{2}},\varphi\right) & = & \lm\intop_{\R}\frac{2a^{3}\varphi(x)\d x}{\pi\left(a^{2}+x^{2}\right)^{2}}=\\
 & = & \lm\frac{2}{\pi}\intop_{\R}\frac{\varphi(at)\d t}{\left(1+t^{2}\right)^{2}}=\\
 & = & \lm\frac{2}{\pi}\intop_{\R}\frac{\left[\varphi(at)-\varphi(0)\right]}{\left(1+t^{2}\right)^{2}}\d t+\frac{2\varphi(0)}{\pi}\intop_{\R}\frac{\d t}{\left(1+t^{2}\right)^{2}}.\end{eqnarray*}

The second integral in the last expression can be computed with the help of the transformation $t=\mbox{tg}\alpha$, $\d t=\left(t^{2}+1\right)\d\alpha$:
\[
\intop_{-\infty}^{+\infty}\frac{\d t}{\left(1+t^{2}\right)^{2}}=\intop_{-\frac{\pi}{2}}^{\frac{\pi}{2}}\cos^{2}\alpha\d\alpha=\frac{\pi}{2}.\]

The first integral can be estimated in the following way.
Let us split the integral into the two integrals:
\begin{eqnarray*}
\lm\frac{2}{\pi}\intop_{\R}\frac{\left[\varphi(at)-\varphi(0)\right]}{\left(1+t^{2}\right)^{2}}\d t & = & \\ \lm\intop_{[-A;A]}\frac{2}{\pi}\frac{\left[\varphi(at)-\varphi(0)\right]}{\left(1+t^{2}\right)^{2}}\d t & + & \lm\intop_{\R\backslash[-A;A]}\frac{2}{\pi}\frac{\left[\varphi(at)-\varphi(0)\right]}{\left(1+t^{2}\right)^{2}}\d t.\end{eqnarray*}

The first integral on the right side of the last equation tends to zero at $a\to0$.
For~the upper limit of the second integral, it is possible to use the value of
\[
\lm\intop_{\R\backslash[-A;A]}\frac{2}{\pi}\frac{2M}{\left(1+t^{2}\right)^{2}}\d t=\frac{4M}{\pi}\intop_{\R\backslash[-A;A]}\frac{\d t}{\left(1+t^{2}\right)^{2}},\]
independent of $a$. At $A\to\infty$, this integral also converges to zero. Thus
\[
\left(\lm\frac{2a^{3}}{\pi\left(a^{2}+x^{2}\right)^{2}},\varphi\right)=\varphi(0).\]

It follows that
\[
\lm\frac{2a^{3}}{\pi\left(a^{2}+x^{2}\right)^{2}}=\delta(x).\]

Recalling this result, eq.~(\ref{eq:eprelim equation})
converges to eq. (\ref{eq:edelta/x}),
which proves the validity of eq.~\textbf{(b)}.

Eq. \textbf{(c)} should take place on the functions from ${\cal D}(\R^{3})$,
i.\,e. from that, the eq. should follow:
\begin{eqnarray}
\intop_{\R^{3}}\delta(\pp)\varphi(\pp)\d\pp & = & \intop_{\R^{3}}\frac{\delta(p)}{2\pi p^{2}}\varphi(\pp)\d\pp,\label{eq:edelta-3}\end{eqnarray}
where $\varphi(\pp)\in {\cal D}(\R^{3})$. Conversing the right part of this equality, we obtain
\begin{eqnarray*} & \displaystyle\intop_{\R^{3}}\frac{\delta(p)}{2\pi p^{2}}\varphi(\pp)\d\pp & = \intop_{\R^{3}}\frac{\delta(p)}{2\pi p^{2}}\varphi(\pp)p^{2}\d p\,\d\Omega=\\
 & = & \intop_{p\geqslant0}2\delta(p)\left(\intop_{\mid\pp\mid=p}\varphi(\pp)\frac{\d\Omega}{4\pi}\right)dp
  =  \intop_{\R}\delta(p)\tilde{\varphi}(p)dp
  =  \tilde{\varphi}(0)=\varphi(0).\end{eqnarray*}

Here, $\Omega$ is solid angle, $\displaystyle\tilde{\varphi}(p)=\intop_{\mid\pp\mid=p}\varphi(\pp)\frac{\d\Omega}{4\pi}$. In the last line, function $\displaystyle\tilde{\varphi}$ was evenly extended to negative $p$.
The integral in the left part of eq. (\ref{eq:edelta-3}) also equals $\varphi(0)$ which proves the validity of that eq. (\ref{eq:edelta-3}) as well as of eq. \textbf{(c)}.

Equation \textbf{(d)} automatically follows from \textbf{(b)} and \textbf{(c)}:
$$\delta_{3}(\pp) = \frac{\delta(p)}{2\pi p^{2}}=
\frac{1}{2\pi}\lim_{a\to0}\frac{2}{\pi}\frac{a}{\left(a^{2}+p^{2}\right)^{2}}=
\lim_{a\to0}\frac{a}{\pi^{2}\left(a^{2}+p^{2}\right)^{2}}.$$

Let us obtain the final approximate equation $(\bigstar)$ for the wave function of the
ground state in Coulomb field using the formula
$\displaystyle\delta(\vec p)=\f{\delta(p)}{2\pi p^2}$, valid for the case of spherical symmetry
and also using the above proved statements. In the momentum representation this function
takes the form:
\[
\varphi_{c}(\pp)=8\pi\alpha\mu\mid\varphi_{c}(r=0)\mid\frac{1}{\left(p^{2}+\alpha^{2}\mu^{2}\right)^{2}},\]
and, using the identity \textbf{(d)}, we obtain
$$ \lim\limits_{\alpha\to 0}\varphi_c(\vec p) = \lim\limits_{\alpha\to 0}
\f{8\pi\alpha\mu|\varphi_c(0)|}{(\vec p^2+\alpha^2\mu^2)^2} =
8\pi^{3}|\varphi_c(r=0)|
\delta(\vec p). $$
Due to the fact that $\alpha\mu\ll1$, the function approximately equals its limit at
$\alpha\mu\to0$.

From the form of Coulomb wave function of the
ground state it follows that the~main contribution into the splitting of the energy levels is due to the momentum from the~ran\-ge that satisfies the condition
${\vec p}^{\,2}\sim\alpha^2\mu^2$. As a result, expansion of the integrand by~$p/m$ would be equivalent to an expansion of the whole integral by $\alpha$
  (under condition that the integral converges).

\section*{13. Heat conduction equation (heat source presents)}
\addcontentsline{toc}{section}{13. HEAT CONDUCTION EQUATION
(HEAT SOURCE PRESENTS)}

\setcounter{equation}0

The temperature on the ends of thin regular rod is maintained constant and equals 
zero. Lateral face of the rod is heat-insulated. The frame of reference is defined so that coordinate axis $x$ is oriented along the rod, its ends have the coordinates $x=0$ and $x=l$.

Thermal
diffusivity coefficient of the material of the rod equals $a^2$.

Find
spatial and time distribution
$T(x,t)$ ($0\leqslant x\leqslant l$, $t\geqslant0$) of temperature
along the rod
in two cases:

\textbf{(a)} at time moment $t=0$ the temperature
of the rod is constant,
$T(x,0)\equiv T_0$ at $0<x<l$;

\textbf{(b)}
in the center of the rod, point source of intensity $Q$ is switched on at time moment $t=0$, and $T(x,0)\equiv0$ at $0\leq x\leq l$.\\

{\bf SOLUTION}

Case \textbf{(a)}. In this case, the decision function is defined as the solution of heat conduction equation
\beq T_t=a^2 T_{xx}, \label{eq:13-1}
\eeq
satisfying supplementary conditions
\beq
T(0,t)=T(l,t)=0 \, \quad (t\geqslant0),\label{eq:13-2}\eeq
\beq T(x,0)=T_0 ~ \quad (0 < x < l).\label{eq:13-3}
\eeq\\

First, let us find the eigenvalues and eigenfunctions of
equation (\ref{eq:13-1}) using separation of variables (Fourier method).

Representing the decision function in the form
$$
T(x,t)=X(x)U(t)
$$
and substituting the expression into (\ref{eq:13-1}), we obtain
$$
X\dot{U}=a^2 U X''.
$$
Hence
$$
\frac{X''}{X}=\frac{\dot{U}}{a^2 U} = -\lambda^2,
$$
\beq
X''+\lambda^2 X=0,\label{eq:13-4}
\eeq
\beq
\dot{U}+a^2 \lambda^2 U=0,\label{eq:13-5}
\eeq
where $\lambda^2=$const. Solving eq. (\ref{eq:13-4}) and (\ref{eq:13-5}) with consideration of boundary conditions~(\ref{eq:13-2}),
 we obtain:
\beq
X_n=\sqrt{\frac{2}{l}} \sin\left(\frac{\pi n x}{l}\right) \quad  \quad \mbox{($0 \leqslant x \leqslant l$, ~ orthonormal system of functions)},\label{eq:13-6}
\eeq
\beq
U_n = C_n \exp\left(-\left(\frac{\pi n a}{l}\right)^2 t\right) \, \quad (t\geqslant0),
\eeq
where $n=1,2, ...$, and $C_n$ -- real coefficients. It is possible to represent the general solution in the form: $\displaystyle T(x,t)=\sum_{n=1}^\infty X_n(x) U_n(t)$. Thus
\beq
T(x,t) = \sum_{n=1}^\infty C_n \sqrt{\frac{2}{l}} \sin\left(\frac{\pi n x}{l}\right) \exp\left( - \left(\frac{\pi n a}{l}\right)^2 t\right).\label{eq:13-8}
\eeq

According to (\ref{eq:13-8}), expansion of the distribution function of initial temperature into series by eigenfunctions (\ref{eq:13-6})
takes the form
$$
T(x,0) = \sum^{\infty}_{n=1} C_n \sqrt{\frac{2}{l}} \sin\left(\frac{\pi n x}{l}\right),
$$
where
$$
C_n = 
\int^{l}_{0} T(\xi,0) \sqrt{\frac{2}{l}} \sin\left(\frac{\pi n \xi}{l}\right) d\xi.
$$
Recalling initial condition (\ref{eq:13-3})
$$ C_n =
\sqrt{\frac{2}{l}} T_0 \int^{l}_{0}\sin\left(\frac{\pi n \xi}{l}\right) d\xi = $$
$$ = \sqrt{\frac{2}{l}}T_0\frac{l}{\pi n}[-\cos\pi n+1] = \frac{\sqrt{2l}T_0}{\pi n}\left(1-(-1)^n\right)=$$
\beq = \left\{ \begin{array}{l}
\displaystyle \frac{2\sqrt{2l}}{\pi (2k+1)}T_0, \quad n=2k+1,\\[2mm]
\vspace{1mm}
\displaystyle 0, \qquad n=2k+2 ~~ ~ (k=0, 1, 2, ...).\end{array}
\right.\label{eq:13-9}\eeq
Substituting the expression (\ref{eq:13-9}) into (\ref{eq:13-8}), we obtain the solution for the case \textbf{(a)}:
$$ T(x,t) = \frac{4T_0}{\pi} \sum^{\infty}_{k=0} \frac{1}{2k+1} \exp\left(-\left(\frac{\pi (2k+1) a}{l}\right)^2 t\right) \sin\left(\frac{\pi (2k+1) x}{l}\right).$$


Case \textbf{(b)}. \underline{Method I}. In this case, the decision function is defined as the solution of heat conductivity equation with the source, i.\,e. boundary problem with
the density of heat generation described by Dirac $\delta$-function:
\beq
T_t=a^2T_{xx}+\frac{Q}{c} \,\delta\left(x-\frac{l}{2}\right), \qquad T(x,0)=0, \quad T(0,t)= T(l,t) = 0.\label{eq:13-11}
\eeq
Here, $c$ is heat capacity of unit length of the thin rod, $\displaystyle Q\delta\left(x-\frac{l}{2}\right)$~-- intensity of heat generation
per unit length.

We shall search for the solution in the form of the sum of a stationary one ($\omega$) with a non-stationary one ($v$):
$$
T(x,t)=\omega(x)+v(x,t).
$$

Stationary solution satisfies
$$
\omega_{xx}=- \,\frac{Q}{a^2c}\delta\left(x-\frac{l}{2}\right),\,\,\, \, \, ~ \omega(0)=\omega(l)=0.
$$
And
non-stationary solution satisfies
$$
v_t=a^2v_{xx}, \,\,\, \, \, ~ v(0,t)=v(l,t)=0, \,\,\, \, v(x,0)=-\,\omega(x).
$$

The common solution of stationary equation
can be
presented in the following form (see \cite{spec4})
$$
\omega=C_1\left|x-\frac{l}{2}\right|+C_2,
$$
where indeed the first constant
can be found from
differential equation
$$
\omega_{xx}=- \,\frac{Q}{a^2c}\delta(x-\frac{l}{2}) ~ \, \Rightarrow ~ \, C_1=-\frac{Q}{2a^2c},
$$
and $C_2$ from boundary conditions
$$
\omega(0)=\omega(l)=0,
$$
exactly
$$
- \,\frac{Q}{2a^2c}\frac{l}{2}+C_2=0 \, \, \Rightarrow \, \, C_2=\frac{Q}{a^2c}\frac{l}{4}.
$$
Finally
$$
\omega(x)=- \,\frac{Q}{2a^2c}\left|x-\frac{l}{2}\right|+
\frac{Q}{a^2c}\frac{l}{4}.
$$

Non-stationary part of the problem is solved using separation of variables, as in the~\textbf{(a)} case, only with the boundary condition $v(x,0)=-\,\omega(x)$. As a result, we obtain $T(x,t)=\omega(x)+v(x,t)$.

Note, that expressions for $\omega(x)$ and $v(x,t)$ can be presented as Fourier series expansions (see details in~\cite{spec4}):
\beq \omega(x) = \f{2Ql}{\pi^2a^2c} \sum_{k=0}^\infty \f{(-1)^k}{(2k+1)^2} \sin\left(\f{\pi (2k+1)x}{l}\right), \label{eq:13-razlogw}\eeq
\beq
v(x,t) = - \,\frac{2Ql}{\pi^2a^2c} \sum^{\infty}_{k=0} \frac{(-1)^k}{(2k+1)^2} \sin\left(\frac{\pi (2k+1) x}{l}\right) \exp\left(-\left(\frac{\pi (2k+1) a}{l}\right)^2 t\right).\label{eq:13-vxt}
\eeq

The answer 
will be
$$
T(x,t) =$$\beq = \frac{2Ql}{\pi^2a^2c} \sum^{\infty}_{k=0} \frac{(-1)^k}{(2k+1)^2} \sin\left(\frac{\pi (2k+1) x}{l}\right) \left[1-\exp\left(-\left(\frac{\pi (2k+1) a}{l}\right)^2 t\right)\right].\label{eq:13-txtsumma}
\eeq

Case \textbf{(b)}. \underline{Method II}.
When solving (\ref{eq:13-11}),
let us expand $T(x,t)$ into Fourier series
$$ T(x,t)=\sum_{n=1}^{\infty}C_{n}(t)\sin\left(\frac{\pi nx}{l}\right),$$
where
$$ C_{n}(t)=\frac{2}{l}\intop_{0}^{l}T(x,t)\sin\left(\frac{\pi nx}{l}\right)dx.$$

In terms of $C_{n}(t)$, the equation (\ref{eq:13-11}) will
take the form (use the expansion of delta-function)
\begin{equation}
\dot{C}_{n}(t)=-\left(\frac{\pi na}{l}\right)^{2}C_{n}(t)+\frac{2Q}{lc}\sin\left(\frac{\pi n}{2}\right).\label{eq:13-14}\end{equation}

The general solution of the eq. (\ref{eq:13-14}) is the sum of the general solution of the uniform equation and partial solution of the non-uniform equation,
i.\,e.
\begin{equation}
C_{n}(t)=A_{n}\exp\left(-\left(\frac{\pi na}{l}\right)^{2}t\right)+\frac{2Q}{l c}\sin\left(\frac{\pi n}{2}\right)\left(\frac{\pi na}{l}\right)^{-2}.\label{eq:13-15}\end{equation}

Recalling the initial condition (\ref{eq:13-11}), we obtain
\begin{equation}
C_{n}(0)=\frac{2}{l}\intop_{0}^{l}T(x,0)\sin\left(\frac{\pi nx}{l}\right)dx=0.\label{eq:13-16}\end{equation}

Considering (\ref{eq:13-16}), from (\ref{eq:13-15}) it follows that
\begin{equation}
A_{n}=-\frac{2Q}{l c}\sin\left(\frac{\pi n}{2}\right)\left(\frac{\pi na}{l}\right)^{-2}.\label{eq:13-17}\end{equation}

Substituting (\ref{eq:13-17}) into (\ref{eq:13-15}), we obtain \begin{equation}
C_{n}(t)=\frac{2Q}{l c}\sin\left(\frac{\pi n}{2}\right)\left(\frac{\pi na}{l}\right)^{-2}\left[1-\exp\left(-\left(\frac{\pi na}{l}\right)^{2}t\right)\right].\label{eq:13-18}\end{equation}

Taking into account (\ref{eq:13-18}), we can write the solution of eq. (\ref{eq:13-11}) in the form
\begin{equation}
T(x,t)=\frac{2Q}{l}\sum_{n=1}^{\infty}\sin\left(\frac{\pi n}{2}\right)\left(\frac{\pi na}{l}\right)^{-2}\left[1-\exp\left(-\left(\frac{\pi na}{l}\right)^{2}t\right)\right]\sin\left(\frac{\pi nx}{l}\right).\label{eq:13-19}\end{equation}

Because
$$ \sin\left(\f{\pi n}{2}\right) = \left\{ \begin{array}{l}
\displaystyle (-1)^{k}, \quad n=2k+1,\\[2mm]
\vspace{1mm}
\displaystyle 0, \qquad n=2k+2,\end{array}
\right. $$
where $k=0, 1, 2, ...$, the solution (\ref{eq:13-19}) coincides with (\ref{eq:13-txtsumma}).

In the limit $t\rightarrow\infty$, the solution (\ref{eq:13-19}) would tend to
the stationary solution $\omega(x)$, determined by (\ref{eq:13-razlogw}).


\section*{14. Heat conduction equation 
with nonlinear add-on}
\addcontentsline{toc}{section}{14. HEAT CONDUCTION EQUATION 
WITH NONLINEAR ADD-ON}

\setcounter{equation}0

\parbox{0.71\textwidth}{Burgers' equation is a fundamental partial differential equation from fluid mechanics and other areas of applied mathematics. It bears the name of the Dutch physicist Johannes Martinus Burgers (1895\,--\,1981). For a given velocity of a fluid $u$ and its viscosity coefficient $\nu$, the general form of Burgers' equation has the following form: $v_t + vv_x =\nu v_{xx}$.}
\hspace{2mm}
\parbox{0.24\textwidth}{\includegraphics[scale=0.41]{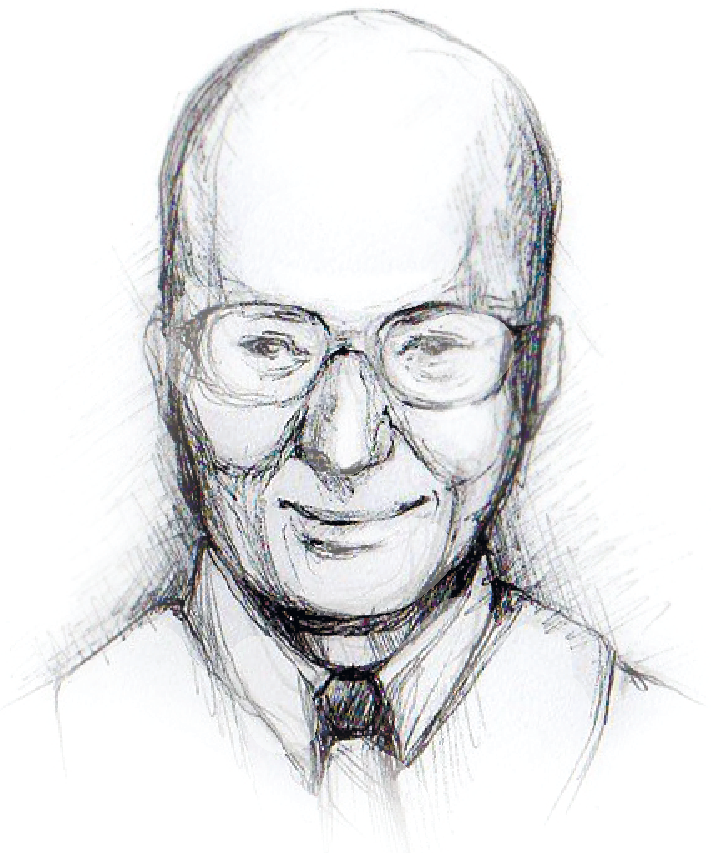}
}\\

Show that it could be linearized by
substitution
$$v=-2\nu\frac{\partial}{\partial x}\ln f,$$
and reduced to the heat conductivity equation $f_t=\nu
f_{xx}$.\\


{\bf SOLUTION}

\setcounter{equation}0

Substitute the
expression
\beq v=-2\nu f_x/f \label{eq:e14p1}\eeq
into the Burgers equation.
The result of such substitution is that the all derivatives in the Burgers equation obtain the form:
\beq \begin{array}{c} v_t=-2\nu \displaystyle\frac{f_{xt}}{f}+2\nu \frac{f_xf_t}{f^2},\, ~ \quad v_x=-2\nu \frac{f_{xx}}{f}+2\nu \frac{f_x^2}{f^2},\,\\[3mm]
v_{xx}=-2\nu \displaystyle\frac{f_{xxx}}{f}+6\nu \frac{f_{xx}f_x}{f^2}-4\nu
\frac{f_x^3}{f^3}.\end{array}\label{eq:e14p2}\eeq

Substituting the expressions
(\ref{eq:e14p2}) into the Burgers equation, we obtain
$$-\,\frac{f_{xt}}{f}+\frac{f_xf_t}{f^2}=\nu
\Bigl(-\,\frac{f_{xxx}}{f}+\frac{f_{xx}f_x}{f^2}\Bigr).$$

The obtained equation can be transformed in the following way:
$$\frac{\partial}{\partial x} \Bigl(\frac{f_t}{f}\Bigr)=\nu \frac{\partial}{\partial
x}\Bigl(\frac{f_{xx}}{f}\Bigr).$$

Then, for $f$ we obtain almost the equation of heat conductivity (or diffusion):
$$f(x,t)_t=\nu f(x,t)_{xx} + {\rm F}(t)f(x,t),$$
where ${\rm F}(t)$ is an arbitrary time function. If ${\rm F}(t)=0$, we really obtain the heat conductivity (or diffusion) equation.\\

\underline{Short reference}.
Suppose that in a certain region of space all particles are moving along straight lines parallel to the $X$ axis.

Let us designate $v=dx/dt$ -- the projection of the medium velocity
(being the function of the coordinate of the point $x$ and time $t$) on the $X$ axis.
The equation of free one-dimensional motion of incompressible fluid
is written in the form:
\begin{equation}v_t+vv_x=0\label{etsirova3}\end{equation}
and, as seen, is non-linear. It has a solution in the form of traveling waves
the front of which is becoming more steep with time
and as a result the wave breaks.
There are many examples of breaking waves from which perhaps the most
visual would be formation of the white caps on the sea surface at strong acceleration of the waves by the wind.

Of course, waves breaking does not always take place.
There are some existing factors that stop process of steeping wave fronts.

One of such factors is viscosity.
If we add the viscosity term to the equation (\ref{etsirova3}) then we obtain the Burgers equation
$$v_t+vv_x=\nu v_{xx}.$$
Here, $\nu $ is the viscosity factor. Within this model, it is possible to
describe the waves in which the competition takes place between the two opposite processes,
steeping wave fronts due to non-linearity and quenching due to viscosity.
As a consequence of such competition, stationary motion can appear.

The point of interest of the Burgers equation is the existence of exact solution built by
Hopf~\cite{hopf}
and Cole~\cite{cole}.
Transformation leading to linearization of the Burgers equation
(recalled in the statement of this problem) is called in literature as Cole--Hopf transformation.\\[10mm]

\begin{center}
\includegraphics[width=11cm]{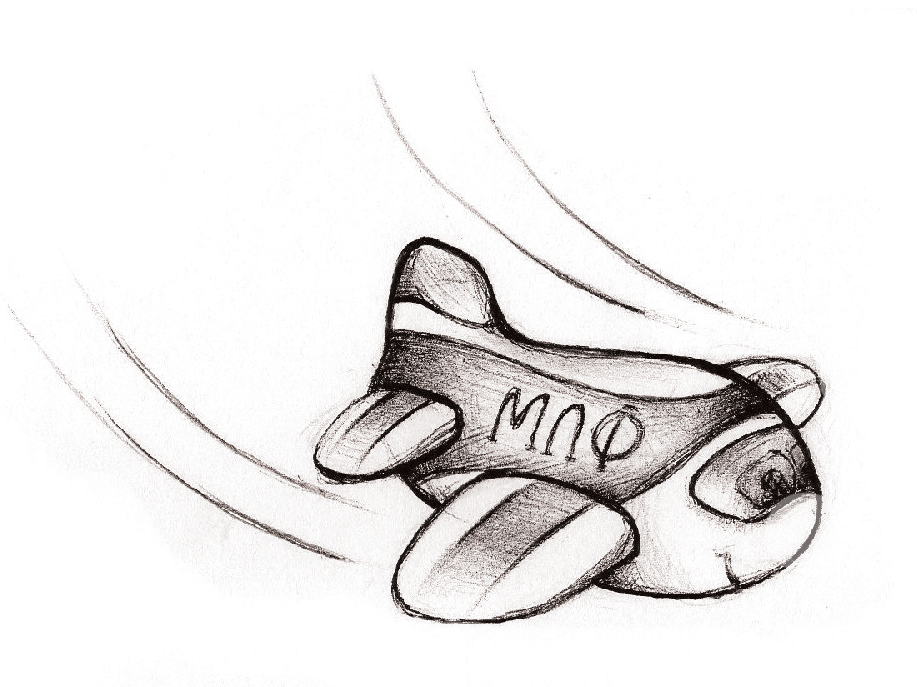}\\[1mm]
\end{center}


%
































\def\bibname







\newpage
{\normalsize




\newpage

$\,$

\vspace{10mm}

\language=0

\begin{center}
{\bf DATA ON AUTHORS}
\end{center}
\addcontentsline{toc}{chapter}{{\itshape Data on authors}}

\fancyhead[LO]{\footnotesize\it \center Mathematical Physics: ~
Data on authors}
\fancyhead[RE]{\footnotesize
\it \center Modern Problems of Mathematical Physics. \,
Special Issue № 3}

\vspace{4mm}


{\bf George Sergeyevitch Beloglazov} -- candidate of physical and mathematical sci\-en\-ces, associate professor, The University of Dodoma - UDOM, Tanzania;
Perm State Pharma\-ce\-u\-tical Academy.

{\bf Bobrick Alexey Leonidovich} -- post-graduate student, Faculty of Science,
Department of Astronomy and Theoretical Physics, Lund University, Sweden.

{\bf Chervon Sergey Viktorovich} -- doctor of physical and mathematical sciences, professor, Ulyanovsk State Pedagogical University.

{\bf Danilyuk Boris Vasilievich} -- Senior Lecturer, Samara State Univer\-sity.

{\bf Dolgopolov Mikhail Vyacheslavovich} -- candidate of physical and mathematical sci\-en\-ces,
associate professor, manager of Samara State University mathe\-ma\-tical physics research Laboratory.

{\bf Ivanov Mikhail
Gennadievich} -- candidate of physical and mathematical sci\-en\-ces,
associate professor, Moscow Institute of Physics and Technology (State University).




{\bf Panina Olga Gennadievna} -- assistant, Samara State Aerospace University.

{\bf Petrova Elena Yurevna} -- student, Samara State University.

{\bf Rodionova Irina Nikolaevna}  --  candidate of physical and mathematical sci\-en\-ces,
associate professor, Samara State University.

{\bf Rykova Elza Nurovna}  -- candidate of physical and mathematical sci\-en\-ces, Senior Lecturer, Samara State University.


{\bf Shalaginov Mikhail Yuryevich} -- post-graduate student, Purdue University, US.

{\bf Tsirova Irina Semyonovna} -- candidate of physical and mathematical sci\-en\-ces,
associate professor, Samara State University.

{\bf Volovich Igor
Vasilievich} -- corresponding member of the Russian Academy of Sciences, head of mathematical physics department of Steklov Mathematical institute of RAS, scientific supervisor of Samara State University mathe\-ma\-tical physics rese\-arch Laboratory.

{\bf Zubarev Alexander Petrovich} -- candidate of physical and mathematical sci\-en\-ces,
leading research fellow
of Samara State University mathe\-ma\-tical physics research Laboratory.

\newpage

\begin{center}
\end{center}



\section*{\phantom{yyyyyyyyyyyyyyyyyyyyyyyyyyyyyyyyyyyyyyyyyyyyy}
{\itshape Annex
}\\[9mm]
\center Statements of the Problems of the Second International Olympiad\\
on Mathematical and Theoretical Physics\\ «Mathematical Physics» \\ {\small September, 4 -- 17, 2010} }
\addcontentsline{toc}{chapter}{{\it Annex:
} Statements of the Problems of the Second International Olympiad}

\vspace{10mm}

\begin{center}
\includegraphics[width=4.2cm]{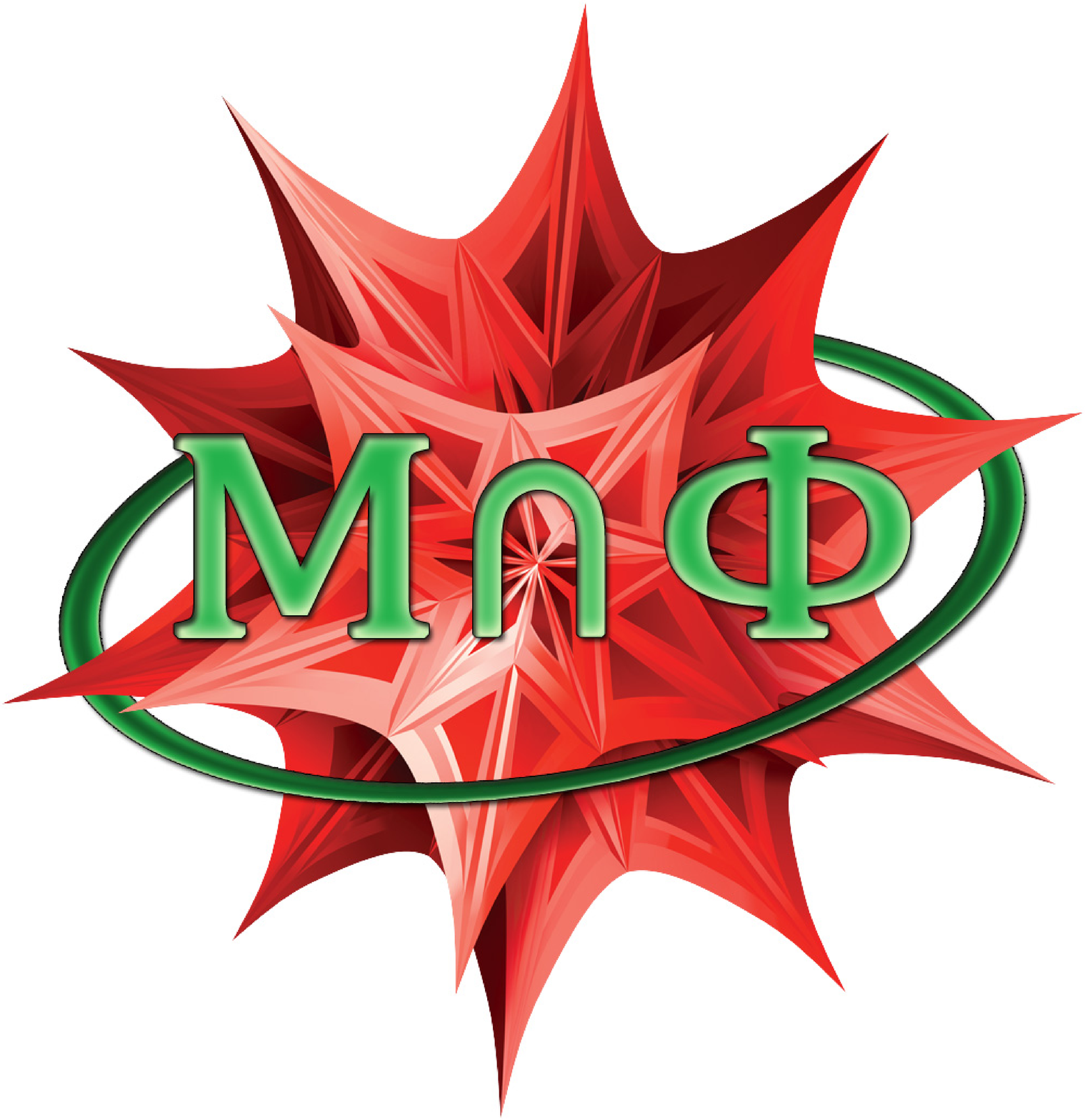}
\end{center}

\begin{flushright}
www.labmathphys.samsu.ru/eng
\end{flushright}

\language=0

\fancyhead[LO]{\footnotesize\it \center Mathematical Physics:
~ Annex 
}
\fancyhead[RE]{\footnotesize\it
\center Series <<Modern Problems in Mathematical Physics>>. Special Issue № 3}

\vspace{7mm}

\setcounter{equation}0

\begin{enumerate}

\item

{\bf "Linear-Nonlinear" \,response}.
Non-linear Burgers equation $v_t+vv_x=\nu v_{xx}$ can be linearized using
Coal--Hopf transformation
\begin{equation}
v=-2\nu\frac{\partial}{\partial x}\ln f. \label{tsiteplo1}
\end{equation}
Here, $v(x,t)$ is the solution of Burgers equation, $f(x,t)$ --
solution of the heat conduction equation $f_t=\nu f_{xx}.$

For the Burgers equation, the initial condition is given as:
$$v(x,0)\bigr|_{t=0}=v_0(x),\quad \int\limits
_{-\infty}^{\infty}v_0(x)dx<\infty.$$

\textbf{(a)}
Using the transformation (\ref{tsiteplo1}) for $v_0(x)$, find the
corresponding function $f_0(x)$
initial condition for the heat
conduction equation.

\textbf{(b)}
The general solution
to the Cauchy problem
for the heat
conduction equation is known:
\begin{equation}
f(x,t)=\frac{1}{\sqrt{4\pi \nu t}}\int \limits _{-\infty}^{\infty
}f_0(y)\exp \Biggl[-\frac{(x-y)^2}{4\nu t}\Biggr]dy,\quad x,y\in
R^1,\quad t\geq 0.\label{tsiteplo2}
\end{equation}

Using transformation (\ref{tsiteplo1}), obtain the solution
$v(x,t)$ of the Burgers equation.

\vspace{5mm}

\underline{Hints}

1. To answer the question \textbf{(b)}
use the result obtained in \textbf{(a)} for the present problem.

2. It is convenient to express the answer to \textbf{(b)} using the function
$$\psi (x,t;y)=\int \limits _0^yv_0(x')dx'+\frac{1}{2t}(x-y)^2.$$

\textbf{(c)}
For Burgers equation, the initial condition is given:
$$v(x,0)
=v_0(x)=\frac{1}{1+(x-5)^2}+\frac{1}{1+(x+5)^2}.$$
Use for this case the solution scheme developed above in paragraphs \textbf{(a)}
and~\textbf{(b)}, find system's response $v(x,t)$.  Follow and analyse
evolution of the obtained solution in time. Utilize the 'Mathematica' package.

\item

{\bf Harmony of a flute}.

\parbox{0.9\textwidth}{ \textit{In woodwind and brass musical instruments, the source of the sound is
 the oscillating column of air. In~a~pipe, the standing waves emerge. Such vibrations occur at
 certain eigen frequencies.}
  }

\vspace{5mm}

 Oscillations of pressure in a pipe
of
length
$L$ are described by the wave equation
$$\frac{\partial ^2p}{\partial x^2}=\frac{\rho _0}{\beta}\frac{\partial ^2p}{\partial
t^2},$$
where $p$ is the overpressure (relative to
the atmospheric), $\rho
_0$
-- density of
air in the pipe, $\beta $   -- modulus of volume elasticity,
$x$
--
coordinate along the pipe axis (see~fig.\,1), $t$ is time.

A specific solution of that equation is the function
$$ p(x,t)=(A\cos kx+B\sin kx) \cos \omega t.$$

\textbf{(a)} Find the values of $A,\ B,\ k,\ \omega $ when both ends of
the pipe are open,
and also the condition
$$ p(x=L/2, t=0)=p_0>0$$
is met.

\begin{center}
\includegraphics{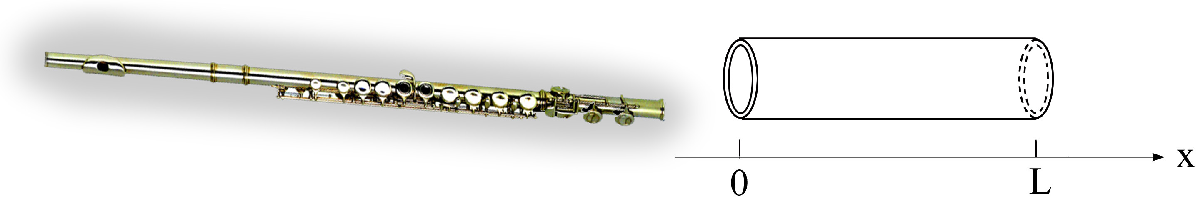}\vspace{-7mm}    \center{\small Fig. 1}
\end{center}

\textbf{(b)} On the basis of the solution obtained, analyse the time evolution of
the~gas pressure $p(x,t)$ in the pipe. For
this purpose, you are encouraged to use the~graphics features of the
'Mathematica' software package.

For your information:\\
1) air density
under normal conditions is $\rho =1,29$ kg/
m$^3$, and the modulus of volume elasticity is $\beta =1,01\times 10^5$Pa;\\
2) the length of a flute may vary widely, so
for
illustration, it is possible to choose $L$=0.5 m.

\textbf{(с)} Illustrate the obtained solution using sound synthesis features of
the 'Mathematica' software package. Stipulate an opportunity to hear the
fundamental tone and some overtones of
the pipe of variable length. How
would the tone of
the pipe depend on the following parameters: $\rho_0, \beta , L$?

\item


\textbf{From the history of LHC: LEP}.

\vspace{1mm}

\parbox{0.9\textwidth}{ \textit{At the end of the XXth century, the colliding beams experiments on electron-positron accelerator have been held
in CERN. Such collider is known as the LEP-collider (Large Electron-Positron).
The detectors
(see figure)
recording collisions of particles with anti-particles were placed at the intersections of the colliding beams.}}

\parbox[b]{0.45\textwidth}{
\begin{center}
\includegraphics{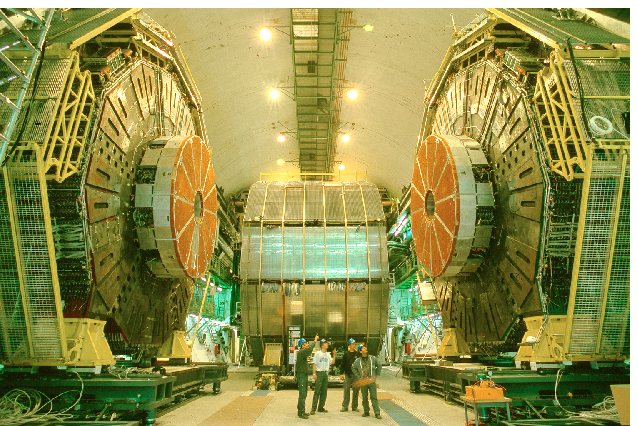}\\
\textit{Mounting
the ALEPH detector} \end{center}\vspace{-8mm}    \center{\small Fig. 2}} \hfill
\parbox[b]{0.45\textwidth}{
\begin{center}
\includegraphics{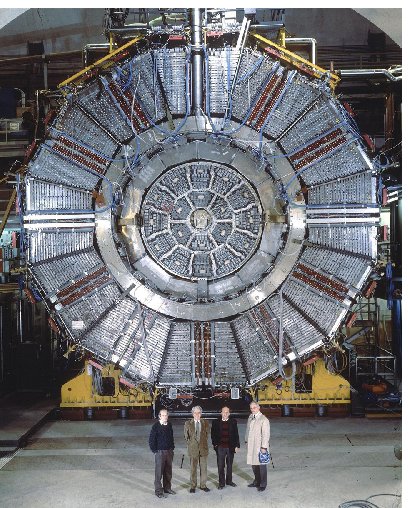}\\
\textit{ALEPH, end view}\end{center}\vspace{-8mm}    \center{\small Fig. 3}}

\textit{To study the collision pattern, it is required not only to
find out which particles
are born but also to measure their characteristics with high precision,
reconstruct the particles' trajectories, find out
their momenta
and energies.}

\begin{center}

\parbox[c]{0.40\textwidth}{\includegraphics{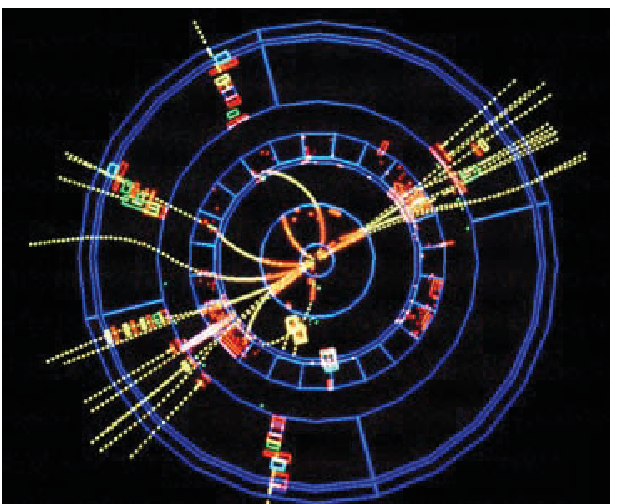}\\
\textit{{\small Fig. 4.} Curvature of the particles tracks} \textit{in a magnetic field}}
\hspace{2mm}
\parbox[c]{0.5\textwidth}{\textit{Such measurements are held with the aid of various types of detectors
that coaxially surround the place of the collision of the
particles. In the area of magnetic field, curvature of a trajectory (see~fig.\,4) enables to
find out the momenta
of the products of a
reaction.}}
\end{center}

\textbf{(a)}\\
\parbox[c]{0.5\textwidth}{
On the figure, the event of the birth of a~neutral
$K^0$-- meson (kaon) is shown. The~length of its trajectory is 0.1542208\,m, the momentum equals
$1.197206\cdot10^{-18}$ kg\,m/s (or 2.240160 GeV/$c$),
the speed of meson is $0.976200c$, where $c$ is the speed of light in vacuum.
Using these data, find out intrinsic lifetime of
$K^0$-- meson, its total and kinetic energies (in GeV).}\hfill
\parbox{0.45\textwidth}{\centering\includegraphics{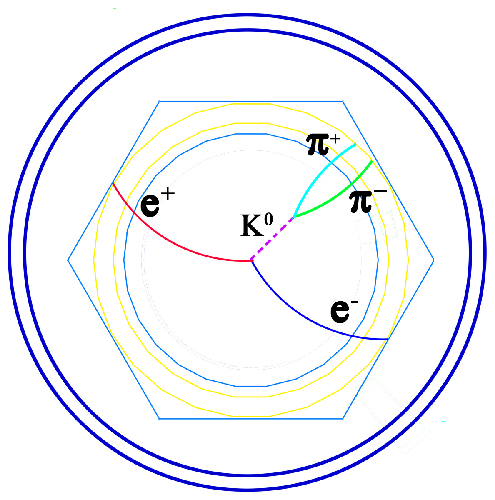}\vspace{-5mm}    \center{\small Fig. 5}
} ~~~~~~~~~~~~~

\textbf{(b)}\\
In a magnetic field with induction $B=1.52$ T,
$K^0$-- meson
decays into $\pi^+$ and $\pi^-$-- mesons with momenta
$5.143114\cdot 10^{-19}$kg m/s and $7.027504\cdot 10^{-19}$kg m/s,
respectively. Analyse maximum possible value
of radii of
the circles of lateral motion (with respect to $\vec B$)
of $\pi ^{\pm}$-mesons. Also, find the angle of their
divergence.
The elementary charge $e_0=1.6\cdot 10^{-19}$ Clmb.\\


\item

{\bf Virial of gravitational
collapse}.
\textit{In classical mechanics of systems executing finite motion, the following
relationship takes place:}
\begin{equation}
\langle K\rangle =-\frac{1}{2}\Big\langle \sum\limits_i\vec F_i
\cdot \vec r_i\Big\rangle .\label{virialtsi}
\end{equation}
 \textit{Here, $\langle K\rangle $ is the mean (for sufficiently long time interval) kinetic energy of the system of point particles
defined by radius vectors
$\vec r_i$ and exposed to the action of
the forces $\vec F_i$.}


 \vspace{2mm}

\noindent\parbox[c]{0.5\textwidth}{ A planet
revolves around the Sun.
Interaction between the planet and the Sun obeys
the law of universal
gravitation. The mass of the Sun is much
larger than the mass of the planet so
the heliocentric reference frame can be considered inertial.

\vspace{2mm}

\textbf{(a)}

Obtain the relationship between the mean kinetic
$\langle K\rangle $ and mean potential $\langle U\rangle $ energies
of the planet directly from the Virial Theorem
(\ref{virialtsi}).}\hfill
\parbox[c]{0.5\textwidth}{
\begin{center}
\includegraphics{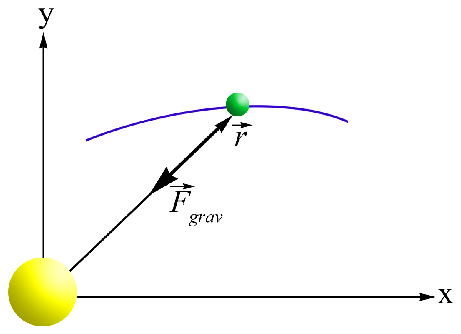}\\
Fig.\,6 to the problem 4) \textbf{(a)}
\end{center}}

\vspace{2mm}

\noindent\parbox[c]{0.5\textwidth}{\textbf{(b)}

A planet
revolves around the Sun along the~circular orbit of the~radius $R$.
Show that the~kinetic $K$ and potential $U$ energies of
the~planet on its circular orbit are related
in the~following way
\begin{equation}K=-\frac{1}{2}U.\label{tsi41}\end{equation}

\vspace{1mm}

\textbf{(c)}

A planet
revolves around the Sun along the~circular orbit of the radius $R$.
If the mass of the Sun would
instantly diminish by 2 times, what will be the
trajectory of the~planet? What relationship would be given by the Virial Theorem in this case?}
\parbox[c]{0.5\textwidth}{
\begin{center}
\includegraphics{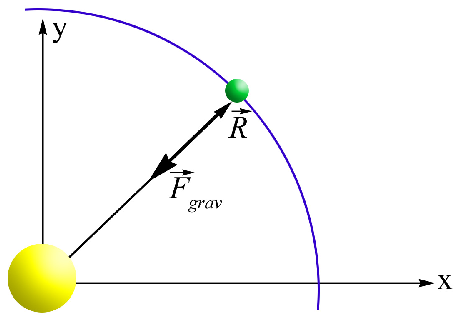}\\
Fig.\,7 to problems 4) \textbf{(b)}, \textbf{(c)}
\end{center}}

\vspace{5mm}

\item

\textbf{Waves on Moebius
strip}.
A Moebius
strip is a rectangular block $0\leq x\leq a$,
$0\leq y\leq b$,
where points
with coordinates
$(0,y)$ and $(a,b-y)$ are glued together~(see fig.).

For sufficiently large ratio $a/b$, the Moebius
strip can be implemented nearly without
stretching as a surface with an edge in
three-dimensional space.

Let the oscillations of the surface of Moebius
strip be described by the wave equation
for the function $u(x,y,t)$
$$
  u_{tt}-\triangle u=0.
$$

The edge of the Moebius
strip is free, and hence Neuman's boundary condition is set (see fig.\,8)
$$
  u_y(x,0,t)=u_y(x,b,t)=0.
$$

\begin{center}
\includegraphics[width=0.4\linewidth]{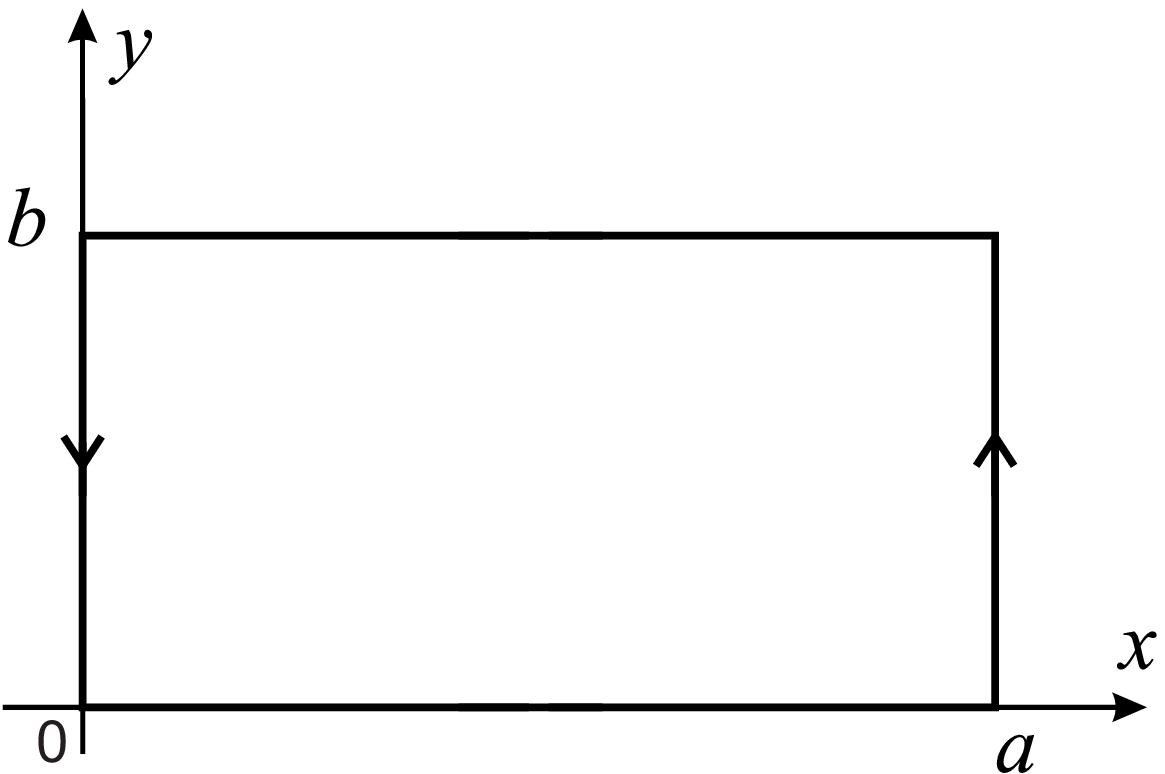}\vspace{-5mm}    \center{\small Fig. 8}
\end{center}

1) State boundary conditions on the gluing line (points with the
coordinates $(0,y)$ and $(a,y)$, $y\in[0,b]$) corresponding to the
longitudinal vibrations ($u$ is a small displacement along the
surface).
Find eigen harmonic oscillations as the solutions of the wave equation with
corresponding boundary conditions.

2) State boundary conditions on the gluing line (points with the coordinates
$(0,y)$ and $(a,y)$, $y\in[0,b]$) corresponding to the
transverse vibrations ($u$ is a~small displacement perpendicular to the
surface). Find eigen harmonic oscillations as the solutions of the wave equation with
corresponding boundary conditions.

\item

{\bf Collapse of a bubble}. Smooth 2-dimensional surface without self-intersections in 3-dimensional space
is topologically equivalent to a sphere. At the initial time moment, the surface
bounds the volume $V$. Points of the surface
are moving with normally oriented variable
velocities. At
each time moment, the projection of the
velocity on the internal normal
equals the Gauss curvature (product of the two main curvatures)
of the surface. Let the surface remain smooth during the process of the motion,
self intersections do not occur. At certain time moment,
the surface collapses into a point. What time will it take
the surface to collapse into a point?

\item

{\bf Random problem}.
Let $\xi_1$ and $\xi_2$ be positive random variables on probability space
$\{\Omega,\cal{F},\mathsf{P}\}$, and such that for all real $p\in[a,b]$, $0<a<b$,
$$
\mathsf{E}\,\xi_1^{p}=\mathsf{E}\,\xi_2^{p}<\infty.
$$

$\mathsf{E}$ is
denoted
as the operator of mathematical expectation value:
$$
\mathsf{E}\xi^p:=\int\limits_\Omega \xi^p\, d\mathsf{P}=\int\limits_{-\infty}^{+\infty} x^p\,dF_\xi(x).
$$

Prove that their distribution functions coincide:
$$
F_{\xi_1}(x)=\mathsf{P}\{\xi_1\leqslant x\}=\mathsf{P}\{\xi_2\leqslant x\}=F_{\xi_2}(x)\mbox{ for all }x\in\mathbb {R}.
$$

\item
{\bf Maximal domain for a matrix}.
Find maximal domain in which the Cauchy problem
\begin{equation} U(x,t)|_{t=0}=T(x), \frac{\partial U}{\partial t}|_{t=0}=N(x),
\end{equation}for the system of equations
\begin{equation}U_{tt}-AU_{xx}=0,\end{equation} with the matrix $A=\left(
                                                             \begin{array}{cc}
                                                               2 & 2 \\
                                                               1 & 3 \\
                                                             \end{array}
                                                           \right),$
                                                           $U=\left(
                                                              \begin{array}{c}
                                                                u_{1} \\
                                                                u_{2} \\
                                                              \end{array}
                                                            \right)
                                                           $
has a unique solution for any $x\in(0,1).$

\item
{\bf Abel's Analogue}.
In 1823 Abel has been working on the generalization of the
Tautochrone Problem
(to find a curve
along which
a heavy particle moving without friction
would reach its lowest position for the same time independing on
its initial position). Abel has
reached
to the equation
\beq
\int\limits_{0}^{x}\frac{f(t)dt}{\sqrt{x-t}}=\varphi(x),\label{eq:abelen}
\eeq
where $f(x)$ is the decision function, $\varphi(x)$ -- given function.

Solution of the equation has the form
$$f(x)=
\frac{1}{\pi}\frac{d}{dx}\int\limits_{0}^{x}\frac{\varphi(t)dt}{\sqrt{x-t}}.$$

In the present problem, it is offered to find a solution of the trigonometric
analogue
of the equation (\ref{eq:abelen})
\begin{equation} \varphi (x)=
\int\limits_{0}^{x}\frac{f(t)dt}{\sqrt{\sin(x-t)}},
0<t<x<\frac{\pi}{2}, \end{equation} where
$\displaystyle \varphi(x)=\frac{1}{\sqrt{\cos x}}.$

\item
{\bf Problem of
energy decomposition}.
Let $u(x,t)\in C^{2}(R\times[0,\infty))$ be a solution of the Cauchy initial value problem
for one-dimensional wave equation
$$u_{tt}-a^{2}u_{xx}=0$$ in $R\times(0,\infty)$, with initial conditions
$$u(x,0)=g(x), ~ u_{t}(x,0)=h(x),$$
where $g(x), h(x)$ are finite functions.

Kinetic energy
$\displaystyle K(t)=\frac{1}{2}\int\limits_{-\infty}^{+\infty}u^{2}_{t}(x,t)dx.$

Potential energy
$\displaystyle P(t)=\frac{1}{2}\int\limits_{-\infty}^{+\infty}u^{2}_{x}(x,t)dx.$

Prove that

a) $K(t)+P(t)a^{2}$ is constant for any $t.$

b) $K(t)=P(t)a^{2}$ for rather large $t.$

\item
{\bf Dirac Problem}.
When deriving so called <<Dirac equation>> in relativistic
quantum mechanics, Dirac has been driven by an idea of <<square-rooting>>
from a~second order differential operator.

Find out in terms of square operator of the first order:

а) a wave one-dimensional operator;

б) Laplace operator in $R^{2}.$

\item
{\bf Certain process for a wave equation}.

Some process is
simulated by a function $u(x,t),$ that satisfies the
initial conditions $$u(x,0)=\left[\begin{array}{c}
                                           \sin^{2}\pi x, ~ 0\leq x\leq 1, \\
                                           0,  ~ \, ~ x<0 ~ \, \mbox{and} \, ~ x>1,
                                         \end{array}\right. ~
~ ~ \frac{\partial u}{\partial t}(x,0)=0.$$ It is known that
even part of
this function $u^{r}(x,t)$ satisfies the wave equation $$u^{r}_{tt}-a^{2}u^{r}_{xx}=0$$
in half plane $t>0$.
Odd part of this function $u^{n}$ satisfies the wave equation
$$u^{n}_{tt}-b^{2}u^{n}_{xx}=0.$$
Find the distance between $x$-coordinates at which
$u(x,T)$ has
minimal values
 at sufficiently
large $T$.

\item
{\bf Maximal domain and a square}.
For the equation \begin{equation}
u_{xx}+\sqrt{y}u_{xy}=0\end{equation}
find maximal domain area on the $x$\,-\,$y$ plane,
where
$$u(x,x)=\varphi(x),\qquad \frac{\partial u}{\partial
x}-\frac{\partial u}{\partial y}=\psi(x),\qquad 0<x<1.$$ Show
that this domain can be divided into 3 parts by straight linear cuts
from which it is possible to make a square block. What will be the area of such square block?

\item
{\bf Evaluation of the solution of the ultrametric diffusion type of equation with fractional derivative}.
When solving equations of the ultrametric diffusion type (such equations
are related
to describing conformation dynamics of compound systems
such as biomacromolecules), the solutions are often represented
in the form of exponent series. One of such series is presented below:
\[S(t)=\mathop{\sum }\limits_{i=0}^{\infty } a^{-i} E_{\beta } (-b^{-i} t^{\beta } ).\]
Here, $\displaystyle E_{\beta } (z)=\sum _{n=0}^{\infty } \frac{z^{n} }{\Gamma (\beta n+1)}$  is Mittag-Leffler function, $0<\beta \leqslant1$, and
$t$~is time, $S(t)$ -- probability
of finding the system in definite state groups,
$a>$1, $b>$1~-- certain parameters.

Study asymptotic behavior of the function  $S(t)$ at $t\to \infty $ and
find its asymptotic evaluation by
$t$-depending elementary functions.

\end{enumerate}

\newpage
\thispagestyle{empty}
\null
\vspace{1cm}
\begin{center}
\end{center}
\vspace{2cm}
\begin{center}
{\large\bf{
MATHEMATICAL PHYSICS}}\\[5mm]
PROBLEMS AND SOLUTIONS\\[6mm]
The Students Training Contest Olympiad\\
in Mathematical and Theoretical Physics\\
({\small On May 21st -- 24th, 2010})\\[8mm]
\end{center}
\begin{center}
Special Issue
№ 3\\
of the Series of Proceedings <<Modern Problems of Mathematical Physics>>\\

\vspace{13mm}

{\bf Authors}:\\
G.S. Beloglazov, A.L. Bobrick, S.V. Chervon, B.V. Danilyuk,\\ 
M.V.~Dolgopolov, M.G.~Ivanov, O.G. Panina, E.Yu. Petrova, I.N. Rodionova,\\
E.N. Rykova, I.S. Tsirova, M.Y. Shalaginov, I.V. Volovich, A.P. Zubarev

\vspace{11mm}

Title Editing ~ \,T.A. Murzinova\\
Computer Design \,\,M.V. Dolgopolov\\
Art drawings \, \,Jy.A. Novikova\\
Cover Art Design \,\,L.N. Zamamykina
\end{center}
\vspace{1cm}
\begin{center}{\small
Signed for printing: 31.11.2010. Format 70$\times$108/16. \\
Paper offset. The press offset.\\
Cond. sheets 5,95; \, acc.publ. 4,25. \, Type family: {\it Times New Roman}. \\
Circulation: 100 copies. Order № 300.
Samara University Press\\ 1 Academic Pavlov st., Samara, 443011.

Tel. +7 846 334-54-23}
\end{center}




\end{document}